\documentclass[onecolumn,noshowpacs,nofootinbib,11pt]{revtex4-1}

\usepackage{float,xcolor,upgreek,yfonts,tikz}
\usepackage{color}
\usepackage{graphicx}
\usepackage{graphicx,float}\usepackage[all]{xy}
\usepackage{amsmath,upgreek}
 
  \newcommand{\blt}{\textcolor{black}}
   \usepackage{caption}
 \usepackage{booktabs,makecell}
\usepackage{subcaption}
\usepackage{amssymb}

\usepackage{alphabeta}
\usepackage{dcolumn}
\usepackage{textgreek}
\usepackage{multirow} 
\usepackage{units}
\usepackage{overpic}

\newcommand{\beq}{\begin{eqnarray}}
\newcommand{\eeq}{\end{eqnarray}}
\newcommand{\be}{\begin{equation}}
\newcommand{\ee}{\end{equation}}

\newcommand{\bea}{\begin{eqnarray}}
\newcommand{\eea}{\end{eqnarray}}

\newcommand{\ba}{\begin{eqnarray}}
\newcommand{\ea}{\end{eqnarray}}
\usepackage[colorlinks,hyperindex,unicode]{hyperref}
\definecolor{green1}{RGB}{0,128,0} 
\hypersetup{hidelinks,backref=true,pagebackref=true,hyperindex=true,colorlinks=true,breaklinks=true,urlcolor= blue}
\hypersetup{%
  colorlinks = true,
  linkcolor  = blue,
  citecolor = cyan,
}
\usepackage{bookmark,textgreek}
\usepackage{hyperref,color,xcolor}
\hypersetup{hidelinks,hyperindex=true,colorlinks=true,breaklinks=true,urlcolor= blue}
\hypersetup{%
  colorlinks = true,
  linkcolor  = blue
}
\newcommand\orcidroldao{{\href{https://orcid.org/0000-0003-3978-532X}{\orcidicon}}}
\newcommand{\orcidicon}{%
	\begin{tikzpicture}
	\draw[lime, fill=lime] (0,0)
		circle [radius=0.16]
		node[white] {{\fontfamily{qag}\selectfont \tiny ID}};
	\draw[white, fill=white] (-0.0625,0.095)
		circle [radius=0.007];
	\end{tikzpicture}	\hspace{-2mm}
}

\begin{document}
\title{Gravitational decoupling of generalized Horndeski hybrid stars}

\author{Roldao da Rocha\orcidroldao\!\!}
\affiliation{Center of Mathematics, Federal University of ABC, 09210-580, Santo Andr\'e, Brazil.}
\email{roldao.rocha@ufabc.edu.br}
\medbreak
\begin{abstract} 

Gravitational decoupled compact polytropic hybrid stars are here addressed in generalized Horndeski scalar-tensor gravity. Additional physical properties of hybrid stars are scrutinized and discussed in the gravitational decoupling setup.
The asymptotic value of the mass function, the compactness, and the effective radius of gravitational decoupled hybrid stars are studied for both cases of a bosonic and a fermionic prevalent core. These quantities are presented and discussed as functions of Horndeski parameters, the decoupling parameter, the adiabatic index, and the polytropic constant. Important corrections to general relativity and generalized Horndeski scalar-tensor gravity, induced by the gravitational decoupling, comply with available observational data. Particular cases involving white dwarfs, boson stellar configurations, neutron stars, and Einstein--Klein--Gordon solutions, formulated in the gravitational decoupling context, are also scrutinized.  

\end{abstract}



\maketitle
\section{Introduction} 

The direct observation of gravitational waves emitted from mergers consisting of neutron stellar configurations has nowadays comprised one of the most important stages of investigation in physics. 
Two orbiting neutron stellar configurations may spiral towards each other and radiate gravitational waves. Neutron stars mergers can lead to the formation of either a more massive neutron star or form the coalescence of a black hole binary \cite{LIGOScientific:2017vwq,Xue:2019nlf}. 
When gravity is investigated in the strong regime, general relativity and generalizations can be experimentally probed by current observations gleaned mainly at LIGO, eLISA, and the Chandra X-ray Observatory. Such experiments can precisely approach extended models that describe gravity when the remnant coalescent binary black hole system does emit gravitational waves pulses while entering into its final state. 

One of the most successful extensions of general relativity, that describes a myriad of stellar configurations, including self-gravitating compact stars, is the gravitational decoupling of Einstein's effective field equations. Anisotropic stellar configurations are naturally addressed by several gravitational decoupling procedures, which is a sharp apparatus that can be used in a wide range to derive new analytical solutions to Einstein's coupled system of field equations \cite{Ovalle:2017fgl,Ovalle:2019qyi,Ovalle:2013xla}. 
Although the gravitational decoupling method is a spontaneous extension of the minimal geometrical deformation \cite{Casadio:2015jva,daRocha:2017cxu,Casadio:2016aum,Ovalle:2018ans,Casadio:2017sze,Fernandes-Silva:2019fez,Ovalle:2013vna,Fernandes-Silva:2018abr,Ferreira-Martins:2019svk}, which is also an important setup to study stars and black holes on fluid branes \cite{Antoniadis:1998ig,Antoniadis:1990ew,daRocha:2012pt}, gravitational decoupled methods do not necessarily depend on any higher-dimensional background, consisting of a genuine and well-established 4-dimensional framework. The gravitational decoupling was comprehensively applied to the study and analysis of several compact stellar configurations, based on an appropriate modification of the stress-energy-momentum tensor that enters Einstein's field equations, beyond the Schwarzschild term \cite{ovalle2007,Casadio:2012rf,Ovalle:2018vmg,Ovalle:2018gic}. Employing the gravitational decoupling of kernel solutions of Einstein's field equations, gravitational field sources can be split into a solution in general relativity and an associate source, which encodes any additional interaction in the theory, including gauge and tidal charges, hairy physical
fields, and extended models of gravity. 
This approach has been generating new solutions that describe a comprehensive list of stellar configurations and include particular cases of coalescing binary neutron stars and black holes \cite{Ovalle:2007bn,Ovalle:2018umz,Estrada:2019aeh,Gabbanelli:2019txr,Pant:2021eit,daRocha:2020rda}, whose acoustic analogs were scrutinized  
 \cite{daRocha:2017lqj,Fernandes-Silva:2017nec}. 
The gravitational decoupling was also used to study gravity coupled to hidden gauge fields \cite{daRocha:2020jdj} and to derive new physical features of superfluid stars \cite{daRocha:2021aww}. Other models involving the gravitational decoupling have been developed \cite{Maurya:2021zvb,Maurya:2021mqx,Singh:2021iwv,Maurya:2021tca,Maurya:2021sju,Jasim:2021kga,Maurya:2021aio,Singh:2020bdv,Maurya:2017jdo,Cavalcanti:2016mbe,Ramos:2021drk,Casadio:2019usg,Rincon:2019jal,Ovalle:2019lbs,Tello-Ortiz:2019gcl,Morales:2018urp,Panotopoulos:2018law,Singh:2019ktp,Maurya:2020djz}, with relevant applications to recently observed anisotropic stellar configurations \cite{Ovalle:2017wqi,Gabbanelli:2018bhs,PerezGraterol:2018eut,Heras:2018cpz,Torres:2019mee,Hensh:2019rtb,Contreras:2019iwm,Stelea:2018cgm,Casadio:2015gea,Maurya:2019hds,Deb:2018ccw,Tello-Ortiz:2021kxg,Zubair:2021zqs,Sharif:2020vvk,Muneer:2021lfz,Contreras:2019mhf,Sharif:2019mzv,Casadio:2013uma}. Also, the quantum entanglement entropy of anisotropic gravitational decoupled black holes revealed important directions towards placing the gravitational decoupling method in gauge/gravity duality \cite{daRocha:2019pla,daRocha:2020gee}. 
Gravitational decoupled anisotropic quark stars, neutron stars, and dwarf remnants constituted mostly of electron-degenerate matter were proposed and scrutinized in Refs. \cite{Contreras:2021xkf,Sharif:2021goe,Sharif:2020lbt,Maurya:2020ebd,Maurya:2021fuy}. Besides, gravitational decoupled black holes with hair were studied \cite{Ovalle:2020kpd,Ovalle:2021jzf,Contreras:2021yxe}. 
Holographic Weyl anomalies have also imposed new directions to the study of hairy gravitational decoupled black holes \cite{Meert:2020sqv,Casadio:2003jc}.

The main physical incitements to study hybrid stars, as solutions of the gravitational decoupling in generalized Horndeski scalar-tensor gravity, are the observations by LIGO, that also corroborate to previsions by extended modified gravity models for black holes and stellar configurations, complying to astrophysical objects with masses up to $ \sim10^7M_\odot$, where $M_\odot$ denotes the Solar mass. Also, several recent observations put forward general relativity ineffectual to evoke gravitational physics at large cosmological scales \cite{Burikham:2016cwz}. Therefore, modifications of gravity are necessary to describe cosmology. With the emergence of gravitational waves detection, new star solutions in modified gravity have become less elusive to pursue \cite{Clifton:2011jh}. 
Since the observational signature of different astrophysical objects can be detected by gravitational waves, one can address compact stellar distributions generated by scalar fields. They constitute successful models to describe boson stars \cite{Brihaye:2020klz,Jetzer:1991jr}, and can interact with surrounding matter of fermionic origin, amalgamating with it and yielding hybrid stellar configurations \cite{Henriques:2003yr,Parisi:2020qfs,Roque:2021lvr}.

The coalescence of compact stars, such as neutron stars, yields environments with huge curvature and very strong gravity. Hence, investigating gravitational decoupled compact stellar configurations enhances the contrast between results obtained by modified gravity theories and general relativity. 
This work is devoted to investigating gravitational decoupled hybrid stars as solutions beyond Horndeski gravity, as a subclass of Gleyzes--Langlois--Piazza--Vernizzi models of gravity~\cite{Gleyzes:2014dya}. This low-energy effective family of field theories~\cite{Georgi:1993mps,Bernal:2009zy,Babichev:2016jom} contains higher derivative operators that do not appear in appropriate limits of Horndeski gravity, such as the Brans-Dicke model, general relativity with a minimally coupled scalar field, covariant galileons, and Gauss--Bonnet gravity, among others \cite{Kuntz:2017pjd}.
Horndeski gravity is usually introduced as a way of explaining the current accelerating expansion of the Universe, implementing modifications of general relativity due to inflatons, at a large scale. The propagation of primordial gravitational waves in Horndeski theories complies with the GW170817 event. The amplitude of the gravitational waves in Horndeski theory can significantly vary, when compared to general relativity \cite{Nunes:2018zot}, showing a completely different physical signature of neutron stars mergers in Horndeski theories and extensions \cite{Tasinato:2014eka,Gripaios:2004ms}. Investigating black hole solutions in scalar-tensor models also permits to study the strong gravity
regime, wherein the non-perturbative sector contributions can generate substantial deviations from general relativity. Also, singularity and no-hair theorems can be further analyzed in different setups, evincing the possibility of new scenarios eventually specified by additional scalar charges and hair.  

 Scrutinizing modifications of gravity that comply with current observations also leads to a more profound knowledge of general relativity, besides perusing further aspects of gravity itself. For example, the Lovelock theorem states that Einstein's field equations are the only possible second-order Euler-Lagrange equations derived from a 4-dimensional Lagrangian that is solely metric-dependent. Hence, to generalize general relativity, one must circumvent the Lovelock theorem, whose assumptions must be therefore relaxed. The most straightforward way of implementing it consists of considering additional objects, as a scalar field, beyond the metric \cite{Deffayet:2013lga}. It is the procedure that will be chosen in this work, approaching modified gravity
and adding a scalar degree of freedom to the gravitational sector. Endowed with these important motivations and new developments, gravitational decoupled hybrid stellar configurations will be here addressed in a low-energy effective theory of gravity. 
This work is organized in the following way: Sec. \ref{s2} is devoted to the gravitational decoupling method, turning to the case of anisotropic polytropic stellar configurations. Three cases are then presented, regarding physically relevant values of the adiabatic index. They encode isothermal self-gravitating compact spheres of gas, encompassing collisionless systems of stars in globular clusters, neutron stars, white dwarfs, and ultrarelativistic degenerate Fermi gases. 
Sec. \ref{220} is dedicated to presenting a family of solutions in the Gleyzes--Langlois--Piazza--Vernizzi approach that generalizes Horndeski scalar-tensor gravity, in a low-energy effective theory with infrared modifications of the gravitational sector. In this setup, Einstein's effective field equations 
are solved, using the gravitational decoupling procedure. The asymptotic value of the mass function, the compactness, and the effective radius of gravitational decoupled compact hybrid stars are then scrutinized and discussed for hybrid stars with a bosonic and a fermionic dominant core. 
Prominent modifications of inherent features of compact hybrid stars are addressed and analyzed. Additional discussion, perspectives, and conclusions compose Sec. \ref{cppp}.

\section{Gravitational decoupling and polytropic stellar configurations}
\label{s2}
\setcounter{equation}{0}
\par
Einstein's field equations relate geometry to the matter content encoded into the stress-energy-momentum tensor, and can be written as 
\begin{equation}
\label{corr2}
R_{\mu\nu}-\frac{1}{2}Rg_{\mu\nu}
=
-a^2T^{\scalebox{0.65}{$\textsc{total}$}}_{\mu\nu}.
\end{equation}
The stress-energy-momentum tensor can be split into two components \cite{Ovalle:2017fgl},
\begin{equation}
\label{emt}
T^{\scalebox{0.65}{$\textsc{total}$}}_{\mu\nu}
=
T^{\scalebox{0.65}{$\textsc{matter}$}}_{\mu\nu}+\upalpha\,\upzeta_{\mu\nu},
\end{equation}
where
\begin{equation}
\label{perfect}
T^{\scalebox{0.65}{$\textsc{matter}$}}_{\mu \nu }=(\rho +p)\,u_{\mu }\,u_{\nu }-pg_{\mu \nu }
\end{equation}
regards the stress-energy-momentum tensor describing ordinary matter, whose thermodynamic properties are determined by the mass density $\rho$ and the thermodynamic pressure $p$. The 4-velocity field of the fluid is normalized, whereas the symmetric tensor $\upzeta_{\mu\nu}$ stands for any source term due to supplementary interactions, such as any gauge charge, tidal charge, hairy physical
fields, or extended models of gravity \cite{Ovalle:2020kpd}, which will be the case to be addressed here. The source $\upzeta_{\mu\nu}$ coupling to gravity is controlled by a non-perturbative constant parameter $\upalpha$, eventually generating anisotropic models for stellar configurations. 
As the perfect fluid term is governed by the Bianchi identity, also the equation
\begin{equation}
\nabla^\mu\,T^{\scalebox{0.65}{$\textsc{total}$}}_{{\mu\nu}}=0,
\label{con11}
\end{equation}
 must hold, accordingly. A static, spherically symmetric, metric 
\begin{equation}
ds^{2}
=
e^{\upnu (r)}dt^{2}-e^{\uplambda (r)}dr^{2}
-r^{2}\, d\theta^{2}-r^2\,\sin ^{2}\theta\, d\phi ^{2},
\label{metric}
\end{equation}
can be employed, where the fluid velocity has components $u^0(r)=e^{-\upnu(r)/2}$ and $u^i=0$, for $0\leq r\leq R$, where $R$ denotes the radius of the stellar configuration.
The metric~(\ref{metric}) must satisfy Einstein's effective field equations~(\ref{corr2}), which can be thus written as the following system of coupled differential equations, \begin{eqnarray}
\label{ec1}
a^2
\left(\rho(r)+\upalpha\upzeta_0^{\ 0}(r)
\right)
&=&
\frac 1{r^2}
-
e^{-\uplambda(r) }\left( \frac1{r^2}-\frac{\lambda'(r)}r\right),
\\
\label{ec2}
a^2
\left(-p(r)+\upalpha\upzeta_1^{\ 1}(r)\right)
&=&
\frac 1{r^2}
-
e^{-\uplambda(r) }\left( \frac 1{r^2}+\frac{\upnu'(r)}r\right),
\\
\label{ec3}
a^2
\left(-p(r)+\upalpha\upzeta_2^{\ 2}(r)\right)
&=&
\frac {e^{-\uplambda(r) }}{4}
\left( -2\,\upnu''(r)-\upnu'^2(r)+\uplambda'(r)\upnu'(r)
-\frac{2}r(\upnu'(r)-\uplambda'(r))\right),
\end{eqnarray}
where $f'$ denotes the derivative of a function $f$ with respect to $r$ and spherical symmetry implies that $\upzeta_3^{\ 3}=\upzeta_2^{\ 2}$.
The conservation equation~(\ref{con11}) is a linear combination of Eqs.~(\ref{ec1}) -- (\ref{ec3}), and yields
\begin{equation}
\label{con1}
p'(r)
+
\frac{\upalpha}{2}\left(\rho(r)+p(r)\right)
-
\upalpha\left(\upzeta_1^{\ 1}(r)\right)'
+
\frac{\upalpha}{2}\left(\upzeta_0^{\ 0}(r)-\upzeta_1^{\ 1}(r)\right)
+
\frac{2\,\upalpha}{r}\left(\upzeta_2^{\ 2}(r)-\upzeta_1^{\ 1}(r)\right)
=
0
,
\end{equation}
The perfect fluid scenario can be recovered when the limit $\upalpha\to 0$ is taken into account. The effective density, the effective radial, and tangential pressures are respectively given by  
\blt{\begin{subequations} 
\begin{eqnarray}
{\check\rho}(r)
&=&
\rho(r)
+\upalpha\,\upzeta_0^{\ 0}(r),
\label{efecden}\\
\check{p}_{r}(r)
&=&
p(r)-\upalpha\,\upzeta_1^{\ 1}(r)
,
\label{efecprera}\\
\check{p}_{t}(r)
&=&
p(r)-\upalpha\,\upzeta_2^{\ 2}(r).
\label{efecpretan}
\end{eqnarray}
\end{subequations}}
\blt{\!As the stellar distribution is usually described by a hydrodynamical fluid, the introduction of the source term $\upzeta_{\mu\nu}$ generates anisotropic models for stellar configurations. The effective tangential pressure occurs whenever the direction of a deforming force is parallel to the cross-sectional area, changing the shape of the stellar distribution out of the perfect sphere fluid description. The effective radial pressure acts towards or away from the central axis in the stellar distribution. 
One must also observe that $\check{\rho}$, $\check{p}_r$, and $\check{p}_t$ are well defined at the star 
center and are regular functions, singularity free throughout
the interior of the stellar configuration.
The effective energy density  has to attain positive values throughout the
stellar configuration, whereas at the star center it must be also finite and monotonically decreasing towards the boundary inside the stellar interior, such that $\frac{d\check\rho}{dr}\leq 0$. Also, the effective radial and tangential pressures must
be positive inside the stellar configuration, whereas the gradient of both the effective radial and tangential pressures  are negative
inside the stellar configuration. At the stellar configuration boundary, the effective radial pressure must 
vanish, however, the effective tangential pressure might not be necessarily equal to zero at
the boundary.  
}

The ancillary source then originates anisotropy, as quantified by the coefficient
\blt{\begin{equation}
\label{anisotropy}
\Pi(r)
=
\check{p}_{t}(r)-\check{p}_{r}(r)
=
\upalpha\left(\upzeta_1^{\ 1}(r)-\upzeta_2^{\ 2}(r)\right).
\end{equation} At the stellar configuration center, both the radial and tangential pressures are equal, meaning that the anisotropy has to vanish at the center, yielding Eq. \eqref{anisotropy} to read $\lim_{r\to0}\Pi(r)
=0.$} 
The metric $g_{\mu\nu}$ that solves the complete Einstein's
field equations~\eqref{corr2} and satisfy the gravitational decoupling has only its radial component influenced by the ancillary source $\upzeta_{\mu\nu}$ \cite{Contreras:2021yxe}. 
This metric $g_{\mu\nu}$ can be derived when Einstein's field equations for the perfect fluid
source $T^{\scalebox{0.65}{$\textsc{matter}$}}_{\mu\nu}$,
\begin{equation}
\label{f1}
{G}_{\mu\nu}
=
-a^2\,T^{\scalebox{0.65}{$\textsc{matter}$}}_{\mu\nu}
,
\qquad
\qquad
\nabla_\mu T^{\scalebox{0.65}{$\textsc{matter}$}\mu\nu}=0
,
\end{equation}
is solved. Thereafter, the remaining quasi-Einstein's equations for the additional source $\upzeta_{\mu\nu}$ 
\begin{equation}
\check{G}_{\mu\nu}
=
-a^2\,{\upzeta}_{\mu\nu}
,
\qquad\qquad
\nabla_\mu\upzeta^{\mu\nu}=0
,
\label{declo}
\end{equation}
are solved, where \cite{Ovalle:2017fgl}
\begin{equation}
\check{G}_{\mu}^{\,\,\nu}
=
{G}_{\mu}^{\,\,\nu}+\frac{1}{r^2}\!\left(\updelta_\mu^{\,\,0}\updelta_0^{\,\,\nu}+\updelta_\mu^{\,\,1}\updelta_1^{\,\,\nu}\right).
\end{equation}
Therefore the solution of Einstein's field equations and the conservation law \eqref{f1} can be expressed as 
\begin{equation}
ds^{2}
=
e^{\chi (r)}\,dt^{2}
-
\frac{dr^{2}}{1-\frac{2M(r)}{r}}
-
r^{2} d\theta^{2}-r^2\sin ^{2}\theta \,d\phi ^{2},
\label{pfmetric}
\end{equation}
where  
\begin{equation}
\label{msh}
M(r)
=
a^2\int_0^r \texttt{r}^2\rho(\texttt{r}) d\texttt{r}
\end{equation}
is the Misner--Sharp--Hernandez mass function.
The outcome of the additional source $\upzeta_{\mu\nu}$, that generates the gravitational decoupling, on the perfect fluid solution can be emulated by the radial component in Eq.~(\ref{pfmetric}). The gravitational decoupled solution can be then implemented when Eq.~\eqref{metric} is taken into account, by the mapping $\upnu(r)\mapsto\chi(r)$, yielding 
\begin{eqnarray}
\label{expectg}
e^{-\uplambda(r)}
\mapsto
1-\frac{2M(r)}{r}+\upalpha\,{\mathsf{f}}^\star(r),
\end{eqnarray}
where ${\mathsf{f}}^\star={\mathsf{f}}^\star(r)$ is the decoupling function to be determined from the use of Eqs.~\eqref{declo}, equivalently written as 
\begin{eqnarray}
\label{ec1d}
a^2\,\upzeta_0^{\ 0}(r)
&=&
-\frac{{\mathsf{f}}^{\star'}(r)}{r}-\frac{{\mathsf{f}}^\star(r)}{r^2},
\\
\label{ec2d}
a^2\,\upzeta_1^{\ 1}(r)
&=&
-{\mathsf{f}}^\star(r)\left(\frac{\chi'(r)}{r}+\frac{1}{r^2}\right)
,
\\
\label{ec3d}
a^2\,\upzeta_2^{\ 2}(r)
=
a^2\,\upzeta_3^{\ 3}(r)
&=&
-\frac{{\mathsf{f}}^\star(r)}{4}\left(2\chi''(r)+\frac{2\chi'(r)}{r}+\chi'^2(r)\right)
-\frac{{\mathsf{f}}^{\star'}(r)}{4}\left(\chi'(r)+\frac{2}{r}\right),
\\\left(\upzeta_1^{\ 1}\right)'&=&
\frac{\chi'(r)}{2}\left(\upzeta_0^{\ 0}(r)-\upzeta_1^{\ 1}(r)\right)
+\frac{2}{r}\left(\upzeta_2^{\ 2}(r)-\upzeta_1^{\ 1}(r)\right).
\end{eqnarray}
One can now derive the decoupling function ${\mathsf{f}}^\star$, iteratively taking the Schwarzschild solution.
The gravitationally decoupled metric can be thus expressed as  
\begin{equation}
\label{Schw}
ds^2
=
\left(1-\frac{2M(r)}{r}\right)dt^2
-\left({\strut\displaystyle{1-\frac{2M(r)}{r}+\upalpha\,{\mathsf{f}}^\star(r)}}\right)^{-1}{dr^2}
-r^{2} d\theta^{2}-r^2\sin ^{2}\theta d\phi ^{2}.
\end{equation}
To obtain the gravitational decoupling ${\mathsf{f}}^\star(r)$, the ancillary source can be considered the one describing a polytropic fluid, satisfying the polytropic Lane--Emden equation of state for the radial pressure, 
\begin{equation}
\label{polyt0}
\check{p_r}
=
K\check\rho^\Upgamma,
\end{equation}
with $\Upgamma=1+1/n$, where $n$ is the polytropic index and $K$ is the polytropic constant encoding the temperature and the entropy per nucleon, as well as the star chemical composition. The energy density reads 
\begin{equation}
	\upepsilon(r)=\left(\frac{\check{p}_r(r)}{K}\right)^{1/\Upgamma}+\frac{\check{p}_r(r)}{\Upgamma-1}.\label{eqrel}
\end{equation} 
In the case where heat flow across the stellar configuration is absent, one can identify $\Upgamma$ to the adiabatic index, namely, the fluid heat capacities ratio at constant
pressure and volume.
The polytropic index quantifies the pressure derivative of the bulk modulus, whose inverse is the fluid compressibility.
The higher the polytropic index, the heavier the density distribution is weighted about the star center. 
Substituting $\rho=p=0$ into Eqs.~\eqref{efecden} and \eqref{efecprera} yields 
$
\upalpha\,\upzeta_1^{\ 1}
=
-K
\left(\upalpha\,\upzeta_0^{\ 0}\right)^\Upgamma.$ 
Now, Eqs.~(\ref{ec1d}) and (\ref{ec2d}) can be replaced into it, implying that \cite{Ovalle:2018umz}
\begin{equation}
\label{poly2}
{{\mathsf{f}}^\star}'(r)
+
\frac{{\mathsf{f}}^\star(r)}{r}
-
\frac1K\left(\frac{Ka^2r}{\upalpha}\right)^{1-1/\Upgamma}
\left(
\frac{{\mathsf{f}}^\star(r)}{r-2M}
\right)^{1/\Upgamma}=0.
\end{equation}
The case $\Upgamma=1$ yields Eq.~(\ref{polyt0}) to represent a barotropic equation of state, representing an isothermal self-gravitating compact sphere of gas.
It also describes a system of stars that do not collide with each other, in a globular cluster. Dense star clusters play the role of formation sites of binary black holes, being realistic sources of binary black holes mergers. It is a prominent case to be scrutinized hereon since binary black holes formed in globular
clusters have a variety of distinct properties when compared to coalescing black holes mergers formation on isolated backgrounds. 
The gravitational decoupling deformation for $\Upgamma=1$ reads 
\begin{equation}\label{kkk}
{\mathsf{f}}^\star(r)
=
\left(1-\frac{2M}{r}\right)^{-1/K}
\left(\frac{\ell}{r}\right)^{1+1/K},
\end{equation}
where $\ell>0$ has dimension of length. Hence, the gravitational decoupled radial component is given by \cite{Ovalle:2018umz}
\begin{equation}
\label{poly}
e^{-\uplambda(r)}
=
\left(1-\frac{2M}{r}\right)
\left[1+\upalpha\,\left(\frac{\ell}{r-2M}\right)^{1+1/K}\right],
\end{equation} when $r>2M$.
Asymptotic flatness at $r\to\infty$ demands that $K\le-1$, whose saturation takes place for the standard Schwarzschild metric. Besides, the effective density reads 
\begin{equation} 
a^2{\rho}(r)
=
\frac{\upalpha}{K\,r^2}
\left(\frac{\ell}{r}\right)^{1+1/K}
\left(1-\frac{2M}{r}\right)^{-1-1/K},
\label{100}
\end{equation}
whereas the effective tangential pressure is given by
\begin{eqnarray}
\blt{\check{p}_{t}(r)}
=
-\upalpha\,\theta_2^{\ 2}(r)
&=&
-\frac{\upalpha\,(K+1)}{2\,K\,r^2}\left(1-\frac{M}{r}\right)
\left(\frac{\ell}{r}\right)^{1+1/K}\left(1-\frac{2M}{r}\right)^{-2-1/K}.
\label{10a}
\end{eqnarray}
Besides the case $\Upgamma=1$, other cases are quite important to the study of gravitational decoupled hybrid stars. The case $\Gamma=2$ will be also considered, as the prototypical case wherein neutron stars are successfully described by self-gravitating compact polytropes \cite{Gleiser:2013mga}. This case might also encompass stars made of Bose--Einstein condensates, obeying the polytropic equation of state \eqref{polyt0} with $\Upgamma=2$. The first gravitational waves event GW170817, originated from a binary neutron star merger, contained a comprehensive network of details regarding neutron stellar configurations \cite{LIGOScientific:2017vwq}. 
In particular, their tidal deformability is one of the most auspicious physical quantities that can be obtained from the detection of gravitational waves. 
 The last case to be addressed, $\Gamma=\frac43$, corresponds to the core of stellar core remnants constituted by electron-degenerate matter. Moreover, ultrarelativistic degenerate Fermi gases can be described by this value of the polytropic constant, which can be also employed to study main-sequence stellar distributions as the Sun in the apparatus of the Eddington model describing stellar structures.

\section{Gravitational decoupled hybrid stellar configurations}
\label{220}

The Gleyzes--Langlois--Piazza--Vernizzi approach generalizes Horndeski gravity and represents a comprehensive set of scalar-tensor theories constituted by a linear combination of the Horndeski Lagrangians. The action that governs the Gleyzes--Langlois--Piazza--Vernizzi model is given by \cite{Gleyzes:2014dya}
\begin{equation}
	S=\int d^{4}x \sqrt{-g}\left(\sum_{k=2}^{5} \mathcal{L}_{k}[\upphi,g_{\mu\nu}]+\mathcal{L}_{\scalebox{0.66}{$\textsc{matter}$}}[\psi,g_{\mu\nu}]\right) \, ,
	\label{eq:lag}
\end{equation}
where $\mathcal{L}_{\scalebox{0.66}{$\textsc{matter}$}}$ is the Lagrangian referring to matter and contains standard model fields; the $\mathcal{L}_k$ regard the gravitational sector with contributions due to the scalar field, and can be expressed as
\begin{eqnarray}\label{eq.Lagrangians}
{}	\!\!\!\!\!\!\!\!\!\!\!\!\!\!\!\!\!\!\!\!\!\!\!	\mathcal{L}_{2}&\!=\!& \scalebox{0.86}{$\textsc{G}$}_{2}(\upphi, X),\label{BH1}\\
{}	\!\!\!\!\!\!\!\!\!\!\!\!\!\!\!\!	\!\!\!\!\!\!\!\mathcal{L}_{3}&\!=\!& \scalebox{0.86}{$\textsc{G}$}_{3}(\upphi, X)\Box\upphi,\\
{}	\!\!\!\!\!\!\!\!\!\!\!\!\!\!\!\!\!\!\!\!\!\!\!	\mathcal{L}_{4}&\!=\!& \scalebox{0.86}{$\textsc{G}$}_{4}(\upphi, X)R\!+\!\scalebox{0.86}{$\textsc{F}$}_{4}(\upphi,X){\epsilon^{\mu\nu\rho}}_{\sigma}\epsilon^{\tau\alpha \beta\sigma}(\nabla_\mu\upphi)(\nabla_{\beta}\upphi)(\nabla_{\nu}\nabla_{\alpha}\upphi)(\nabla_{\rho}\nabla_{\beta}\upphi),\nonumber\\
&&\qquad\qquad\qquad\qquad\qquad\qquad-2\scalebox{0.86}{$\textsc{G}$}_{4,X}(\upphi,X)\left[(\Box\upphi)^{2}\!-\!(\nabla_{\mu}\nabla_{\nu}\upphi)(\nabla^{\mu}\nabla^{\nu}\upphi)\right]\label{eq.quartic}\\
{}	\!\!\!\!\!\!\!\!\!\!\!\!\!\!\!\!\!\!	\!\!\!\!\!\mathcal{L}_{5}&\!=\!& \scalebox{0.86}{$\textsc{G}$}_{5}(\upphi, X)\left(R_{\mu\nu}\!-\!\frac{1}{2}Rg_{\mu\nu}\right)\nabla^\mu\nabla^\nu\upphi\!\nonumber\\
{}		&&\quad\quad\!\!\!\!+\,\scalebox{0.86}{$\textsc{F}$}_{5}(\upphi,X)\epsilon^{\mu\nu\rho\sigma}\epsilon^{\tau\alpha\beta\zeta}(\nabla_\mu\upphi)(\nabla_\tau\upphi)(\nabla_{\nu}\nabla_{\alpha}\upphi)(\nabla_{\rho}\nabla_{\beta}\upphi)\nabla_{\sigma}\nabla_{\zeta}\upphi\nonumber\\
{}		&&\quad\quad\!\!\!\!+\frac{1}{3}\scalebox{0.86}{$\textsc{G}$}_{5,X}(\upphi,X)\left[(\Box\upphi)^{3}\!-\!3\Box\upphi (\nabla_{\mu}\nabla_{\nu}\upphi)(\nabla^{\mu}\nabla^{\nu}\upphi)\!+\!2(\nabla_{\mu}\nabla_{\nu}\upphi)(\nabla^{\mu}\nabla^{\sigma}\upphi)\nabla^{\nu}\nabla_{\sigma}\upphi\right],\label{BH2}
	\end{eqnarray}
where $R$ is the Ricci scalar curvature, and $\scalebox{0.86}{$\textsc{G}$}_2$, $\scalebox{0.86}{$\textsc{G}$}_3$, $\scalebox{0.86}{$\textsc{G}$}_4$, $\scalebox{0.86}{$\textsc{G}$}_5$, $\scalebox{0.86}{$\textsc{F}$}_4$, $\scalebox{0.86}{$\textsc{F}$}_5$ are arbitrary functions depending on the scalar field $\upphi$ that generates bosonic scalar matter; the kinetic term, $X= g^{\mu\nu}\nabla_\mu\upphi\nabla_\nu\upphi$, is the leading dimension-four operator that rules the
scalar field dynamics, and $\Box\upphi = g_ {\mu\nu}\nabla^\mu\nabla^\nu\upphi$, whereas $\scalebox{0.86}{$\textsc{G}$}_{i, X}$ denotes the partial derivative with respect to the kinetic term. 
When $\scalebox{0.86}{$\textsc{G}$}_4=M_{\scalebox{0.66}{$\textsc{p}$}}^2/2$, where $M_{\scalebox{0.66}{$\textsc{p}$}}$ denotes the Planck mass, and $\scalebox{0.86}{$\textsc{G}$}_2=-M_{\scalebox{0.66}{$\textsc{p}$}}^2\Uplambda$, for all other functions vanishing, one recovers general relativity with cosmological constant. Also, making $\scalebox{0.86}{$\textsc{G}$}_2=-\frac{1}{2}X-V(\upphi)$ yields the Einstein--Klein--Gordon theory. 

Suitably choosing the $\mathcal{L}_k$, in such a way that low-energy effective gravity sets in, yields \cite{Barranco:2021auj,Brihaye:2016lin}
\begin{eqnarray}\label{lagra}
{}	\mathcal{L}_{\scalebox{0.66}{$\textsc{grav}$}}&=& \frac{1}{2}M_{\scalebox{0.66}{$\textsc{p}$}}^2 R - X -m^2\upphi\bar{\upphi}+ \frac{M_{\scalebox{0.66}{$\textsc{p}$}}}{\Uplambda^3}\left\{c_4\left[XR- 2\Box\upphi\Box\bar\upphi-(\nabla^{\mu}\nabla^{\nu}\upphi)(\nabla_{\mu}\nabla_{\nu}\bar\upphi)\right]\right.\nonumber\\&&\qquad\qquad\qquad\qquad\qquad\left.	+\frac{d_4}{X}{\epsilon^{\mu\nu\rho}}_{\sigma}\epsilon^{\tau\alpha \beta\sigma}(\nabla_\mu\upphi)(\nabla_\tau\bar\upphi)(\nabla_{\nu}\nabla_{\alpha}\upphi)\nabla_{\rho}\nabla_{\beta}\bar\upphi
	\right\}.
\end{eqnarray}
The Lagrangian \eqref{lagra} encodes generalized Horndeski scalar-tensor gravity, containing a subgroup of families that contain infrared modifications of the gravitational sector.
The scalar field $\upphi$ has a mass parameter denoted by $m$. It is worth mentioning that as the $\scalebox{0.86}{$\textsc{F}$}_5$ sector in Eq. (\ref{BH2}) includes operators that have mass dimension at least equal to nine, it is therefore suppressed at low energies. The contributions from $\mathcal{L}_{3}$ and $\mathcal{L}_{5}$ vanish when $\mathbb{Z}_2$ symmetry acting on $\upphi$ is taken into account. 
Hereon, the range $m<\Uplambda\ll M_{\scalebox{0.66}{$\textsc{p}$}}$ will be regarded, such that the Lagrangian \eqref{lagra} holds \cite{Roque:2021lvr}. 

The variation of the action (\ref{lagra}) with respect to the metric yields 
\begin{eqnarray}\label{acc1}
R_{\mu\nu}\!-\!\frac{1}{2}Rg_{\mu\nu}+\frac{c_4}{2M_{\scalebox{0.66}{$\textsc{p}$}}\Uplambda^{3}}\mathsf{H}_{\mu\nu}=\frac{1}{M_{\scalebox{0.66}{$\textsc{p}$}}^2}\left(T_{\mu\nu}^{\scalebox{0.66}{$\textsc{f}$}}-T_{\mu\nu}^{\upphi}\right),
\end{eqnarray}
where the tensor $\mathsf{H}_{\mu\nu}$ encodes the generalized Horndeski model correction to general relativity, having the form \cite{Roque:2021lvr}
\begin{eqnarray}\label{acc2}
\!\!\!\!\!\mathsf{H}_{\mu\nu}&\!=\! &\left(R_{\mu\nu}\!-\!\frac{1}{2}Rg_{\mu\nu}\right)X\!+\!g_{\mu\nu}\left( (\nabla_{\alpha}\nabla_\rho\bar\upphi)\nabla^{\alpha}\nabla^\rho\upphi-\Box\upphi\Box\bar\upphi\!+\!2R_{\alpha\rho}\nabla^\alpha\upphi\nabla^\rho\bar\upphi\right) \!-\!\nabla^\alpha\upphi\nabla^\rho\bar\upphi R_{\mu(\alpha|\nu|\rho)}\nonumber\\
{}	&&+\left[\nabla_\mu \upphi\left( \frac{R}{2}\nabla_\nu\upphi-R_{\nu\alpha}\nabla^\alpha\upphi\right)+(\nabla_\mu\nabla_\nu\bar\upphi)\Box\upphi-R_{\mu\alpha}\nabla_{\nu}\bar\upphi\nabla^\alpha\upphi-(\nabla_\mu\nabla_\alpha\bar\upphi)\nabla_{\nu}\nabla^{\alpha}\upphi\right],
\end{eqnarray}
with correspondent complex conjugate terms, where $T_{\mu\nu}^{\scalebox{0.66}{$\textsc{f}$}}$ and $T_{\mu\nu}^{\upphi}$ denote the energy-momentum tensor of the fermionic and bosonic fields, respectively,
\begin{eqnarray}
	T_{\mu\nu}^{\scalebox{0.66}{$\textsc{f}$}}&=&-\frac{2}{\sqrt{-g}}\frac{\updelta \mathcal{L}_{\scalebox{0.66}{$\textsc{matter}$}}}{\updelta g^{\mu\nu}},\label{acc3}\\
	T_{\mu\nu}^{\upphi}&=&g_{\mu\nu}\left(X+m^2\upphi\bar{\upphi} \right)-\nabla_\mu\upphi\nabla_\nu\bar\upphi-\nabla_\mu\bar\upphi\nabla_\nu{\upphi}.\label{acc4}
\end{eqnarray}  
Varying the action (\ref{lagra}) with respect to the complex conjugate $\bar\upphi$ yields the following equation of motion, 
\begin{equation}\label{field_eq}
	\Box\upphi-m^{2}\upphi+	\frac{2 c_4M_{\scalebox{0.66}{$\textsc{p}$}}}{\Uplambda^{3}}\left(R_{\mu \nu}-{1 \over 2}g_{\mu \nu }R\right)\nabla^\mu\nabla^\nu\upphi=0.
\end{equation}
The case $c_4=0$, or equivalently $\Uplambda\to\infty$, yields the usual case regarding the Klein-Gordon equation. 

The gravitational decoupled metric (\ref{Schw}, \ref{kkk}, \ref{poly}) can be now taken into account, together with the harmonic splitting of the scalar field,
\begin{equation}
	\upphi(t,r)=\upsigma(r)e^{i \omega t} . \label{ansat}
\end{equation}
Therefore the equations of motion~\eqref{acc1} -- \eqref{field_eq} are led to the coupled system of ODEs,  
\begin{eqnarray}
{}	 \frac{(1+\Xi)}{r}\uplambda'e^{-\uplambda}+\frac{(1-\kappa_0)}{M_{\scalebox{0.66}{$\textsc{p}$}}^2}e^{-\uplambda}\upzeta^2-\left[m^2+(1-\upgamma){\omega^2 e^\upnu}\right]\frac{\upsigma^2}{M_{\scalebox{0.66}{$\textsc{p}$}}^2}+\left(1+e^{-\uplambda}\right)\frac{1}{r^2}&=&\frac{T_{00}}{ M_{\scalebox{0.66}{$\textsc{p}$}}^2 }e^{-\upnu},\label{coupled1}\\[0.2cm]  
{}	 \frac{(1+\Xi)}{r}\upnu e^{-\uplambda}-\frac{\left(1-\kappa_1\right)}{M_{\scalebox{0.66}{$\textsc{p}$}}^2}e^{-\uplambda}\upzeta^2+\left[m^2-(1-\upgamma){\omega^2 e^\upnu}\right]\frac{\upsigma^2}{M_{\scalebox{0.66}{$\textsc{p}$}}^2}-\left(1+e^{-\uplambda}\right)\frac{1}{r^2}&=&\frac{T_{11}}{M_{\scalebox{0.66}{$\textsc{p}$}}^2 }e^{-\upnu},\label{coupled2}\\[0.2cm]
{}	 (1+\upxi)\upzeta'+\left[\frac{(1-\upeta)}{2}\left({\upnu'}-\uplambda'\right)+\frac{2(1+\upzeta)}{r}\right]\upzeta-
	e^{\uplambda}\left(m^2-(1+\uptheta){e^{-\upnu}\omega^2}\right)\upsigma&=& 0,\label{coupled3}
\end{eqnarray}
where the dimensionless functions are respectively given by the sets
\begin{subequations}
\begin{eqnarray} 
	\Xi(r)&=&\frac{2 c_4\upsigma^2(r)}{M_{\scalebox{0.66}{$\textsc{p}$}}\Uplambda^3}\left[ {\omega^2e^{-\upnu(r)}}-\frac{3\upsigma^{\prime2}(r)}{\upsigma^2(r)}e^{-\uplambda(r)}\right],\label{fea}\\\label{kappa}
	\kappa_0(r)&=&\frac{2 c_4M_{\scalebox{0.66}{$\textsc{p}$}}}{\Uplambda^3 r^2}\left[1-e^{-\uplambda(r)}+\frac{4r\upsigma''(r)\upnu'(r)}{\upsigma'(r)}\right], \\
	\upgamma(r)&=&\frac{2 c_4M_{\scalebox{0.66}{$\textsc{p}$}}}{\Uplambda^3 r^2}\left[1-e^{-\uplambda(r)}\right],\\
	\kappa_1(r)&=&\frac{2 c_4M_{\scalebox{0.66}{$\textsc{p}$}}}{\Uplambda^3 r^2}\left[1+3e^{-\uplambda(r)}-\frac{4\omega^2\upsigma(r)}{\upsigma'(r)}re^\upnu(r)\right],\label{feb} 	
\end{eqnarray}
\end{subequations}	and 
	\begin{subequations}
\begin{eqnarray} 
	\upxi(r)&=&-\frac{2 c_4M_{\scalebox{0.66}{$\textsc{p}$}}e^{\uplambda}(r)}{\Uplambda^3r^2}\left[1+r\upnu'(r)+e^\uplambda(r)\right], \\
	\upeta(r)&=&-\frac{2 c_4M_{\scalebox{0.66}{$\textsc{p}$}}}{\Uplambda^3 r^2}\left[3+e^{-\uplambda(r)}\right] , \label{fec}\\
	\upzeta(r)&=&-\frac{c_4M_{\scalebox{0.66}{$\textsc{p}$}}}{\Uplambda^3}e^{-\uplambda(r)}\left[\upnu''(r)+ \upnu^{\prime2}(r)-\frac{6\upnu'(r)}{\uplambda'(r)}\right] ,\\
	\uptheta(r)&=&\frac{2 c_4M_{\scalebox{0.66}{$\textsc{p}$}}}{\Uplambda^3 r^2}\left[1+e^{\uplambda(r)}-{r\uplambda'(r)}\right]. \label{fec}
\end{eqnarray}
\end{subequations}
Gravitationally decoupled hybrid stars profiles can be then derived by noticing that the gravitationally decoupled metric is a solution of the coupled system of equations \eqref{coupled1} -- \eqref{coupled3}, with conservation equation $\nabla^{\mu}T_{\mu\nu}=0$ and fermionic matter modelled by Eq. (\ref{perfect}). This description of fermionic matter can be alternatively implemented in the fluid/gravity correspondence \cite{Meert:2018qzk}. The resulting system can be thus solved numerically, with boundary conditions,
\begin{eqnarray}
\!\!\!\!\!\!\!\!\!\!\!\!\!\!\lim_{r\to0}\uplambda(r)= 0,\qquad \lim_{r\to0}\upnu(r)=\upnu_{0},\qquad \lim_{r\to0}p(r)=p_0,\qquad 
\lim_{r\to0}\upsigma(r)= \upsigma_{0},\qquad\lim_{r\to0}\upsigma'(r)=0,\label{coes}
\end{eqnarray}
where $p_0$ is the central fermionic pressure, whereas $\upnu_0$ is related to the lapse function at the center of gravitational decoupled hybrid star configuration, since the only non-null component of the fluid velocity is the temporal one, given by $u^0(r)=e^{-\upnu(r)/2}$. 
\blt{The conditions \eqref{coes} match regular spacetime configurations with no divergent curvature invariants.}

Gravitationally decoupled hybrid stars profiles can be then derived by noticing that the gravitationally decoupled metric is a solution of the coupled system of equations \eqref{coupled1} -- \eqref{coupled3}, with conservation equation $\nabla^{\mu}T_{\mu\nu}=0$ and fermionic matter modeled by Eq. (\ref{perfect}). 
Besides, the asymptotic radial limits must be adopted,
\begin{equation}\label{bounday2}
 \!\!\!\!\!\!\!\!\!\! \!\!\!\!\!\!\!\!\!\! \lim_{r\to\infty}p(r)= 0,\qquad\quad
 \lim_{r\to\infty}\upsigma(r)= 0,\qquad\quad
 \lim_{r\to\infty}\upnu(r)= \upnu_{\infty}\in\mathbb{R}^+,\qquad \quad
 \lim_{r\to\infty}\uplambda(r)= 0,
\end{equation}
where the equality \blt{$\upnu_{\infty} = -\lim_{r\to\infty}\uplambda(r)$} is derived when $\upalpha=0$ in Eq. (\ref{emt}), corresponding to the Schwarzschild solution.

The coupled system of equations (\ref{fea}) -- (\ref{fec}) can be hereon solved by taking into account the polytropic equation (\ref{polyt0}) and by suitable expressing them with respect to dimensionless quantities,  
\begin{eqnarray}
\!\!\!\!\!\!\!\!\!\!\!\!\!\!\!	\tilde{r}= m r, \qquad\tilde{\omega}=\frac{\omega}{m}, \qquad \tilde{\upsigma}=\frac{\upsigma}{M_{\scalebox{0.66}{$\textsc{p}$}}},\qquad \tilde{\Uplambda}=\frac{\Uplambda}{(M_{\scalebox{0.66}{$\textsc{p}$}}m^{2})^{1/3}},\qquad\tilde{\upepsilon}=\frac{\upepsilon}{m^{2}M_{\scalebox{0.66}{$\textsc{p}$}}^2}\qquad\tilde{p}=\frac{p}{m^{2}M_{\scalebox{0.66}{$\textsc{p}$}}^2}. \label{quan}
\end{eqnarray}
As posed in Ref. \cite{Roque:2021lvr}, this rescaling makes the equations of motion~\eqref{acc1} -- \eqref{field_eq} to be independent of $m$ and $M_{\scalebox{0.66}{$\textsc{p}$}}$, also unifying the two energy scales, $m$ and $\Uplambda$, into a single one given by $\tilde{\Uplambda}$.

The profiles of gravitational decoupled hybrid stars can be derived when one solves numerically the coupled system (\ref{fea}) -- (\ref{fec}) under the Neumann and Dirichlet boundary conditions~\eqref{coes} and \eqref{bounday2}, employing the shooting method \cite{daRocha:2020rda,daRocha:2020jdj}. \blt{One must observe that given a 2-tuple $(\upsigma_0, p_0)$, the scalar field ground state must be chosen to circumvent the eventual appearance of nodes in the scalar field profiles. The first step of the numerical analysis is to reduce the second-order differential equations to first-order ones, the number of resulting differential equations matches the boundary conditions \eqref{coes}
 and the asymptotic radial limits \eqref{bounday2}. The shooting method involves finding solutions to the initial value problem for different initial conditions until one finds the solution that also satisfies the boundary conditions of the boundary value problem. The central values in the boundary conditions \eqref{coes} must be consistent with the asymptotic radial limits. The shooting method is then implemented by the multidimensional and globally convergent Newton--Raphson method. Hence one solves the coupled system of differential equations with boundary conditions \eqref{coes}. Thereafter the asymptotic limit $r\to\infty$ to the numerical solutions is calculated and the difference between the asymptotic limit $r\to\infty$ to the numerical solutions and the asymptotic radial limits given by \eqref{bounday2} is minimized \cite{Press:1992zz}. This can be implemented by shooting out trajectories in different directions from the boundaries at $r=0$ until one finds the trajectory that hits the asymptotic conditions at $r\to\infty$. 
The Newton--Raphson method is employed to derive the adjustment of the boundary conditions at $r=0$ that make the discrepancies, between the subsequent values of the respective derived functions at $r\to\infty$ and the chosen asymptotic values \eqref{bounday2} to be less than the numerical acceptable error $\epsilon=10^{-10}$}. One can consider phenomenological values for the polytropic index in Eq. (\ref{polyt0}), chosen to cover a large range of stellar configurations, as discussed at the end of Sec \ref{s2}.
In what follows gravitational decoupled hybrid stellar configurations with either a bosonic core or a fermionic one will be separately studied. For implementing it, generalized Horndeski models with the values $c_4\in\{0,\pm0.5\}$ and $d_4=0$ will be employed, regarding the low-energy effective Lagrangian (\ref{lagra}). Also, the parameter $\upalpha$ that governs the gravitational decoupling strength in Eq. (\ref{emt}), defining the polytropic gravitational decoupled radial component in Eq. (\ref{poly}), will be considered in what follows attaining the values $\upalpha=0.1$ and $\upalpha=0.5$, to a deeper analysis. Other values of $\upalpha$ have been numerically investigated and the results are qualitatively analogous. The numerical range of $\omega$ was chosen to ensure that the scalar field decays exponentially fast. It is worth noticing that the case involving the Einstein--Klein--Gordon system is equivalent to considering $c_4=0$ and $\upalpha=0$. 
Also, one must observe that the arbitrary parameter $\ell$ naturally emerges from the solution encoded in Eqs (\ref{poly}) by dimensional reasons. Since the constant terms in Eqs. (\ref{poly}) -- (\ref{10a}) regard $\upalpha\ell^{1+1/K}$, one can absorb the constant 
$\upalpha\ell^{1+1/K}$ into $\upalpha$, for the sake of conciseness at numerical calculations. 

\begin{figure}[H]\centering
 \begin{subfigure}[b]{0.32\textwidth}
\!\!\!\!\!\!\!\!\!\!\!\!\!\!\!\!\!\!\!\! \includegraphics[width=1.5\textwidth]{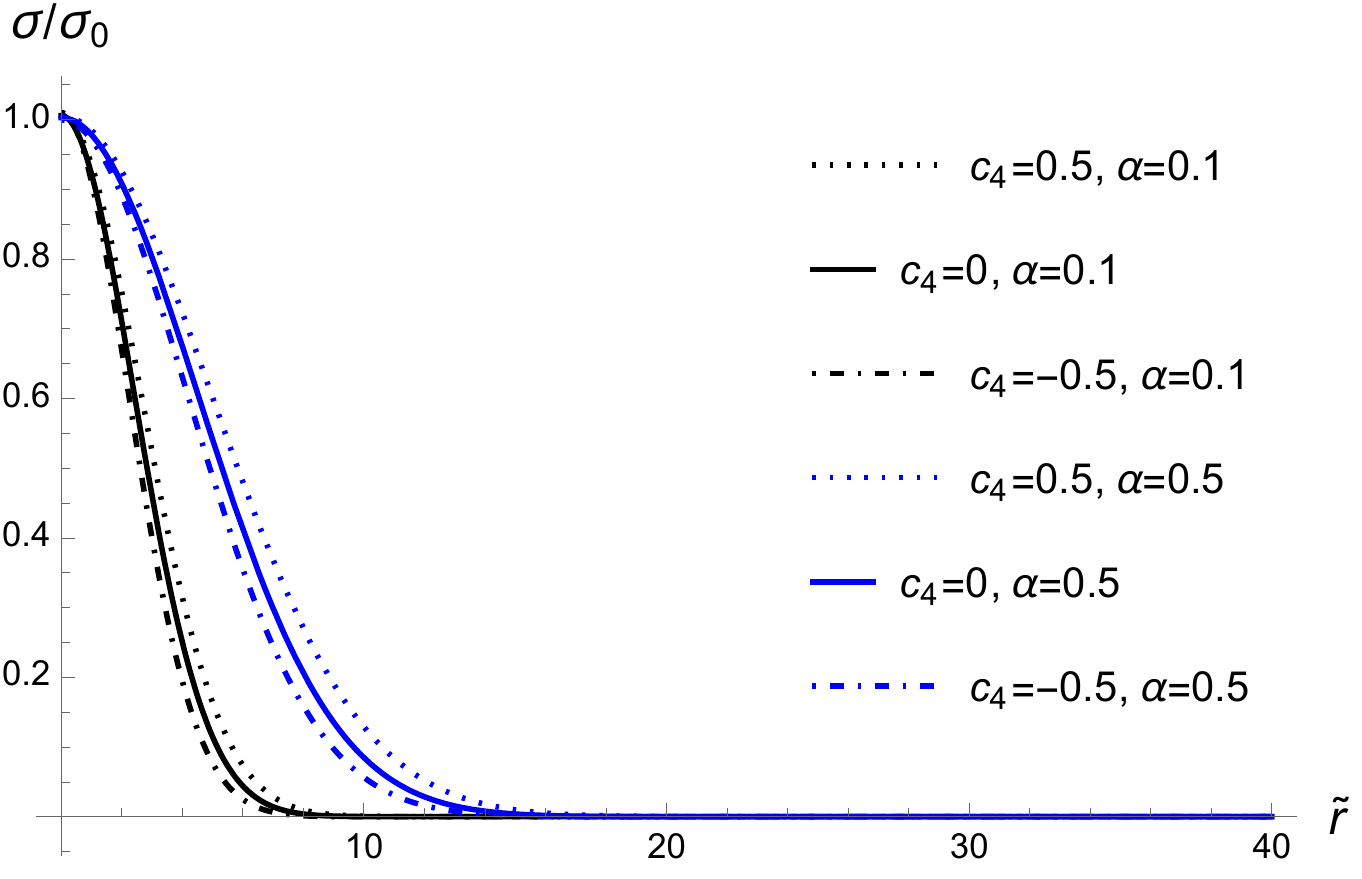}
 \caption{\footnotesize {$\Upgamma=1$.}}
 \label{1}
 \end{subfigure}\quad\qquad\qquad\quad\quad
 \begin{subfigure}[b]{0.32\textwidth}
 \includegraphics[width=1.5\textwidth]{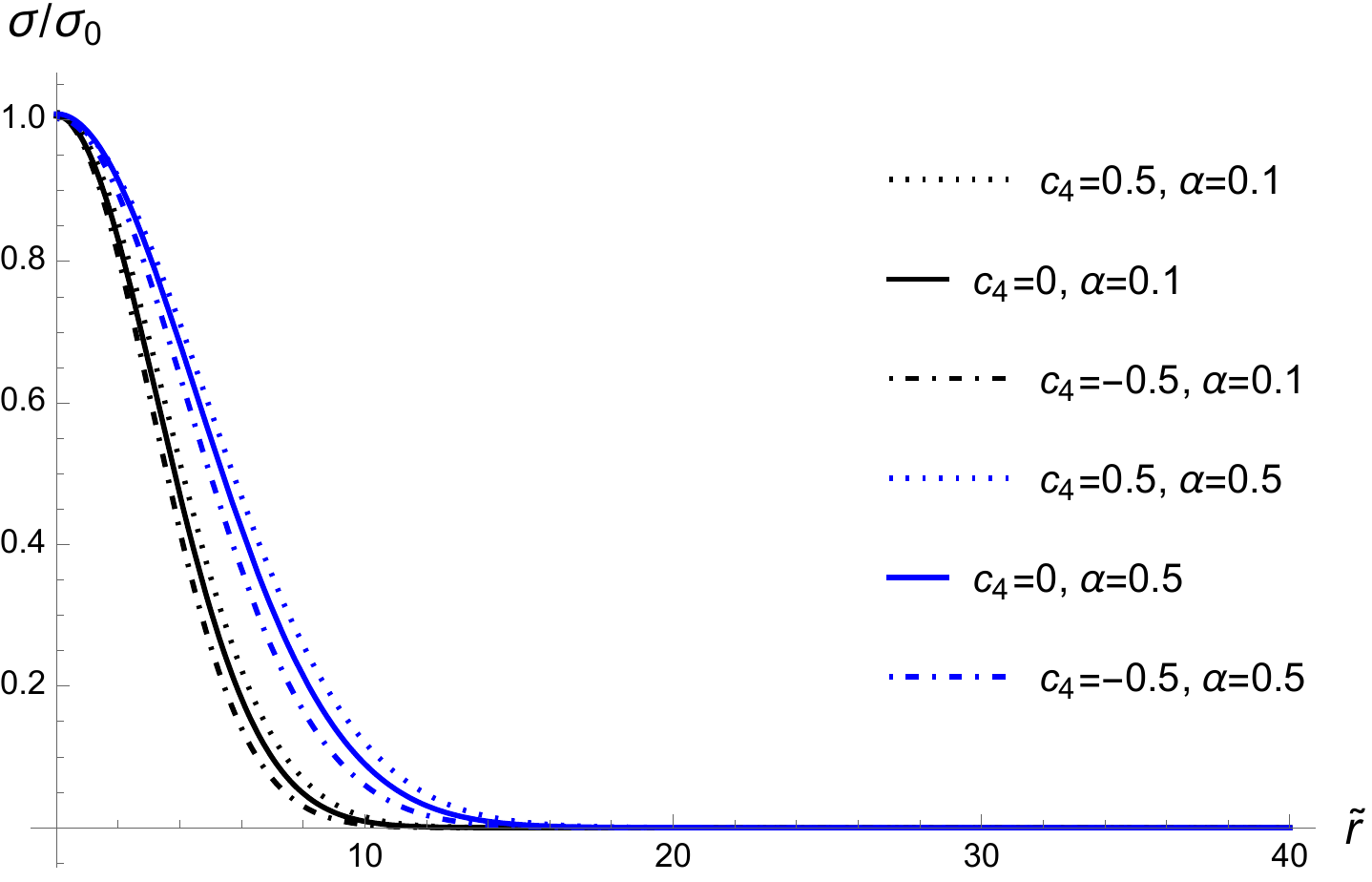}
 \caption{\footnotesize {$\Upgamma=2$.}}
 \label{2}
 \end{subfigure}\newline
 \hfill
 \begin{subfigure}[b]{0.325\textwidth}
 \includegraphics[width=1.5\textwidth]{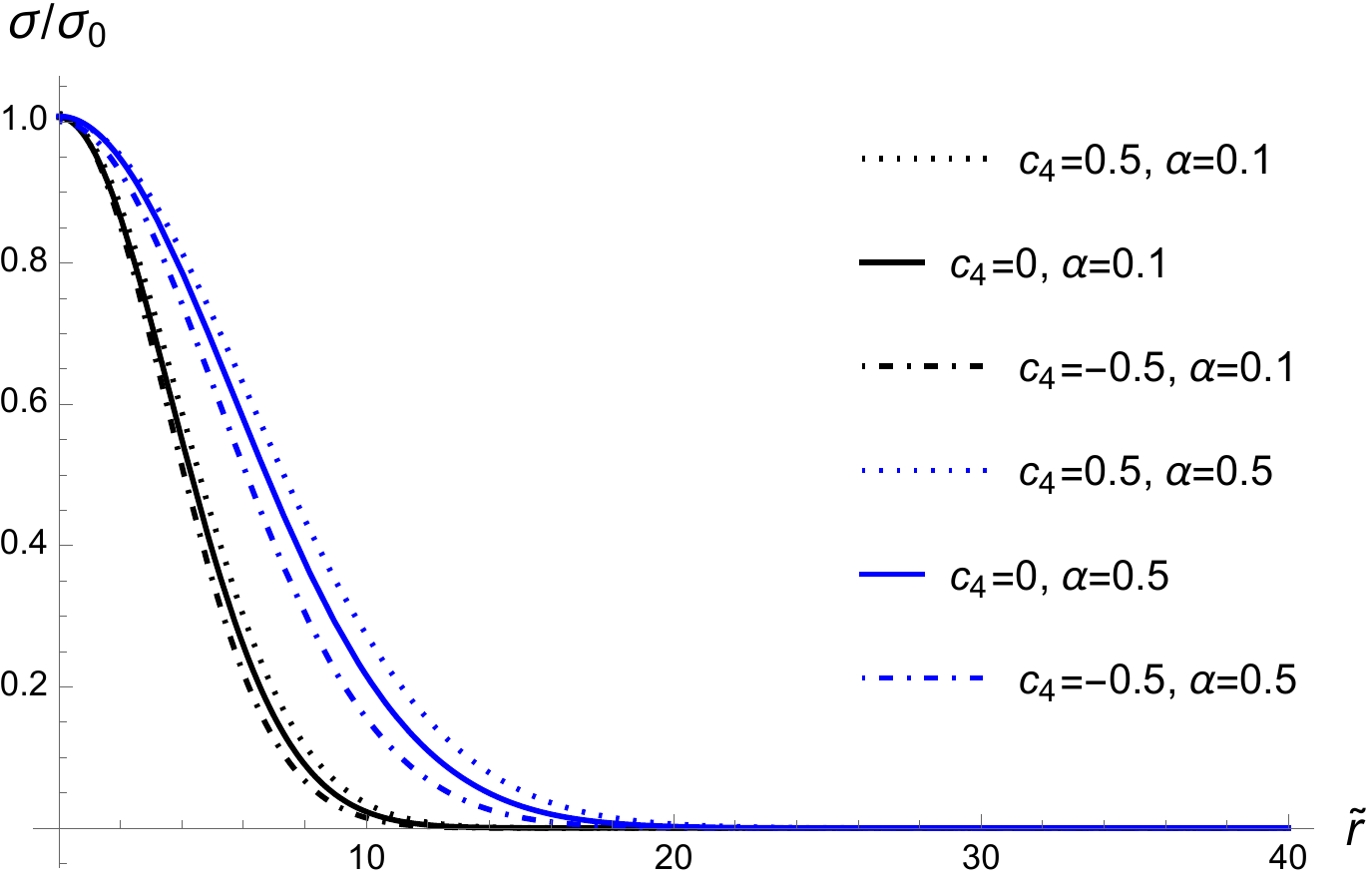}
 \caption{\footnotesize {$\Upgamma=\frac43$.}}
 \label{3}
 \end{subfigure}
 \caption{\footnotesize {Scalar field $\upphi$ generating gravitational decoupled hybrid stellar configurations with a bosonic core, in generalized Horndeski gravity for three values of $c_4$ and two values of $\upalpha$; $\Uplambda=1.5(M_{\scalebox{0.66}{$\textsc{p}$}}m^{2})^{1/3}$, and a fermionic fluid with $K=10^2\, m^{-2}M_{\scalebox{0.66}{$\textsc{p}$}}^{-2}$ and $\upsigma_0=0.15M_{\scalebox{0.66}{$\textsc{p}$}}$ is considered. Fig. \ref{1} shows the case $\Upgamma=1$ and Fig. \ref{2} illustrates the case where $\Upgamma=2$, whereas Fig. \ref{3} shows the case $\Upgamma=\frac43$.}}
 \label{fig100}
\end{figure}

\begin{figure}[H]
 \centering
 \begin{subfigure}[b]{0.32\textwidth}
 \centering
 \!\!\!\!\!\!\!\!\!\!\!\!\!\!\!\includegraphics[width=1.5\textwidth]{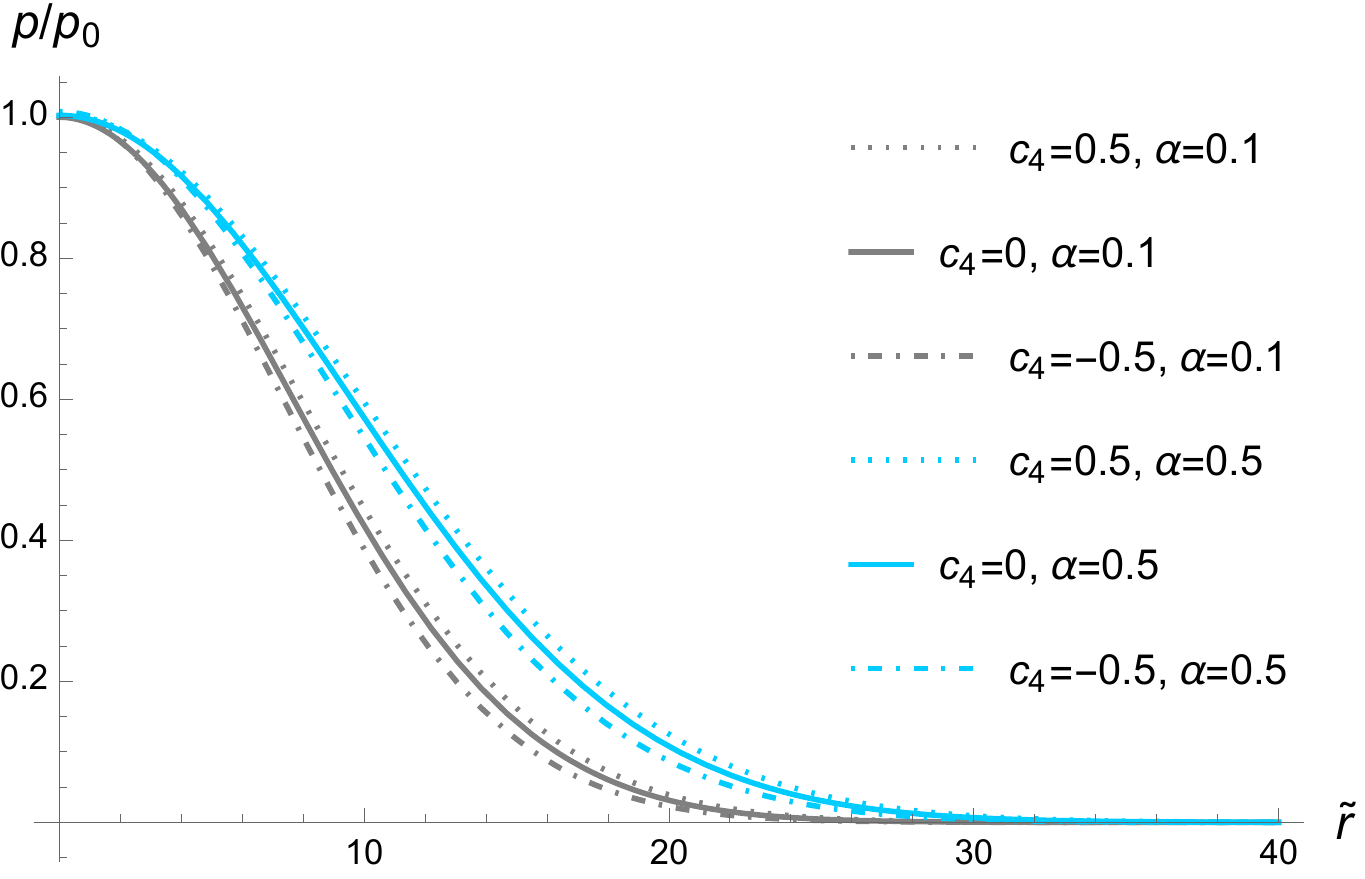}
 \caption{\footnotesize {$\Upgamma=1$.}}
 \label{4}
 \end{subfigure}\qquad\qquad\qquad\qquad
 \begin{subfigure}[b]{0.32\textwidth}
 \centering
 \includegraphics[width=1.5\textwidth]{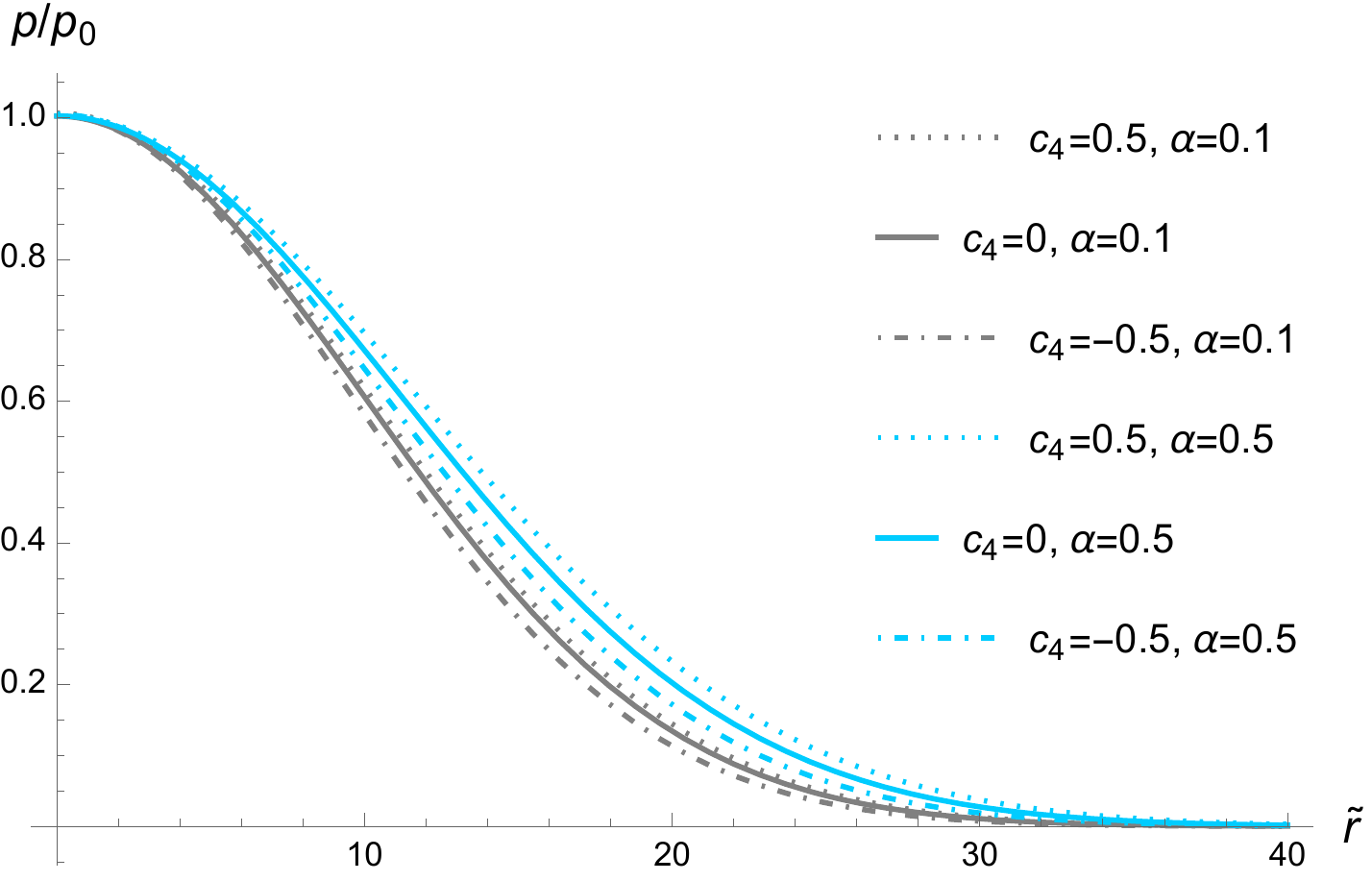}
 \caption{\footnotesize {$\Upgamma=2$.}}
 \label{5}
 \end{subfigure}\newline
 \hfill
 \begin{subfigure}[b]{0.325\textwidth}
 \centering
 \includegraphics[width=1.5\textwidth]{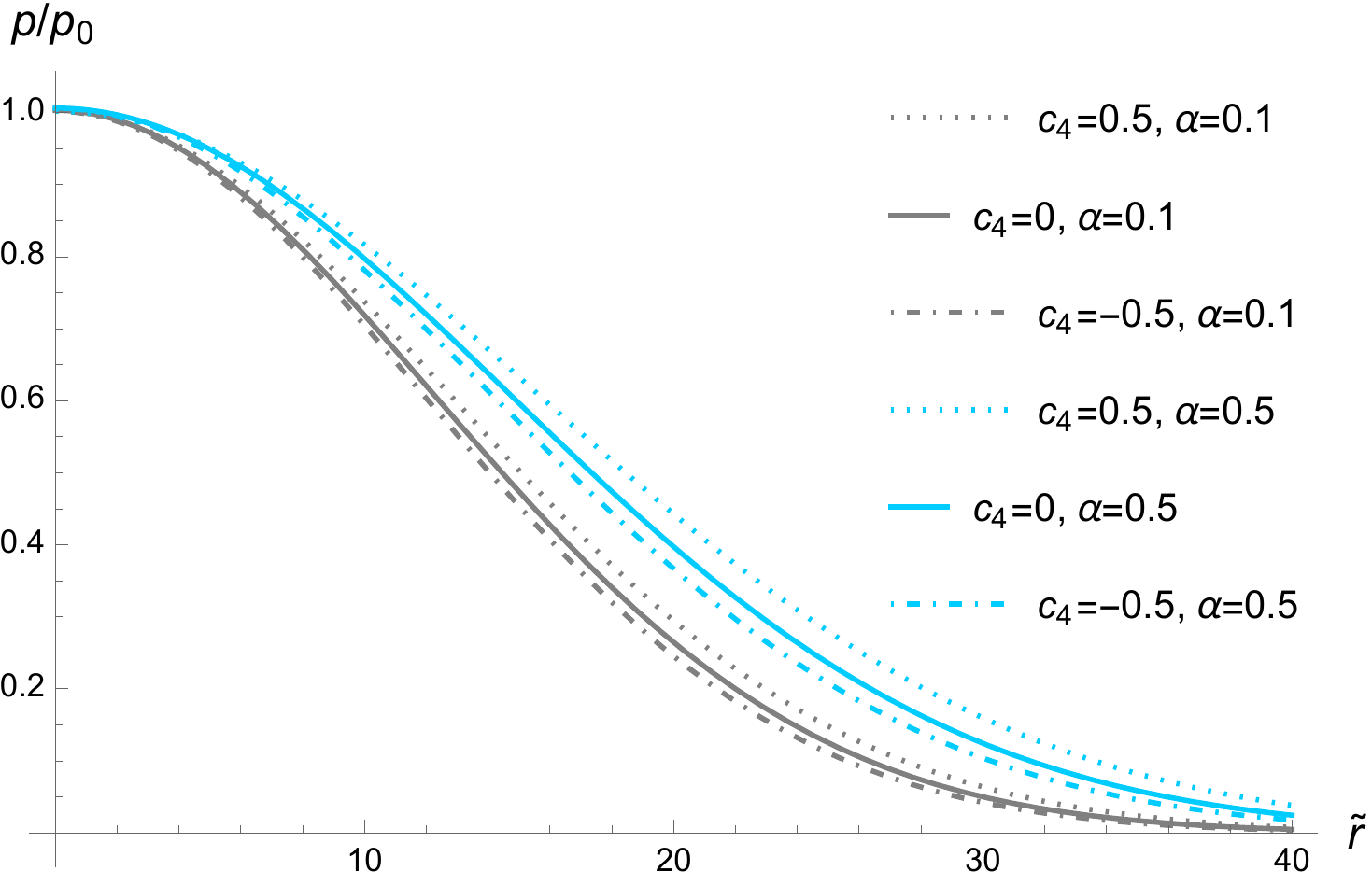}
 \caption{\footnotesize {$\Upgamma=\frac43$.}}
 \label{6}
 \end{subfigure}
 \caption{\footnotesize {Fermionic fluid pressure $p$ of gravitational decoupled hybrid stars with a bosonic core in generalized Horndeski gravity, for three values of $c_4$ and two values of $\upalpha$; $\Uplambda=1.5\,(M_{\scalebox{0.66}{$\textsc{p}$}}m^{2})^{1/3}$, and a fermionic fluid with $K=10^2\, m^{-2}M_{\scalebox{0.66}{$\textsc{p}$}}^{-2}$ and $p_0=10^{-2}m^2M^2_{\scalebox{0.66}{$\textsc{p}$}}$ is considered. Fig. \ref{4} shows the case $\Upgamma=1$ and Fig. \ref{5} illustrates the case where $\Upgamma=2$, whereas Fig. \ref{6} shows the case $\Upgamma=\frac43$.}}
 \label{fig101}
\end{figure}

\begin{figure}[H]
 \centering
 \begin{subfigure}[b]{0.32\textwidth}
 \centering
 \!\!\!\!\!\!\!\!\!\!\!\!\!\!\!\!\!\!\includegraphics[width=1.5\textwidth]{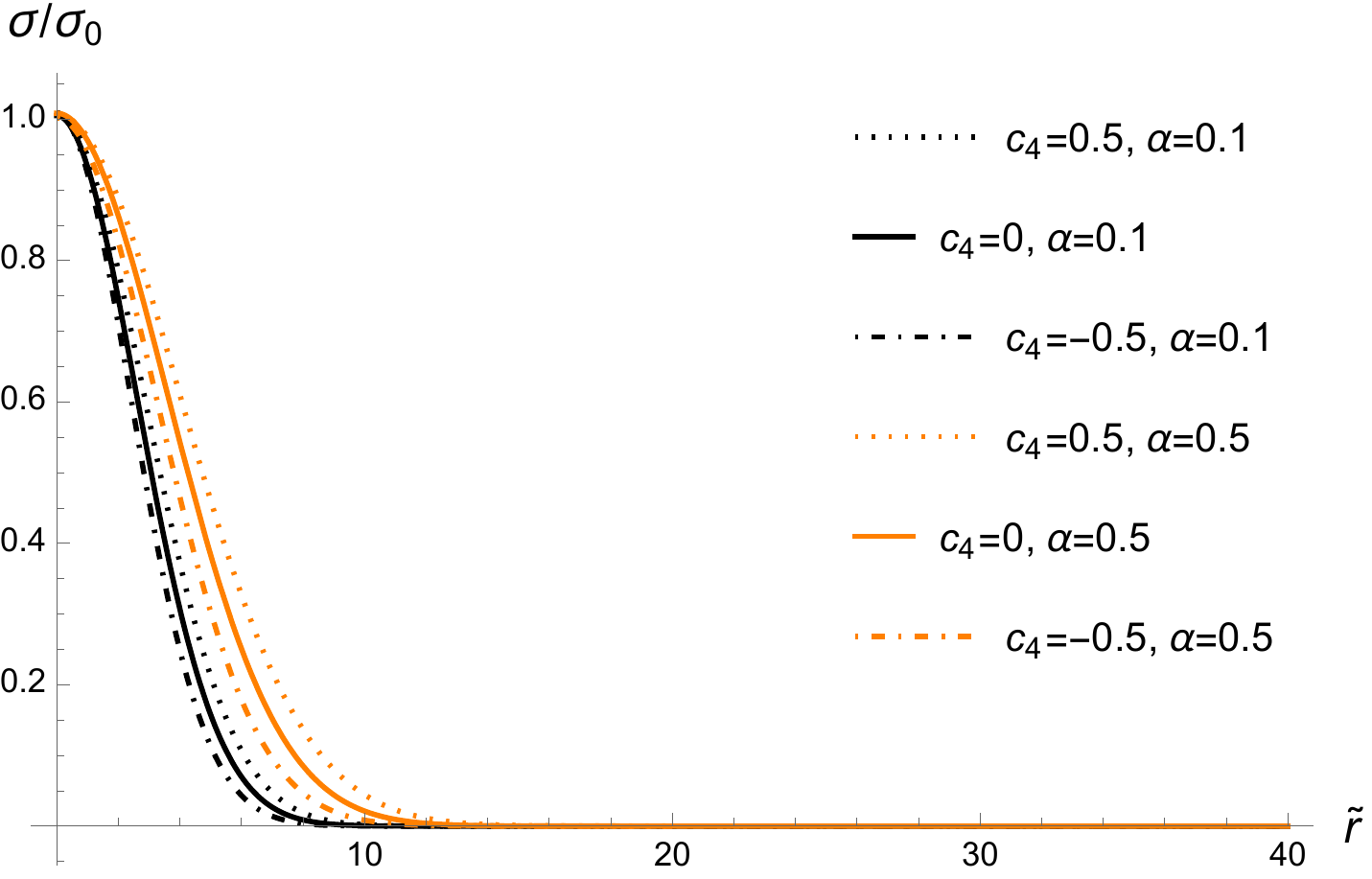}
 \caption{\footnotesize {$\Upgamma=1$.}}
 \label{7}
 \end{subfigure}\quad\qquad\quad\quad\quad
 \begin{subfigure}[b]{0.32\textwidth}
 \centering
 \includegraphics[width=1.5\textwidth]{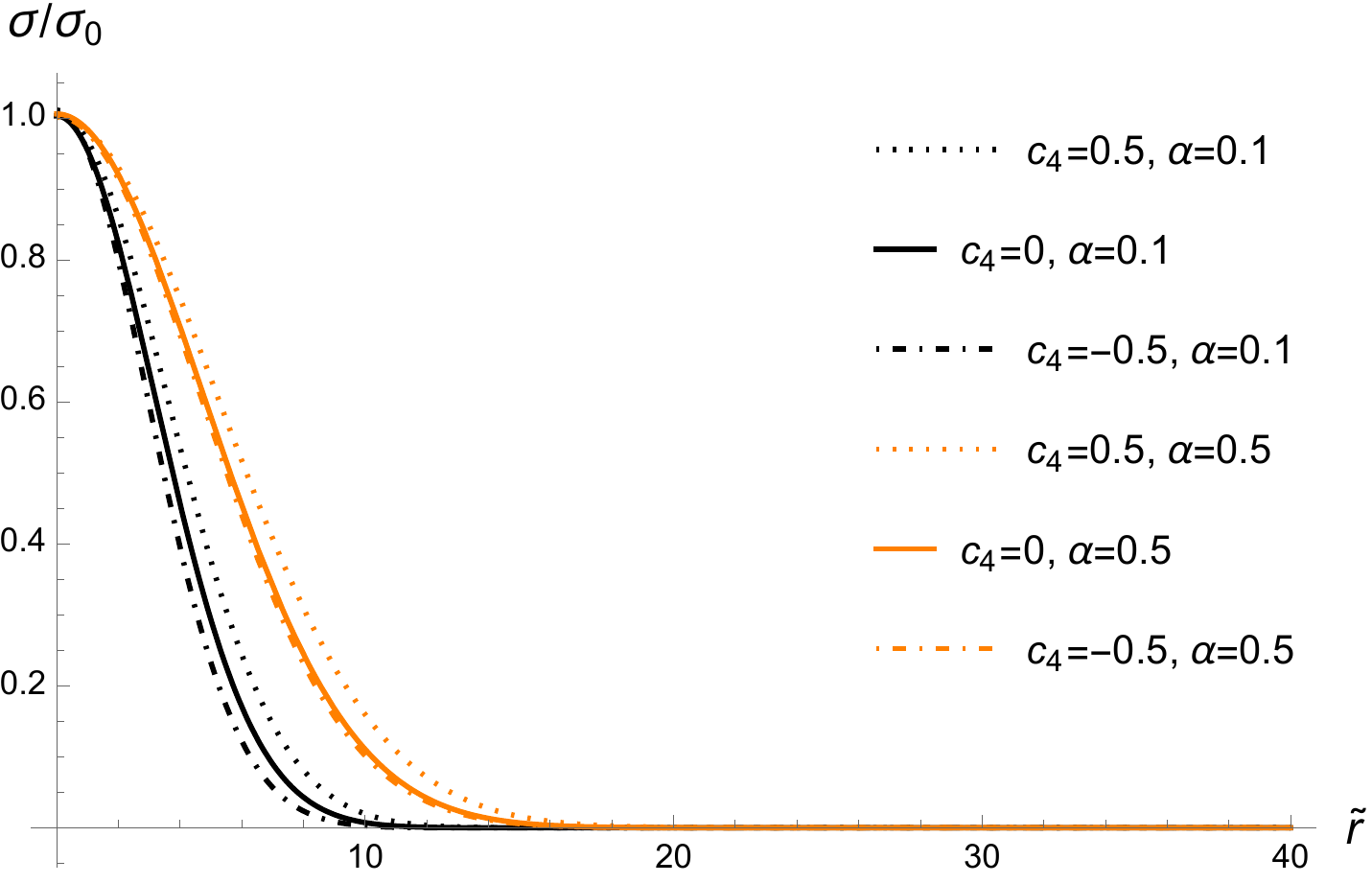}
 \caption{\footnotesize {$\Upgamma=2$.}}
 \label{8}
 \end{subfigure}\newline
 \hfill
 \begin{subfigure}[b]{0.325\textwidth}
 \centering
 \includegraphics[width=1.5\textwidth]{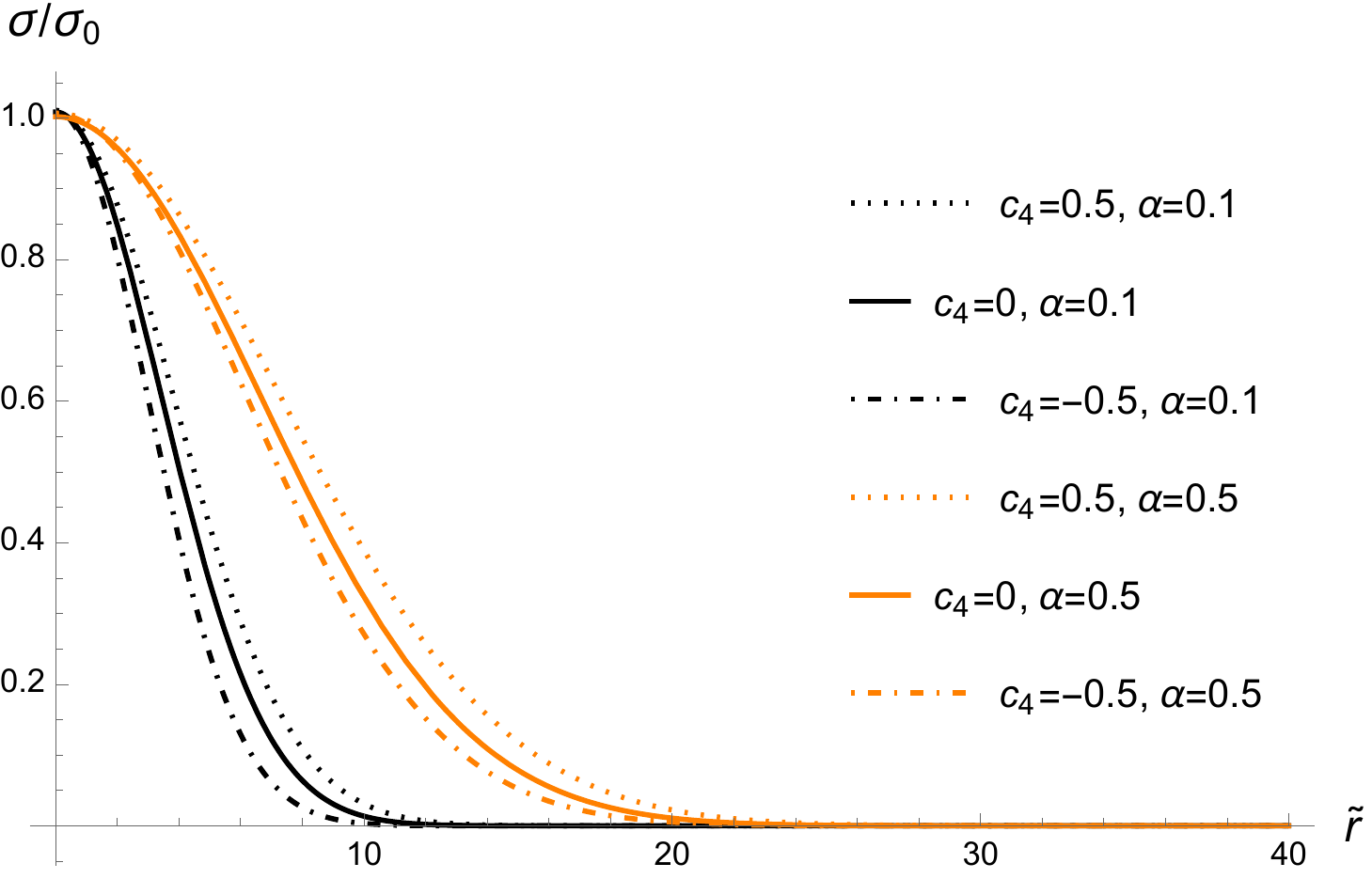}
 \caption{\footnotesize {$\Upgamma=\frac43$.}}
 \label{9}
 \end{subfigure}
 \caption{\footnotesize {Scalar field $\upphi$ generating gravitational decoupled hybrid stars with a fermionic core in generalized Horndeski gravity, for three values of $c_4$ and two values of $\upalpha$; $\Uplambda=1.5(M_{\scalebox{0.66}{$\textsc{p}$}}m^{2})^{1/3}$, and a fermionic fluid with $K=10^2\, m^{-2}M_{\scalebox{0.66}{$\textsc{p}$}}^{-2}$ and $\upsigma_0=0.26M_{\scalebox{0.66}{$\textsc{p}$}}$ is considered. Fig. \ref{7} shows the case $\Upgamma=1$ and Fig. \ref{8} illustrates the case where $\Upgamma=2$, whereas Fig. \ref{9} shows the case $\Upgamma=\frac43$.}}
 \label{fig102}
\end{figure}

\begin{figure}[H]
 \centering
 \begin{subfigure}[b]{0.32\textwidth}
 \centering
\!\!\!\!\!\!\!\!\!\!\!\!\!\!\!\!\!\!\!\!\includegraphics[width=1.5\textwidth]{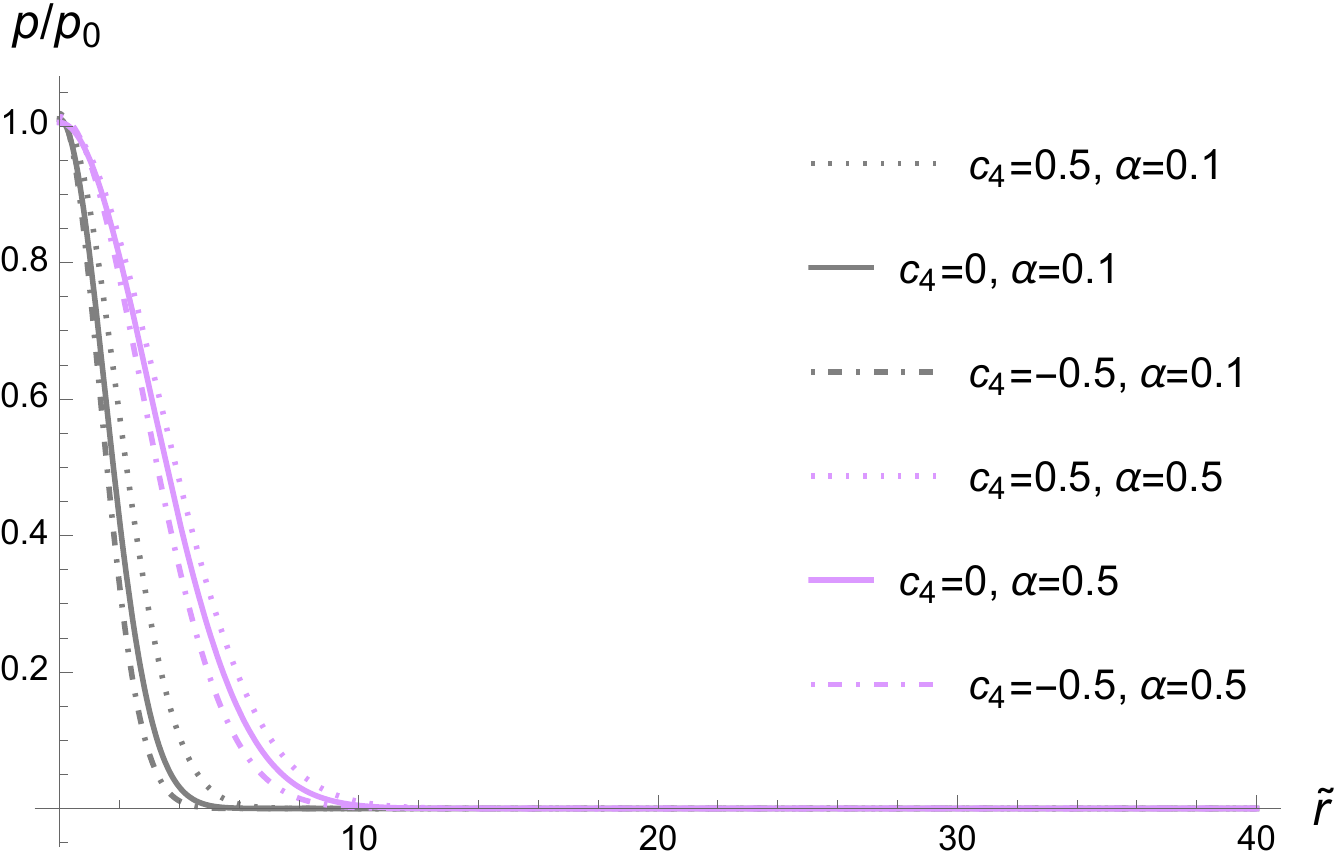}
 \caption{\footnotesize {$\Upgamma=1$.}}
 \label{10}
 \end{subfigure}\quad\qquad\quad\qquad
 \begin{subfigure}[b]{0.32\textwidth}
 \centering
 \includegraphics[width=1.5\textwidth]{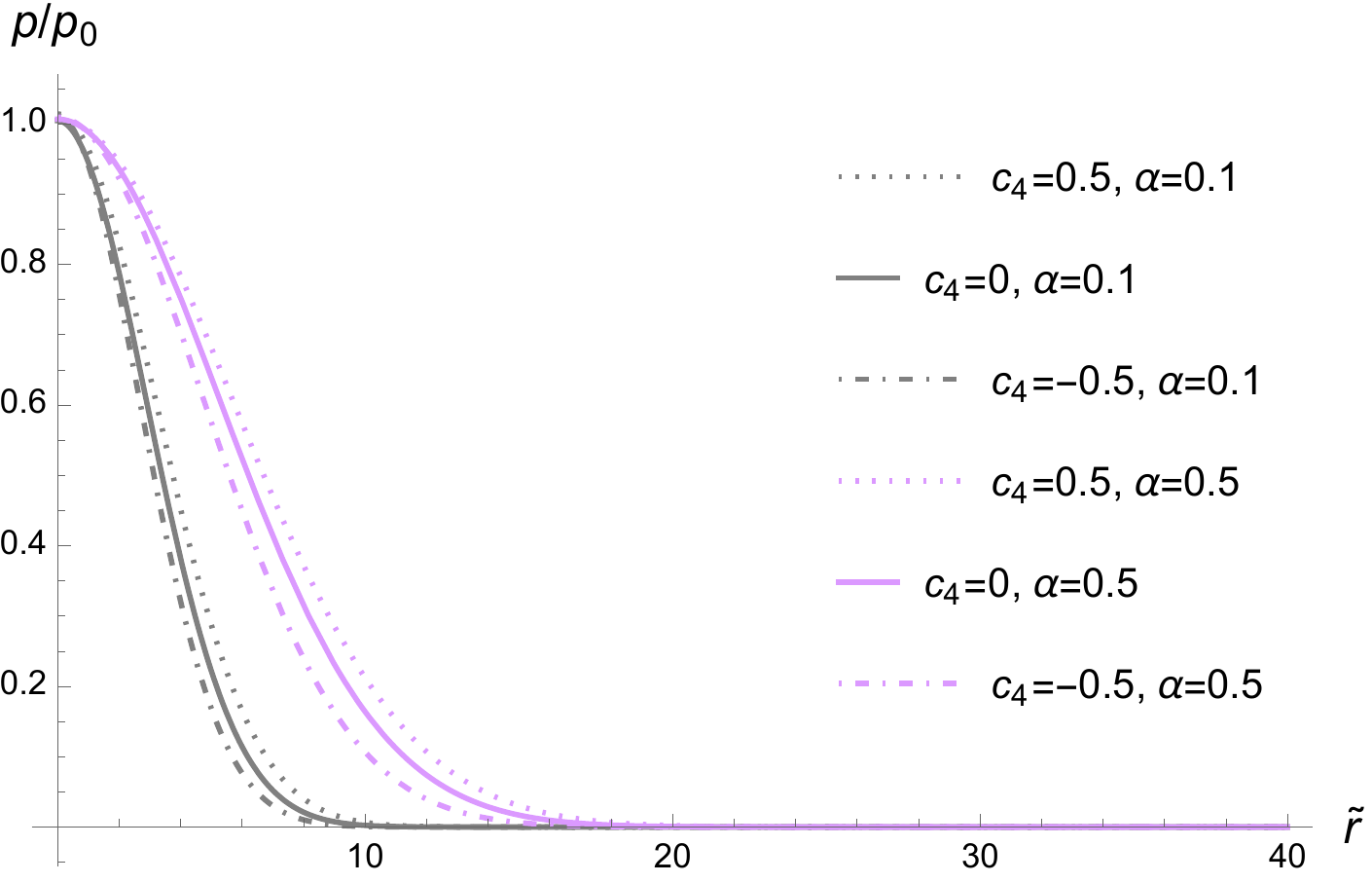}
 \caption{\footnotesize {$\Upgamma=2$.}}
 \label{11}
 \end{subfigure}\newline
 \hfill
 \begin{subfigure}[b]{0.325\textwidth}
 \centering
 \includegraphics[width=1.5\textwidth]{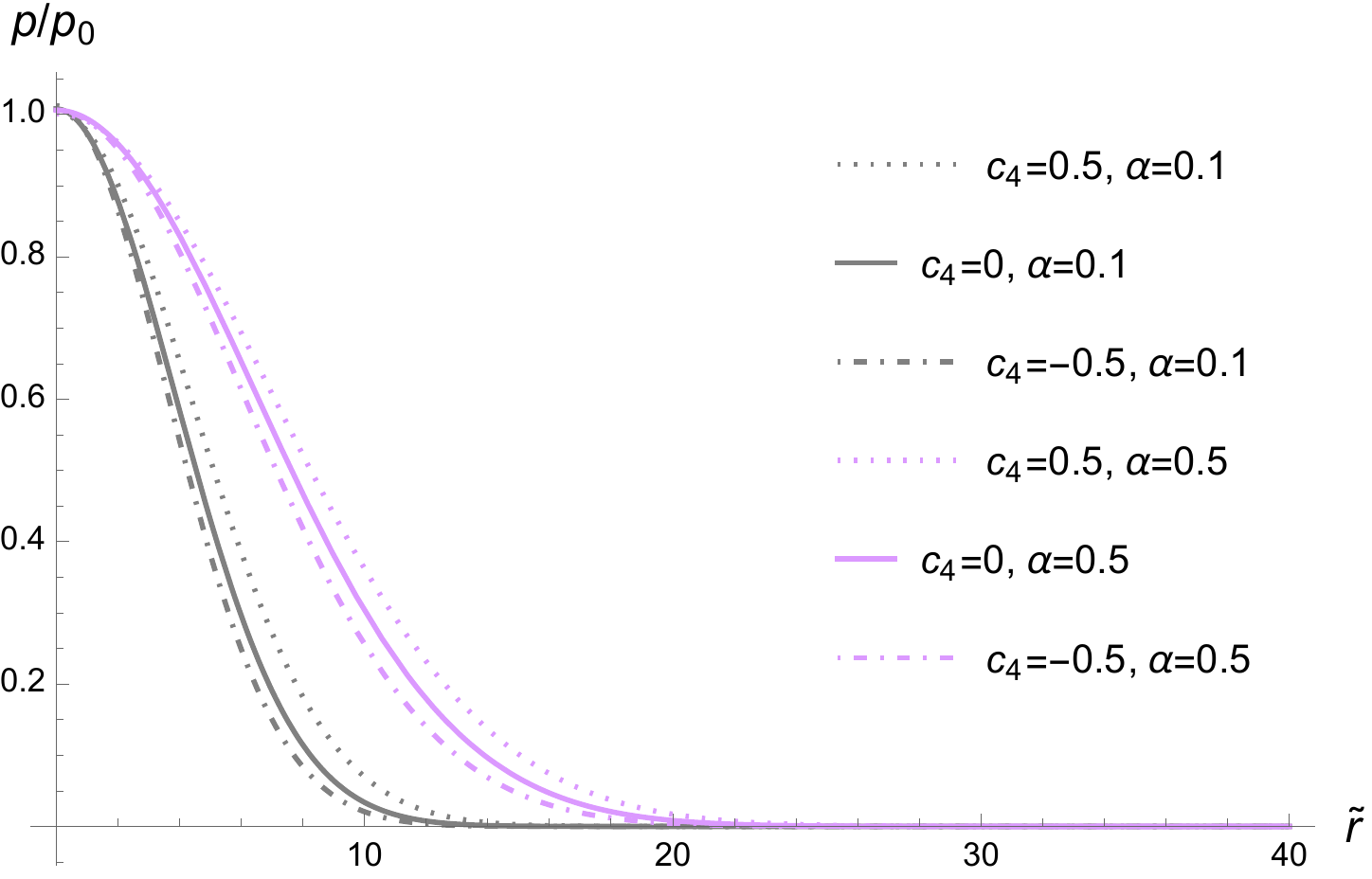}
 \caption{\footnotesize {$\Upgamma=\frac43$.}}
 \label{12}
 \end{subfigure}
 \caption{\footnotesize {Fermionic fluid pressure $p$ of gravitational decoupled hybrid stars with a fermionic core in generalized Horndeski gravity, for three values of $c_4$ and two values of $\upalpha$; $\Uplambda=1.5\,(M_{\scalebox{0.66}{$\textsc{p}$}}m^{2})^{1/3}$, and a fermionic fluid with $K=10^2 m^{-2}M_{\scalebox{0.66}{$\textsc{p}$}}^{-2}$ and $p_0=2.3\times 10^{-4}\,(mM_{\scalebox{0.66}{$\textsc{p}$}})^2$ is considered. Fig. \ref{10} shows the case $\Upgamma=1$ and Fig. \ref{11} illustrates the case where $\Upgamma=2$, whereas Fig. \ref{12} shows the case $\Upgamma=\frac43$.}}
 \label{fig103}
\end{figure}

First, the stationary splitting of the 
scalar field in Eq. (\ref{ansat}) and the fermionic fluid pressure $p$ are addressed as functions of the radial coordinate, for hybrid stellar configurations in Horndeski generalized models, for different values of the parameters $c_4$, $\upalpha$, and the adiabatic index $\Upgamma$. The plots in Figs. \ref{fig100} and \ref{fig101} illustrate gravitational decoupled hybrid stars with a bosonic core, whereas Figs. \ref{fig102} and \ref{fig103} regard gravitational decoupled hybrid stars with a fermionic core. The type of core in hybrid stars is encoded by the choice of the parameters $\Uplambda$, $\upsigma_0$, and $p_0$ \cite{Barranco:2021auj,Chagoya:2018lmv}. 
In all plots that follow, the polytropic constant 
$K=10^2\, m^{-2}M_{\scalebox{0.66}{$\textsc{p}$}}^{-2}$ is employed, corroborating to phenomenological data. The involved parameters were taken in the parameter space derived in Refs. \cite{Barranco:2021auj,Chagoya:2018lmv}, in such a way that fermionic matter emulates typical astrophysical neutron stellar configurations, with central density $\rho_0=\displaystyle{\lim_{r\to0}}\rho(r)\sim10^{12} \rho_{\odot}$ and central pressure $p_0\sim10^{18}$ $p_{\odot}$, \blt{denoting the Solar average density by $\rho_{\odot} = 1.412\times 10^3$ kg/m${}^3$ and the Solar core average pressure by $p_{\odot} = 2.653 \times 10^{15}$ Pa.} 

Fig. \ref{fig100} shows the behavior of the scalar field, in gravitational decoupled hybrid stars with a bosonic core, as a function of the radial coordinate. For each fixed value of the adiabatic index $\Upgamma$ and fixed gravitational decoupling parameter $\upalpha$, the higher the $c_4$ parameter, the less sharp the scalar field decreases along the radial coordinate. When $\upalpha$ varies and $c_4$ is kept fixed, the lower the $\upalpha$ parameter, the steeper the scalar field decreases as a function of the radial coordinate. Now, for fixed $\upalpha$ and $c_4$, making the adiabatic index $\Upgamma$ to vary yields an interesting result: the higher the adiabatic index $\Upgamma$, the slower the scalar field reaches its asymptotically null value. It is also important to emphasize that for all values of $c_4$, $\upalpha$, and $\Upgamma$ the scalar field attains the asymptotically null value, however at different rates. Remembering that $c_4=0$ represents the gravitational decoupled Einstein--Klein--Gordon system, for values $c_4>0$ [$c_4<0$], the scalar field profile is smoother [sharper] along the radial coordinate when compared to the Einstein--Klein--Gordon system. For the plots \ref{1} -- \ref{3}, both the asymptotic conditions $\displaystyle{\lim_{{r}\to\infty}}\upsigma({r})=0$ and $\displaystyle{\lim_{{r}\to\infty}}\upsigma'({r})=0$ hold, for all values of $\upalpha$, $\Upgamma$, and $c_4$.

\noindent Fig. \ref{fig101} illustrates the profile of the fermionic fluid pressure, regarding gravitational decoupled hybrid stars with a bosonic core, as a function of the radial coordinate. Although the results are qualitatively similar to the one depicted in the plots of Fig. \ref{fig100}, the main difference resides in the evanescence rate of the fermionic fluid pressure. For each fixed value of the adiabatic index $\Upgamma$ and fixed decoupling parameter $\upalpha$, for the studied values of $c_4$, the fermionic fluid pressure dampens to its asymptotically null value in a slower rate, when compared to the respective scalar field profiles. Besides, still comparing to Fig. \ref{fig101} to the plots in Fig. \ref{fig100} with the same respective value of the adiabatic index $\Upgamma$, one can realize that the rate at which the scalar field profile decays to its asymptotically null value is around four times steeper than the fermionic fluid pressure profile. Also, for values $c_4>0$ [$c_4<0$], the fermionic fluid pressure is smoother [sharper] along the radial coordinate, when compared to the results coming from the gravitational decoupled Einstein--Klein--Gordon solution, which corresponds to $c_4=0$. In addition, in Fig. \ref{fig101} the asymptotic limits $\displaystyle{\lim_{{r}\to\infty}}p({r})=0$ and $\displaystyle{\lim_{{r}\to\infty}}p'({r})=0$ also hold.

Now, Fig. \ref{fig102} portrays the profile of the scalar field related to gravitational decoupled hybrid stars with a fermionic core. 
For each fixed value of the adiabatic index $\Upgamma$ and fixed gravitational decoupling parameter $\upalpha$, the higher the $c_4$ parameter, the smoother the scalar field decreases along the radial coordinate. For fixed $\upalpha$ and $c_4$ and making the adiabatic index $\Upgamma$ to vary, the higher the 
adiabatic index $\Upgamma$, the broader the graphic is. Besides, the scalar field in Fig. \ref{fig102} reaches its asymptotically null value at a higher value of the radial coordinate, when compared to the respective cases of hybrid stars with a bosonic core in Fig. \ref{fig100}, for each fixed $\Upgamma$. Again a recurrent feature can be observed: for positive [negative] values of $c_4$, the scalar field fades in a smoother [sharper] rate, along the radial coordinate, when compared to the gravitational decoupled Einstein--Klein--Gordon system.
However, contrary to the scalar field profile for hybrid stars with a bosonic core in Fig. \ref{fig100}, the plots in Fig. \ref{fig102} show a peculiar behaviour regarding gravitational decoupled hybrid stars with a fermionic core: allowing $\upalpha$ to vary and $c_4$ to be fixed, the higher the $\upalpha$ parameter, the steeper the scalar field decreases along the radial coordinate. Similarly to the case of a bosonic core, here also for fermionic cores both the asymptotic conditions $\displaystyle{\lim_{{r}\to\infty}}\upsigma({r})=0$ and $\displaystyle{\lim_{{r}\to\infty}}\upsigma'({r})=0$ are valid.

Fig. \ref{fig103} illustrates the fermionic fluid pressure profile, regarding gravitational decoupled hybrid stars with a fermionic core, as a function of the radial coordinate. For each fixed value of the adiabatic index $\Upgamma$ and fixed gravitational decoupling parameter $\upalpha$, for the respective values of the $c_4$ parameter, the fermionic fluid pressure undergoes a sharper approach to its asymptotically null value, when compared to the respective scalar field profiles. This behavior is opposite to the one regarding hybrid stars with a bosonic core. Besides, for values $c_4>0$ [$c_4<0$], the fermionic fluid pressure is smoother [sharper] along the radial coordinate, when compared to the results coming from the gravitational decoupled Einstein--Klein--Gordon solution. One can conclude that regarding gravitational decoupled hybrid stars with a fermionic core, fermionic fluid pressure profiles in Fig. \ref{fig103} are even sharper than the one for 
hybrid stars with a bosonic core in Fig. \ref{fig101}. For the plots \ref{10} -- \ref{12}, the asymptotic limits $\displaystyle{\lim_{{r}\to\infty}}p({r})=0$ and $\displaystyle{\lim_{{r}\to\infty}}p'({r})=0$ are verified, irrespective of the values of $\upalpha$, $\Upgamma$, and $c_4$.

Therefore, pairwise comparing Fig. \ref{fig100} to Fig. \ref{fig102}; and Fig. \ref{fig101} to Fig. \ref{fig103}, in the gravitational decoupling context, hybrid stars with a bosonic core present the scalar field fading more sharply along the radial coordinate than the fermionic fluid pressure profile, whereas in hybrid stars with a fermionic core the fermionic fluid pressure profile evanesces more steeply along the radial coordinate than the scalar field profile. Besides, for fixed $\Upgamma$ and $\upalpha$, for $c_4>0$ [$c_4<0$] both the scalar field and the fermionic fluid pressure are smoother [sharper] along the radial coordinate, when compared to the gravitational decoupled Einstein--Klein--Gordon system, which corresponds to $c_4=0$.
Hence, one can conclude that the gravitational decoupling setup emulates standard results in extended Horndeski scalar-tensor gravity \cite{Roque:2021lvr,Barranco:2021auj}, also corroborating to the conjecture that those negative [positive] couplings correspond to attractive [repulsive] self-interaction. 
Comparing the plots in Figs. \ref{fig100} -- \ref{fig103}, one can realize that in gravitational decoupled hybrid stars with a bosonic core, the scalar field evanesces faster than the pressure profile, as a function of the radial coordinate. On the other hand, in gravitational decoupled hybrid stellar configurations containing a fermionic core, the pressure profile plunges sharper than the scalar field. All these fields vanish asymptotically. 
	
\begin{figure}[H]
\centering
 \begin{subfigure}[b]{0.32\textwidth}
 \!\!\!\!\!\!\!\!\!\!\!\!\!\!\!\includegraphics[width=1.5\textwidth]{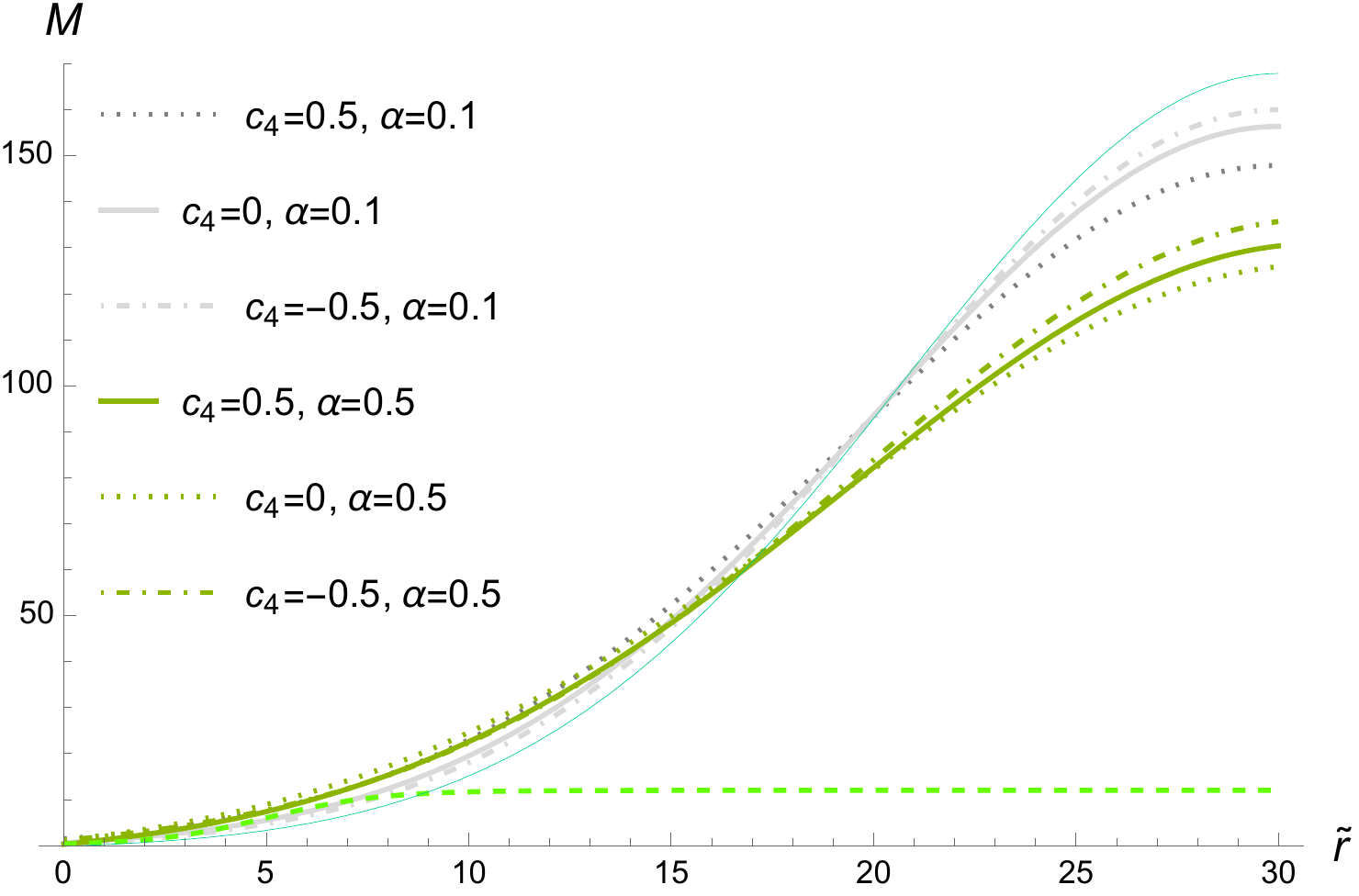}
 \caption{\footnotesize {$\Upgamma=1$.}}
 \label{13}
 \end{subfigure}\qquad\qquad\qquad\qquad
 \begin{subfigure}[b]{0.32\textwidth}
 \includegraphics[width=1.5\textwidth]{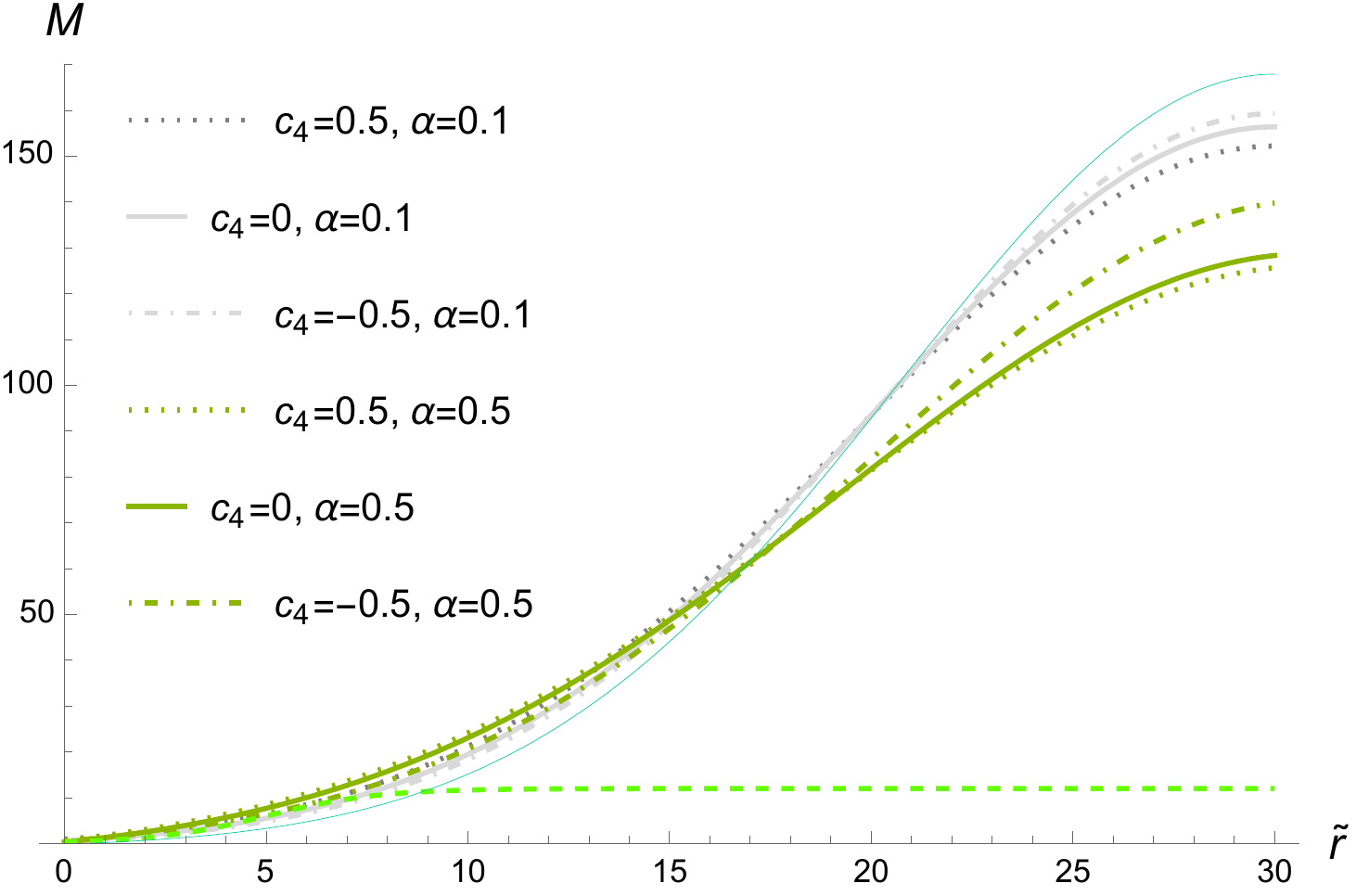}
 \caption{\footnotesize {$\Upgamma=2$.}}
 \label{14}
 \end{subfigure}\newline
 \hfill
 \begin{subfigure}[b]{0.325\textwidth}
 \includegraphics[width=1.5\textwidth]{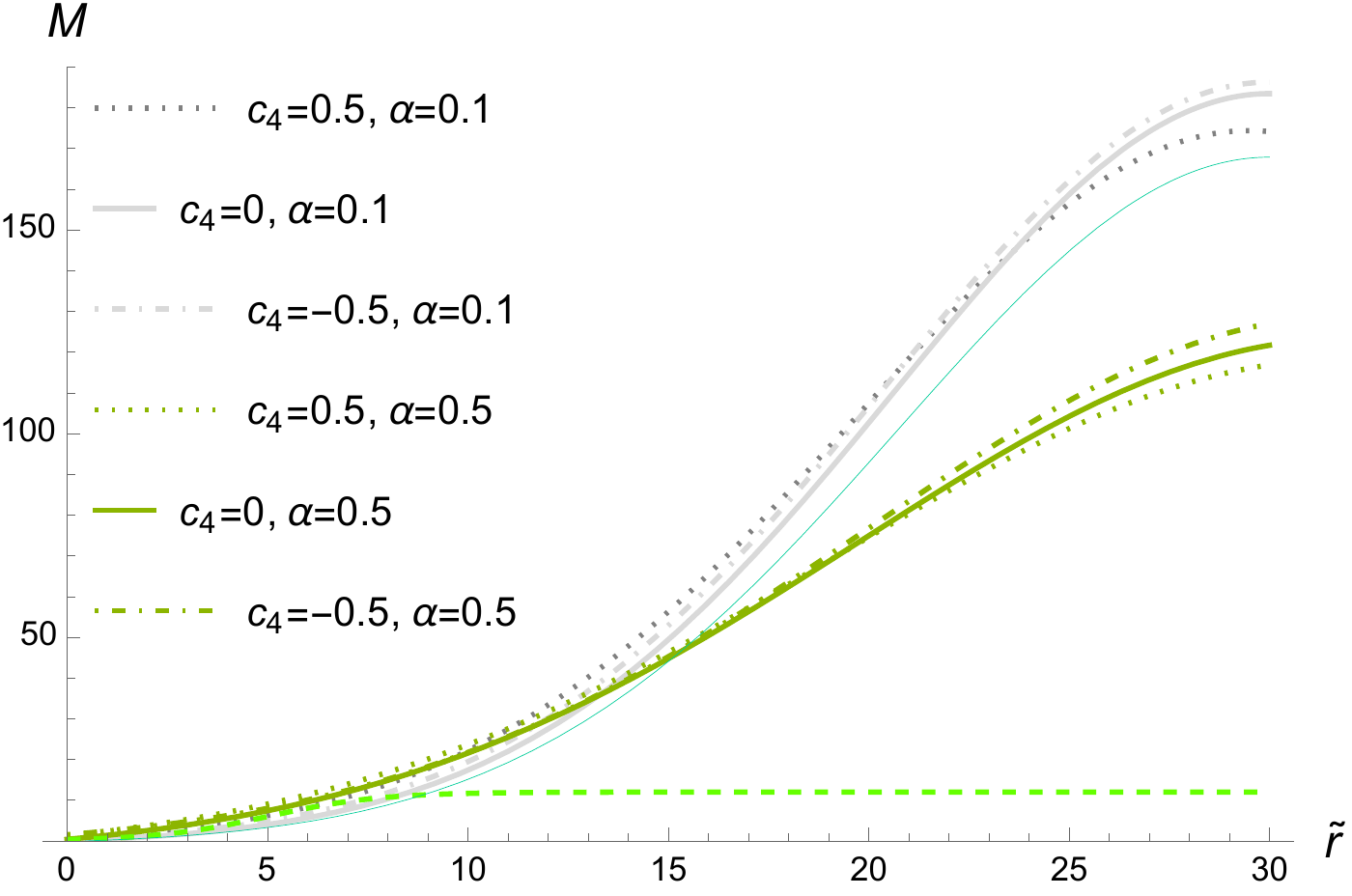}
 \caption{\footnotesize {$\Upgamma=\frac43$.}}
 \label{15}
 \end{subfigure}
 \caption{\footnotesize {Mass function $M(r)$ of gravitationally decoupled hybrid stars with a bosonic core, in generalized Horndeski gravity, for $c_4\in\{0,\pm 0.5\}$ and two values of $\upalpha$, for $\Uplambda=1.5(M_{\scalebox{0.66}{$\textsc{p}$}}m^{2})^{1/3}$, and a fermionic fluid with polytropic constant $K=10^2\, m^{-2}M_{\scalebox{0.66}{$\textsc{p}$}}^{-2}$ and $\upsigma_0=0.15M_{\scalebox{0.66}{$\textsc{p}$}}$ is considered. 
 Lightgray curves represent the case where $\upalpha = 0.1$ and apple green curves regard the case where $\upalpha = 0.5$. Gravitational decoupled neutron stars with $\upsigma=0$ are depicted as the continuous thin Caribbean green curves, whereas the case where the fermionic fluid pressure is null represents a gravitationally decoupled boson star, displayed as the bright green dashed curve. Fig. \ref{13} shows the case $\Upgamma=1$ and Fig. \ref{14} illustrates the case where $\Upgamma=2$, whereas Fig. \ref{15} shows the case $\Upgamma=\frac43$.}}
 \label{fig104}
\end{figure}

\begin{figure}[H]
 \centering
 \begin{subfigure}[b]{0.32\textwidth}
 \centering
 \!\!\!\!\!\!\!\!\!\!\!\!\!\!\! \!\!\!\!\!\!\!\!\!\!\!\!\!\!\!\!\!\!\!\!\!\!\!\!\!\includegraphics[width=1.5\textwidth]{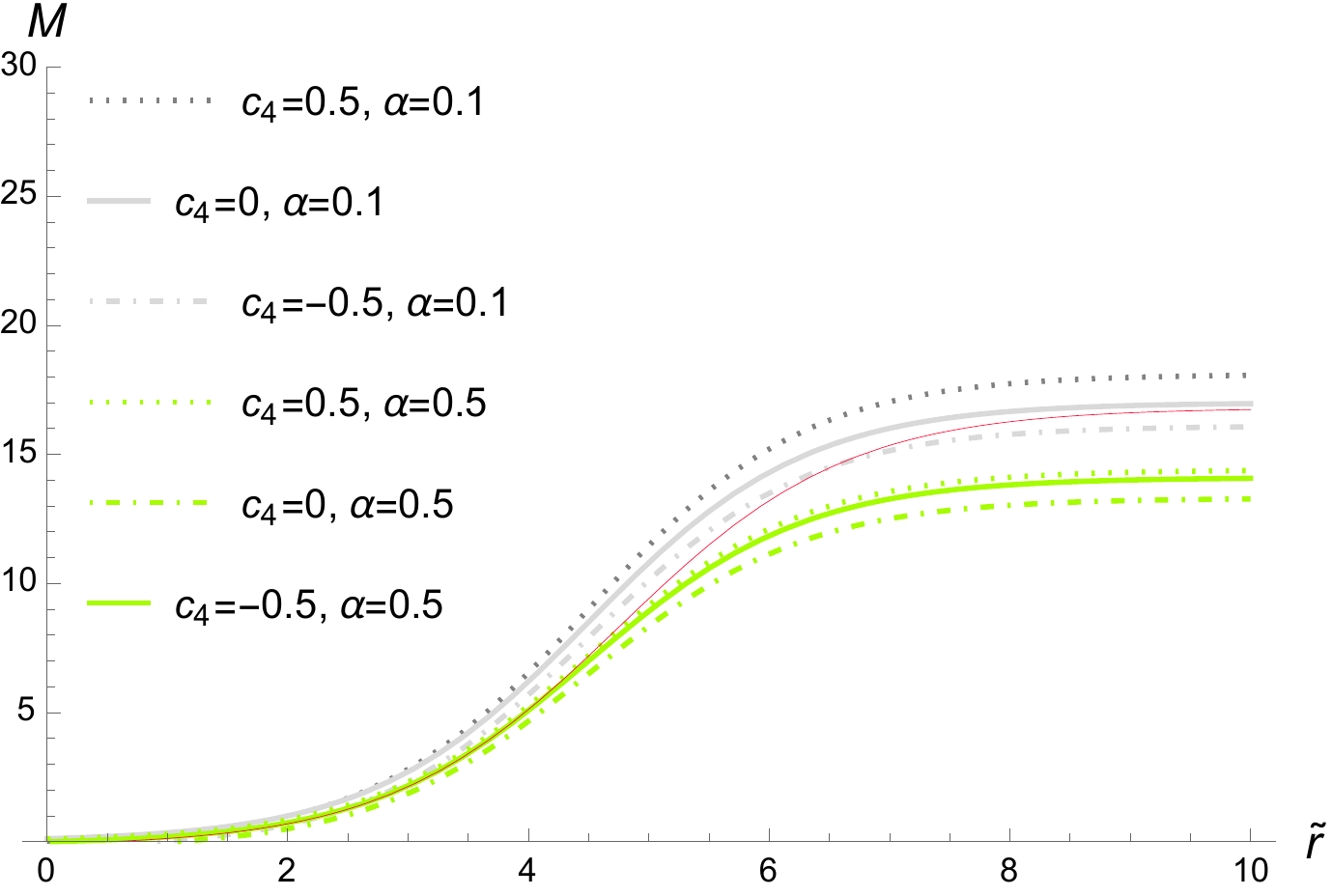}
 \caption{\footnotesize {$\Upgamma=1$.}}
 \label{16}
 \end{subfigure}\qquad\qquad\qquad\qquad\qquad
 \begin{subfigure}[b]{0.32\textwidth}
 \centering
 \includegraphics[width=1.5\textwidth]{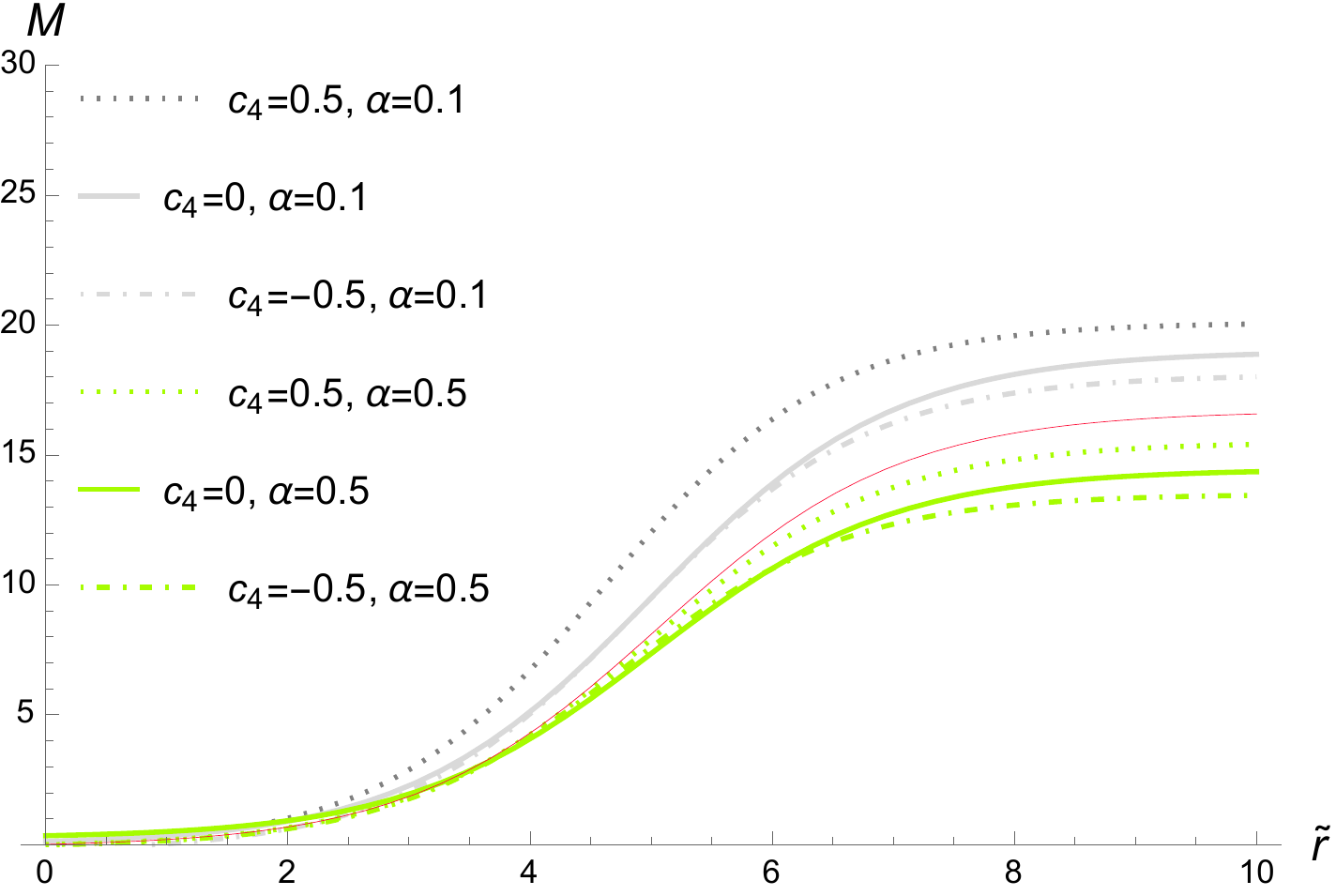}
 \caption{\footnotesize {$\Upgamma=2$.}}
 \label{17}
 \end{subfigure}\newline
 \hfill
 \begin{subfigure}[b]{0.325\textwidth}
 \centering
\!\!\!\!\!\!\!\!\!\! \includegraphics[width=1.5\textwidth]{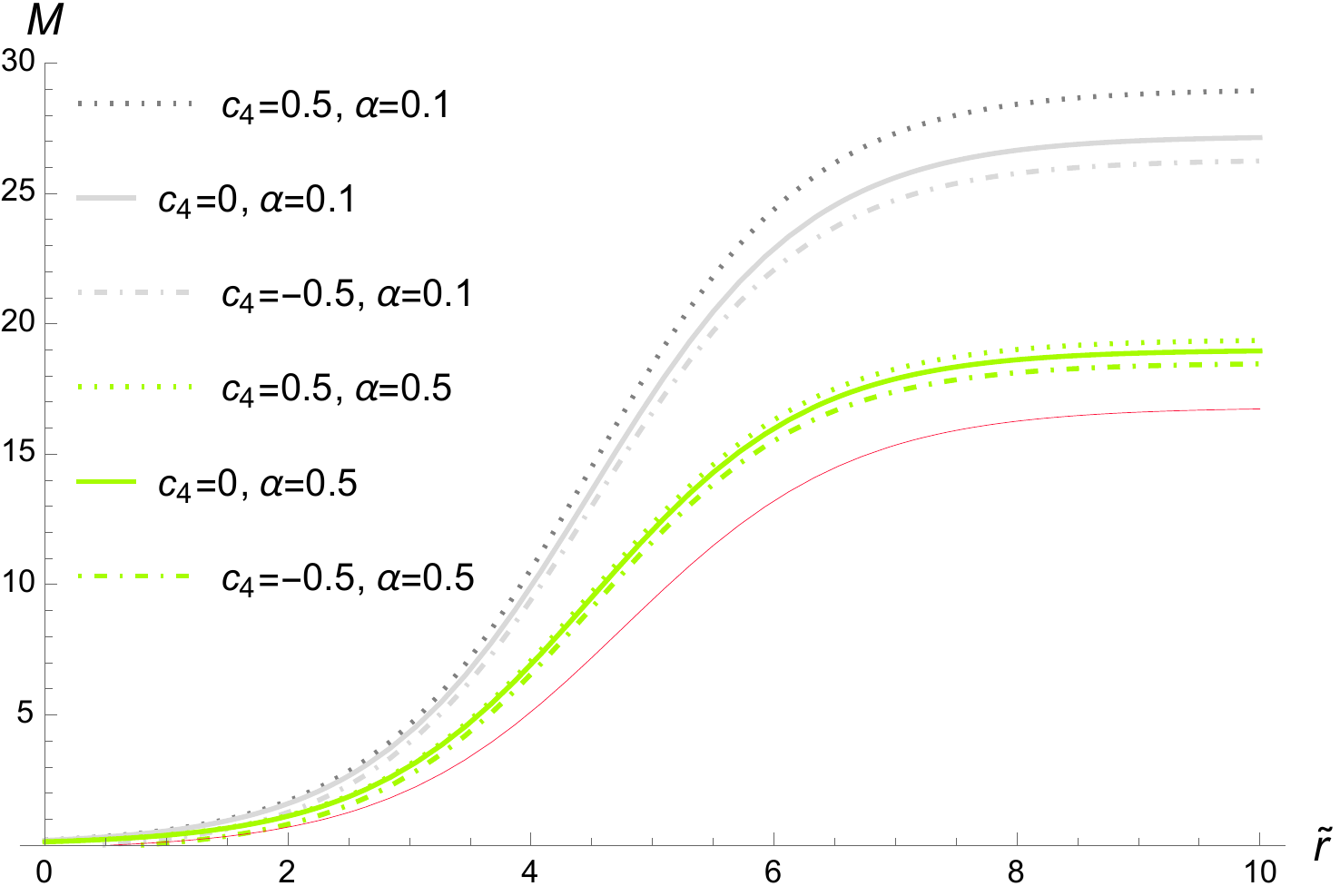}
 \caption{\footnotesize {$\Upgamma=\frac43$.}}
 \label{18}
 \end{subfigure}\qquad\qquad\qquad\qquad\qquad
 \begin{subfigure}[b]{0.325\textwidth}
 \centering
 \includegraphics[width=1.35\textwidth]{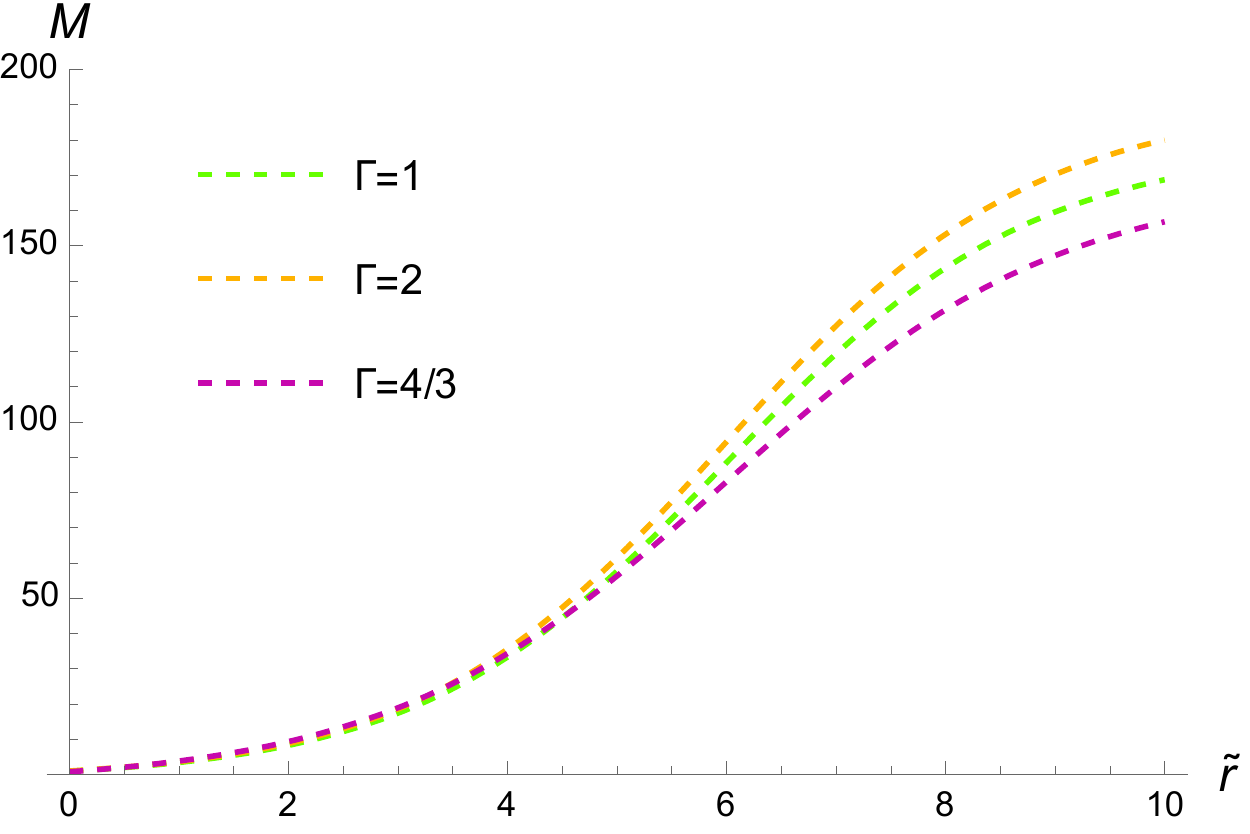}
 \caption{\footnotesize {The case where the fermionic fluid pressure is null, representing a gravitational decoupled boson star, for each value of the adiabatic index $\Upgamma$. }}
 \label{18meio}
 \end{subfigure}
 \caption{\footnotesize {Mass function $M(r)$ of gravitationally decoupled hybrid stars with a fermionic core, in generalized Horndeski gravity, for three values of $c_4\in\{0,\pm0.5\}$ and two values of $\upalpha$; $\Uplambda=1.5(M_{\scalebox{0.66}{$\textsc{p}$}}m^{2})^{1/3}$, and a fermionic fluid with $K=10^2\, m^{-2}M_{\scalebox{0.66}{$\textsc{p}$}}^{-2}$ and $\upsigma_0=0.26M_{\scalebox{0.66}{$\textsc{p}$}}$ is considered. 
 Light-gray curves represent the case where $\upalpha = 0.5$ and apple green curves regards the case where $\upalpha = 0.1$. Gravitational decoupled isothermal compact spheres of gas, with $\upsigma=0$ and $\Gamma=1$, correspond to the red curve in Fig. \ref{16}. 
 Neutron stars, with $\upsigma=0$ and $\Gamma=2$, are depicted as the red curve in Fig. \ref{17}. Gravitational decoupled white dwarfs, with $\upsigma=0$ and $\Gamma=\frac43$, are represented by the red curve in Fig. \ref{18}. }}
 \label{fig105}
\end{figure}

\begin{figure}[H]
 \centering
 \begin{subfigure}[b]{0.32\textwidth}
 \centering
 \!\!\!\!\!\!\!\!\!\!\!\!\!\!\!\includegraphics[width=1.45\textwidth]{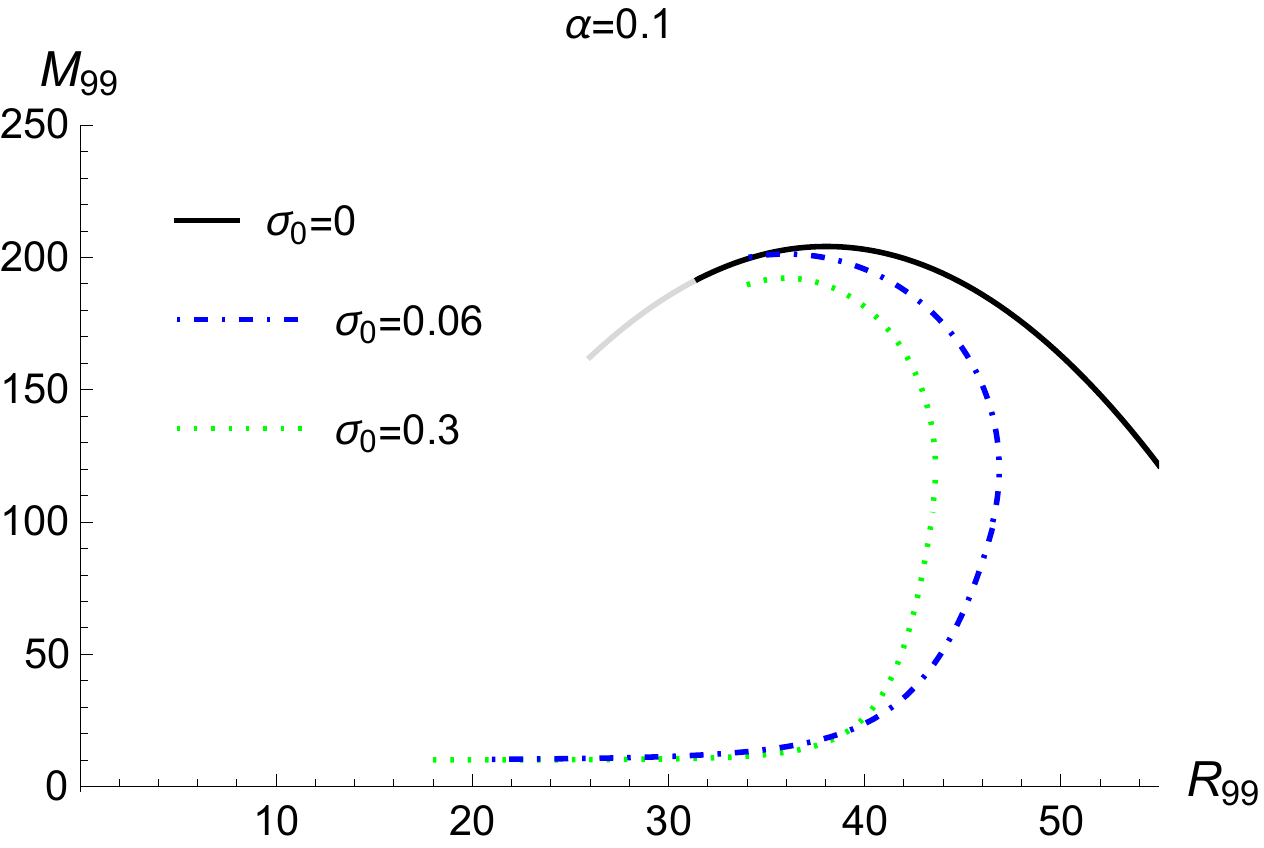}
 \caption{\footnotesize {$\Upgamma=1$.}}
 \label{22}
 \end{subfigure}\qquad\qquad\qquad
 \begin{subfigure}[b]{0.32\textwidth}
 \centering
 \includegraphics[width=1.5\textwidth]{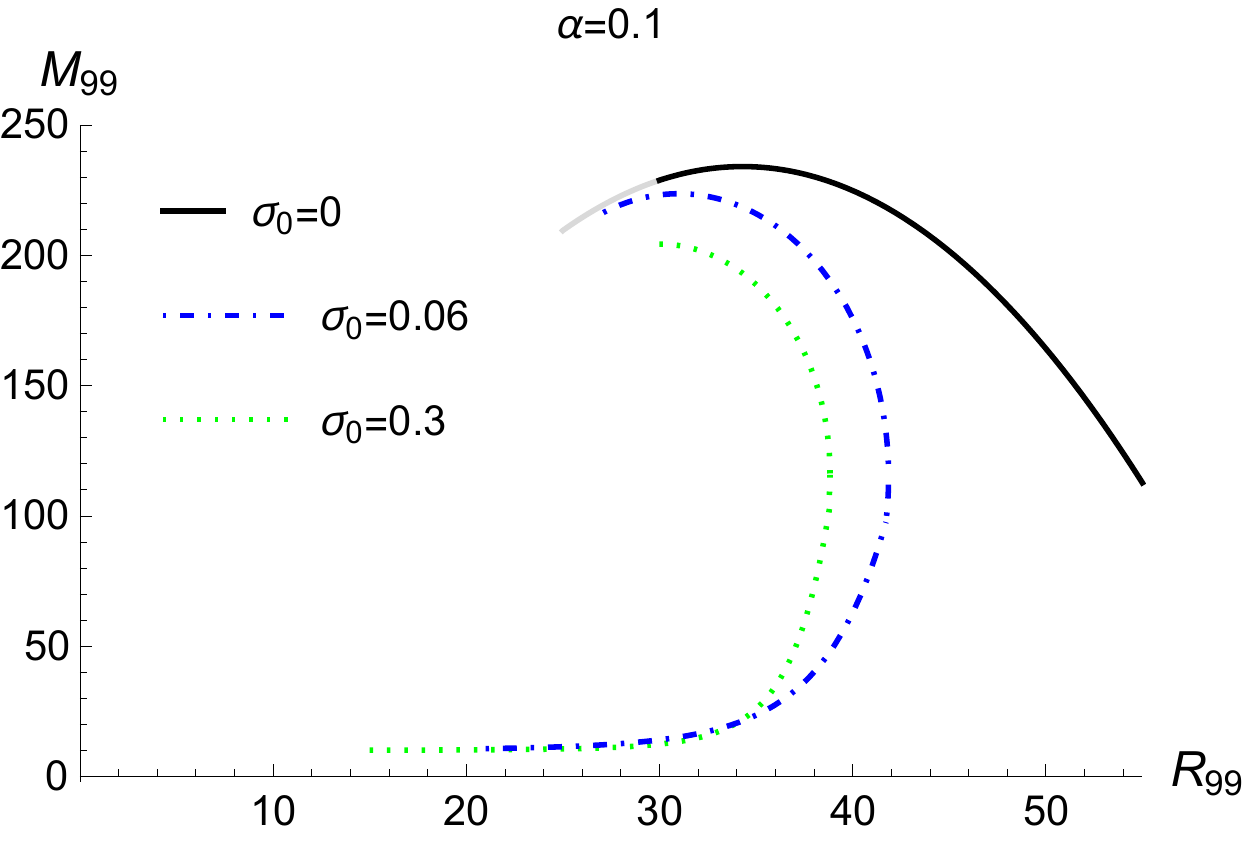}
 \caption{\footnotesize {$\Upgamma=2$.}}
 \label{23}
 \end{subfigure}\newline
 \vfill \vfill \vfill
 \begin{subfigure}[b]{0.325\textwidth}
 \centering
 \includegraphics[width=1.5\textwidth]{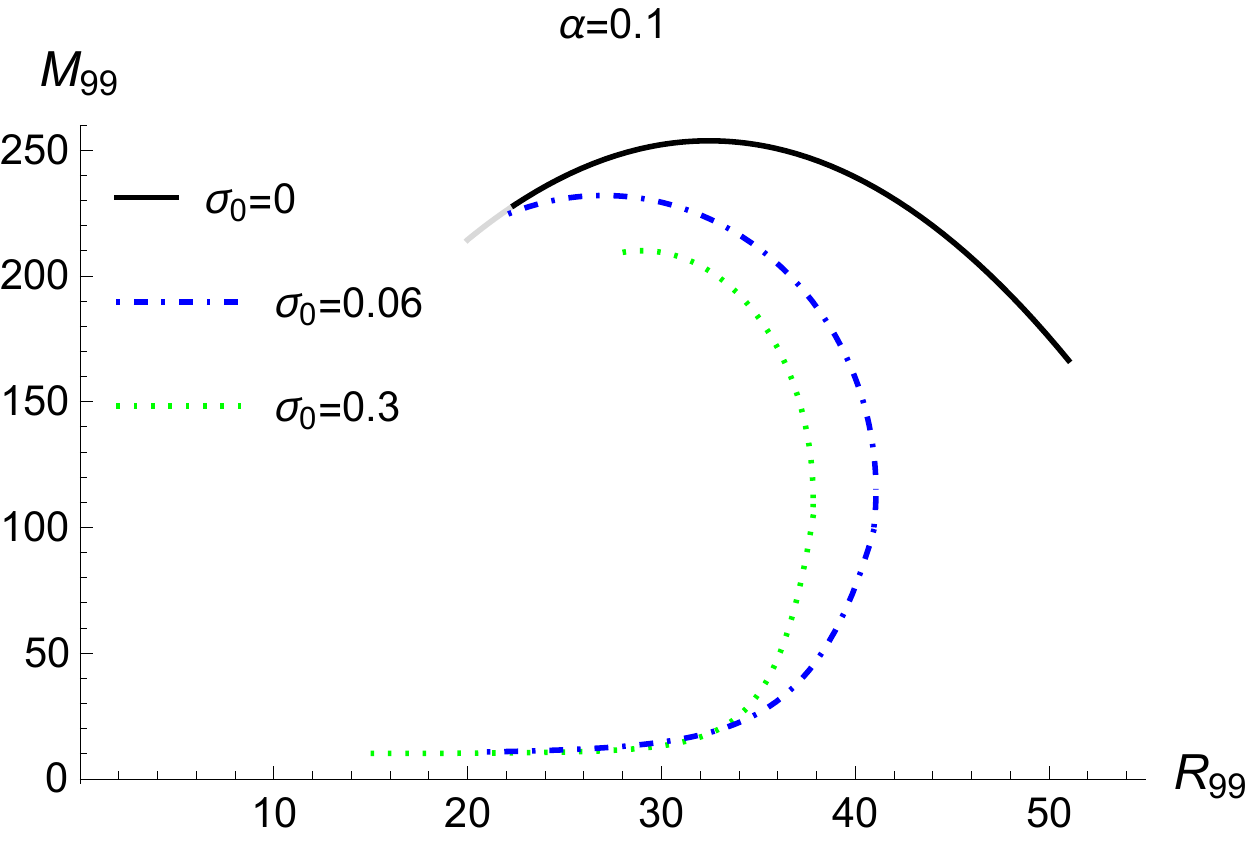}
 \caption{\footnotesize {$\Upgamma=\frac43$.}}
 \label{24}
 \end{subfigure}
\caption{\footnotesize {The mass portion $M_{99}$ as a function of the effective radius $R_{99}$ of gravitationally decoupled hybrid stars, for $\upalpha=0.1$ and $\Uplambda= 1.5(M_{\scalebox{0.66}{$\textsc{p}$}}m^{2})^{1/3}$. Each curve was derived by the variation of $p_0$ in the range from $ 10^{-6} m^{2}M_{\scalebox{0.66}{$\textsc{p}$}}^2$ to $10^{-2} m^{2}M_{\scalebox{0.66}{$\textsc{p}$}}^2$. The green dotted curve takes into account $\upsigma_0=0.3 M_{\scalebox{0.66}{$\textsc{p}$}}$ and the blue dot-dashed curve regards $\upsigma_0=0.06 M_{\scalebox{0.66}{$\textsc{p}$}}$, whereas $\upsigma_0=0$ regards the case for the fermionic [bosonic] core indicated by the black [light-gray] curve. }}\label{fig106}
\end{figure}

\begin{figure}[H]
 \centering
 \begin{subfigure}[b]{0.32\textwidth}
 \centering
 \!\!\!\!\!\!\!\!\!\!\!\!\!\!\!\includegraphics[width=1.3\textwidth]{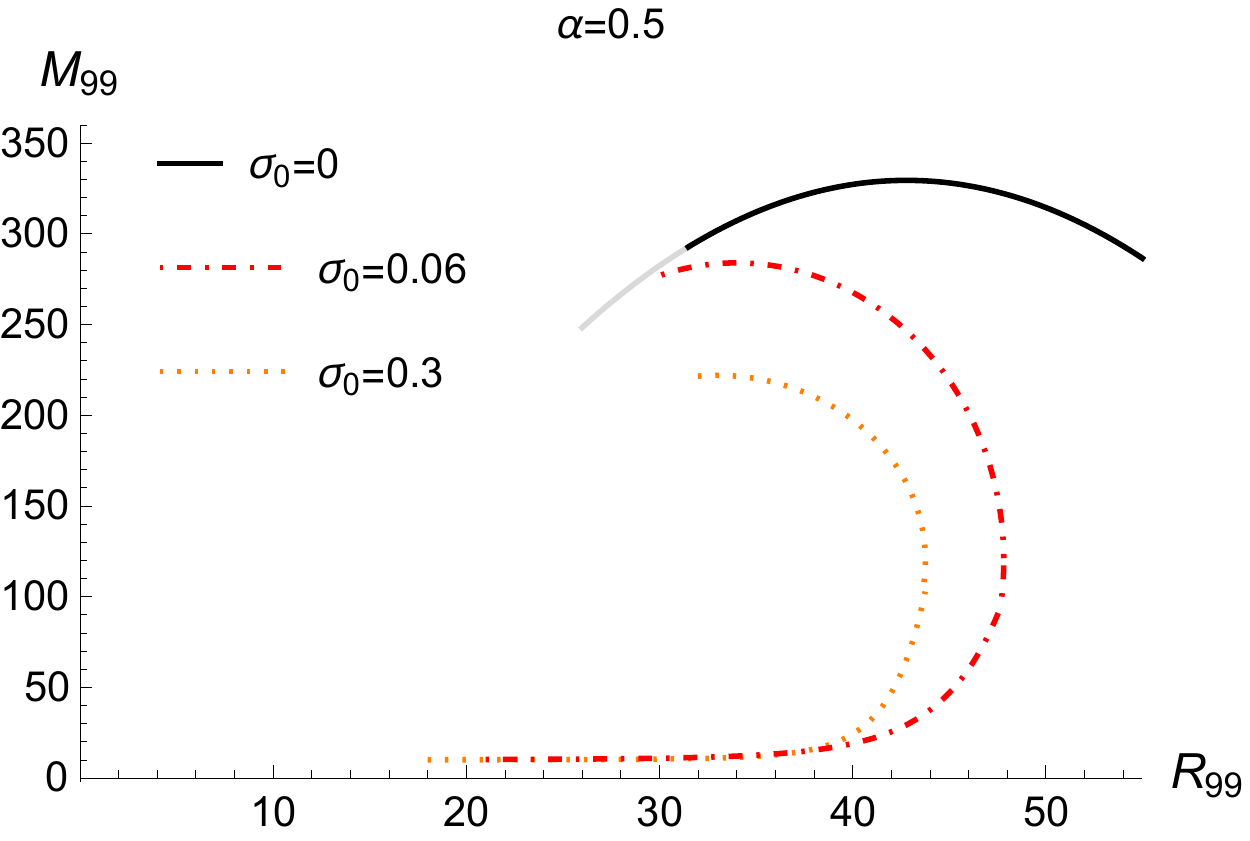}
 \caption{\footnotesize {$\Upgamma=1$.}}
 \label{25}
 \end{subfigure}\qquad\qquad\qquad
 \begin{subfigure}[b]{0.32\textwidth}
 \centering
 \includegraphics[width=1.3\textwidth]{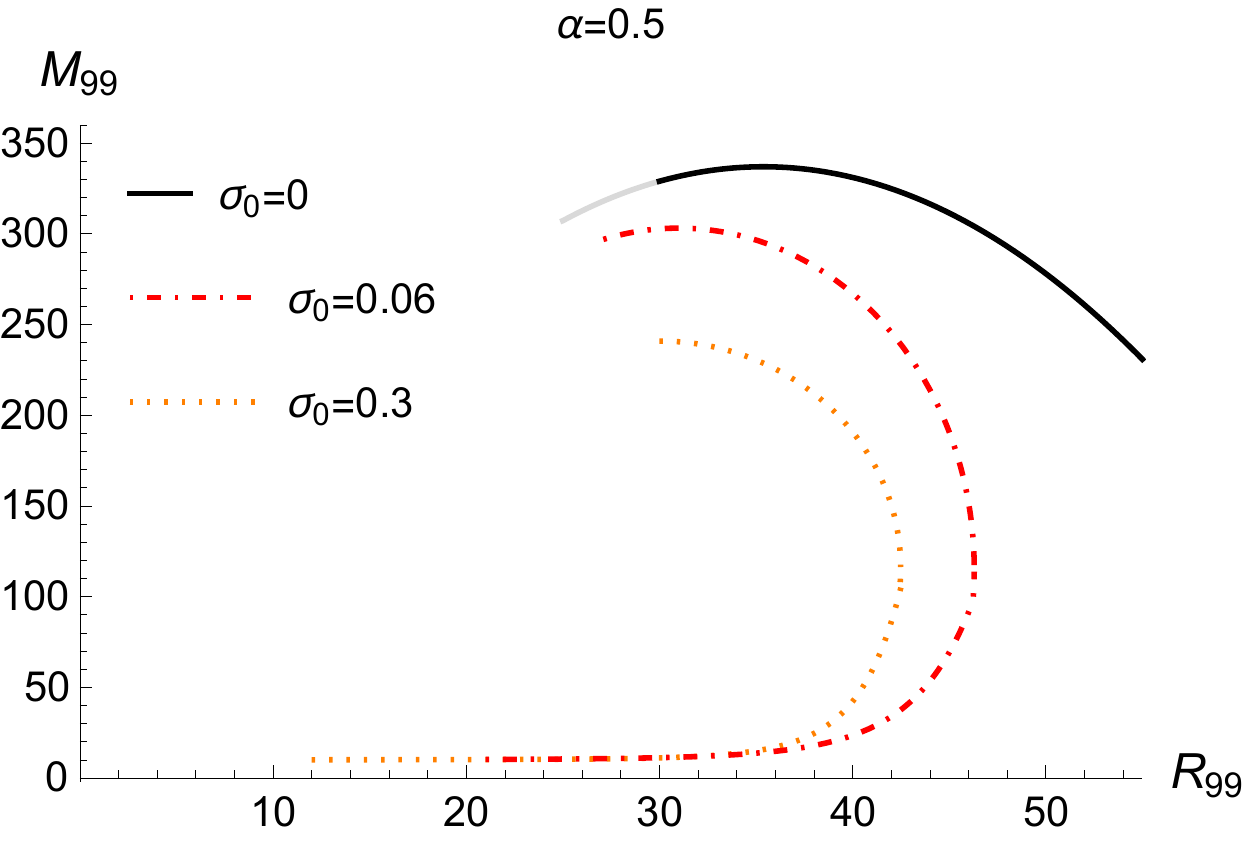}
 \caption{\footnotesize {$\Upgamma=2$.}}
 \label{26}
 \end{subfigure}\newline
 \hfill
 \begin{subfigure}[b]{0.325\textwidth}
 \centering
 \includegraphics[width=1.3\textwidth]{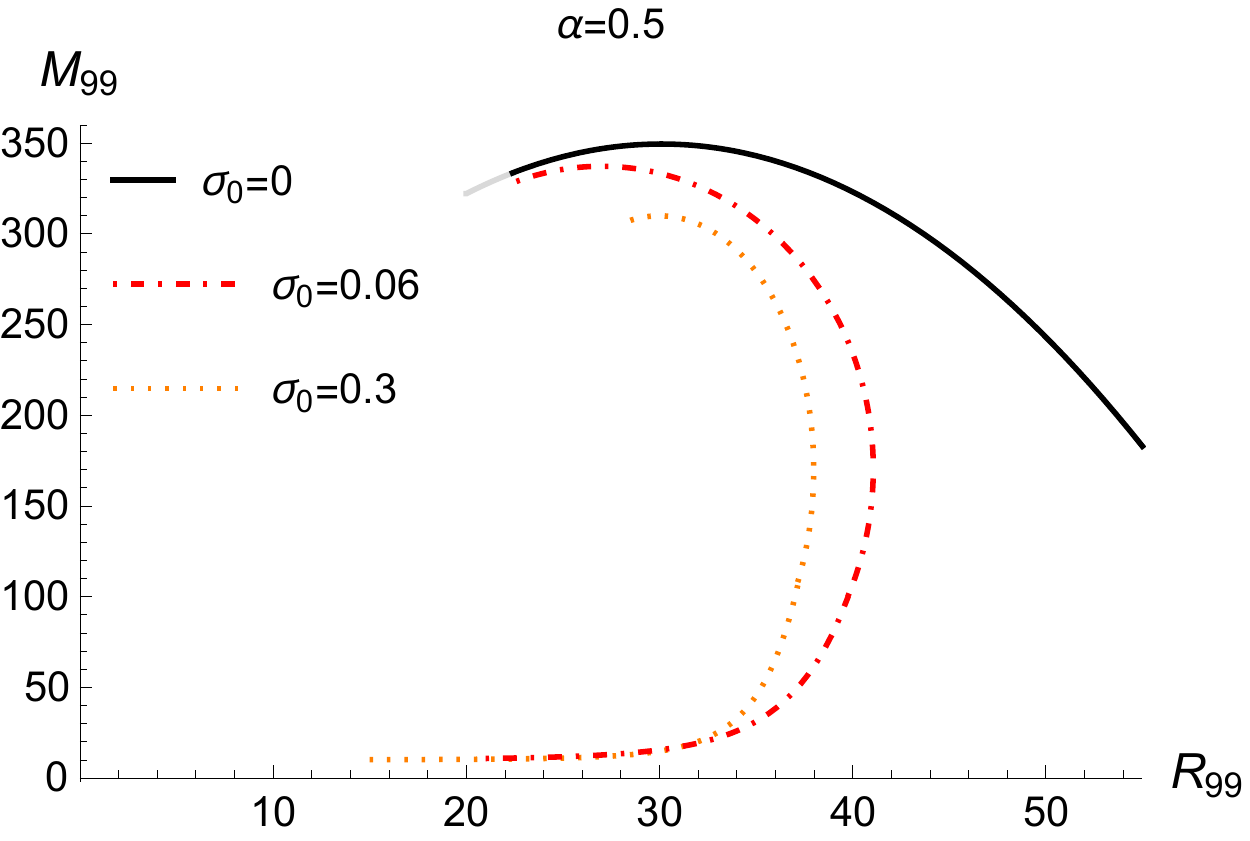}
 \caption{\footnotesize {$\Upgamma=\frac43$.}}
 \label{27}
 \end{subfigure}
\caption{\footnotesize {The maximal mass portion $M_{99}$ as a function of the effective radius $R_{99}$ of gravitationally decoupled hybrid stars, for $\upalpha=0.5$ and $\Uplambda= 1.5\, (M_{\scalebox{0.66}{$\textsc{p}$}}m^{2})^{1/3}$. Each curve was derived by the variation of $p_0$ in the range from $ 10^{-6} (mM_{\scalebox{0.66}{$\textsc{p}$}})^2$ to $10^{-2} (mM_{\scalebox{0.66}{$\textsc{p}$}})^2$. The orange dotted curve takes into account $\upsigma_0=0.3 M_{\scalebox{0.66}{$\textsc{p}$}}$ and the red dot-dashed curve regards $\upsigma_0=0.06 M_{\scalebox{0.66}{$\textsc{p}$}}$, whereas $\upsigma_0=0$ regards the case for the fermionic [bosonic] core indicated by the black [light-gray] curve. }}\label{fig107}
\end{figure}

	One can define the asymptotic maximal value of the mass function,
\beq
\label{avms}
M_\infty = \lim_{r\to\infty}M(r).
\eeq
Since Eq. (\ref{quan}) defined $\tilde{r}= m r$, taking the limit ${r\to\infty}$ is equivalent to ${\tilde{r}\to\infty}$.
The Misner--Sharp--Hernandez mass function $M(r)$ in Eq. (\ref{msh}) will be studied in the graphics in Figs. \ref{fig104} -- \ref{fig107} as a function of the radial coordinate, where a rescaling $M\mapsto{M}M_{\scalebox{0.66}{$\textsc{p}$}}^2/m$ is regarded. 
The plots in Fig. \ref{fig104} illustrate the mass profile of gravitational decoupled hybrid stars with a bosonic core, for which the scalar field behaviour is sharper towards its asymptotic null value than the respective fermionic fluid pressure profile, for each fixed value of $\Upgamma$, $c_4$, and $\upalpha$. Therefore the fermionic fluid density contributes more to the asymptotic maximal mass value. 
For the three values of $\Upgamma$ studied
in Figs. \ref{13} -- \ref{15}, the mass function profiles of gravitational decoupled hybrid stars with a bosonic core are very similar to the case of a gravitational decoupled neutron star, corresponding to $\upsigma=0$ and $\Gamma=2$, which is depicted as the continuous thin Caribbean green curves in Figs. \ref{13} -- \ref{15}, for the same value of $p_0$. 
Also, for $c_4>0$ [$c_4<0$], the mass $M(r)$ is greater  (lower) than in the gravitational decoupled Einstein-Klein-Gordon case corresponding to $c_4=0$. 
The plots in Figs. \ref{13} -- \ref{15} differ by the respective value of the adiabatic index $\Upgamma$ used to derive the curves. The Horndeski parameter $c_4$ and the decoupling parameter $\upalpha$ induce the hybrid star to have an asymptotic maximal mass that is lower when compared to the gravitationally decoupled neutron star case, for the adiabatic index $\Upgamma=2$; when compared to the isothermal sphere of gas emulating globular clusters, when $\Upgamma=1$; and when compared to the ultrarelativistic degenerate Fermi gas case, for $\Upgamma=\frac43$, depicted by the respective Caribbean green curves.
The higher the decoupling parameter $\upalpha$, the lower the hybrid star maximal mass is, irrespective of the value of $c_4$. Another important feature is the fact that the gravitational decoupled hybrid stars with $c_4=-0.5$ have asymptotic maximal mass higher than the gravitational decoupled Einstein--Klein--Gordon solution ($c_4=0$), whose asymptotic maximal mass value is higher than hybrid stars with $c_4=0.5$, for both values of the decoupling parameter $\upalpha$ and all values of the adiabatic index $\Upgamma$. Also, the minimal mass represented by the bright green dashed curves in Figs. \ref{13} -- \ref{15} regard a gravitational decoupled boson star solution.

Fig. \ref{fig105} displays the mass function profile of gravitationally decoupled hybrid stars with a fermionic core, for which the scalar field behavior is smoother towards its asymptotic null value than the respective fermionic fluid pressure profile, for each fixed value of $\Upgamma$, $c_4$, and $\upalpha$. In this case, the fermionic fluid density is suppressed, when compared to the scalar field contribution to the value of the asymptotic maximal mass. 
For the three values of $\Upgamma$ studied
in Figs. \ref{16} -- \ref{18}, the mass function profiles of gravitational decoupled hybrid stars with a fermionic core are qualitatively similar to the case of gravitational decoupled neutron stars, corresponding to $\upsigma=0$ and $\Gamma=2$, which is depicted as the thin magenta curves in Figs. \ref{16} -- \ref{18} for different values of the adiabatic index. However, their asymptotic maximal values of the mass function, as well as their rate of increment, are completely different, quantitatively. For $c_4>0$ [$c_4<0$], the mass function is higher (lower) than the gravitational decoupled Einstein-Klein-Gordon counterpart, which can be obtained by considering $c_4=0$. 

The plots in Figs. \ref{16} -- \ref{18} differ by the respective value of the adiabatic index $\Upgamma$ used to derive the curves. Similarly to the case of gravitational decoupled hybrid stars with bosonic core, both the Horndeski parameter $c_4$ and the decoupling parameter $\upalpha$ yield a maximal value of the Misner--Sharp--Hernandez mass function that can be lower or intermediary than the asymptotic maximal mass of the gravitational decoupled boson stars, depicted by the respective magenta curves.
The higher the decoupling parameter $\upalpha$, the lower the hybrid star maximal mass is, irrespective of the value of the Horndeski parameter $c_4$. Besides, the higher the value of the adiabatic index $\Upgamma$, the bigger the gap between the group of light-gray curves, corresponding to $\upalpha=0.1$, and the group of curves, corresponding to $\upalpha=0.5$, is. It means that the asymptotic maximal mass value of hybrid stars decreases at a higher rate, for higher values of the decoupling parameter $\upalpha$. 
Now, contrary to the bosonic core case, gravitational decoupled hybrid stars with fermionic core with $c_4=-0.5$ have maximal mass higher than the gravitational decoupled Einstein--Klein--Gordon solution ($c_4=0$), whose maximal mass value is higher than hybrid stars with $c_4=0.5$, for both values of the decoupling parameter $\upalpha$ and all values of the adiabatic index $\Upgamma$. Also contrary to the bosonic core case, the gravitational decoupled hybrid star with fermionic core has mass function displayed in Fig. \ref{18meio}. It shows that the mass function of the gravitational decoupled hybrid star with the fermionic core is much lower, when compared to the isothermal sphere of gas, when $\Upgamma=1$ (green dashed curve); when compared to the gravitationally decoupled neutron star case, for the adiabatic index $\Upgamma=2$ (orange dashed curve); and when compared to ultrarelativistic degenerate Fermi gas case, for $\Upgamma=\frac43$ (purple dashed curve).
For gravitational decoupled isothermal compact sphere of gas, with $\upsigma=0$ and $\Gamma=1$ in Fig. \ref{16}, although its mass function increases with a rate that is lower than all cases concerning $\upalpha=0.1$, its asymptotic value $M_\infty$ converges to the gravitational decoupled Einstein--Klein--Gordon case $c_4=0$, $\upalpha=0.1$.  
Next, regarding gravitational decoupled neutron stars, with $\upsigma=0$ and $\Gamma=2$ in Fig. \ref{17}, this case splits the range of $M_\infty$, associated with gravitational decoupled hybrid stars, into two domains. The first one represents the $\upalpha=0.1$ case, whose asymptotic values of the mass function are higher than $M_\infty$ associated with gravitational decoupled neutron stars. The second domain represents the 
 $\upalpha=0.5$ case, whose asymptotic values of the mass function are lower than $M_\infty$ of gravitational decoupled neutron stars, irrespectively of the value of $\upalpha$. 
Now, a relevant feature regards gravitational decoupled white dwarfs, with $\upsigma=0$ and $\Gamma=\frac43$, in Fig. \ref{18}. For this case, the asymptotic value of the mass function is lower than all other cases. The case where the fermionic fluid pressure is null represents a gravitationally decoupled boson star, separately displayed in Fig. \ref{18meio} for each value of the adiabatic index $\Upgamma$. All cases in Fig. \ref{fig105} show that the decoupling parameter $\upalpha$ induces a decrement of $M_\infty$.

Tables \ref{ta1} -- \ref{ta3} display the respective values of the asymptotic value of the mass function \eqref{avms} regarding gravitationally decoupled hybrid stars with a bosonic core, for different values of the involved parameters. The values can be read off Fig. \ref{fig104}
\begin{table}[H]
\parbox{.45\linewidth}{
\centering
\medbreak
\begin{tabular}{||c||c|c|c||}
\hline\hline& \;$c_4=-0.5$\; & \;$c_4=0$ \;&\; $c_4 = 0.5$\; \\
 \hline\hline
\; $\upalpha = 0.1$\;&\;\;\;160.13 \;&\; 156.17\;&\; 145.78 \;\\ \hline
 \;$\upalpha = 0.5$\;&\; 133.40\;&\; 129.59\;&\; 124.13 \;\\ \hline
\hline
\end{tabular}
\caption{\footnotesize {Asymptotic value of the mass function, $M_\infty$, for $\Gamma=1$, in Fig. \ref{13}}.} \label{ta1}}
\hfill
\parbox{.45\linewidth}{
\centering
\medbreak
\begin{tabular}{||c||c|c|c||}
\hline\hline& \;$c_4=-0.5$\; & \;$c_4=0$ \;&\; $c_4 = 0.5$\; \\
 \hline\hline
\; $\upalpha = 0.1$\;&\;\;\;157.11 \;&\; 155.62\;&\; 149.35 \;\\ \hline
\; $\upalpha = 0.5$\;&\; 138.16\;&\; 129.48\;&\; 124.10 \;\\ \hline
\hline
\end{tabular}
\caption{\footnotesize {Asymptotic value of the mass function, $M_\infty$, for $\Gamma=2$, in Fig. \ref{14}.} }\label{ta2}}\vfill\medbreak
\parbox{.45\linewidth}{
\centering
\medbreak
\begin{tabular}{||c||c|c|c||}
\hline\hline& \;$c_4=-0.5$\; & \;$c_4=0$ \;&\; $c_4 = 0.5$\; \\
 \hline\hline
 \;$\upalpha = 0.1$\;&\;\;\;184.49 \;&\; 181.12\;&\; 172.67 \;\\ \hline
\; $\upalpha = 0.5$\;&\; 127.30\;&\; 120.09\;&\; 110.85 \;\\ \hline
\hline
\end{tabular}
\caption{\footnotesize {Asymptotic value of the mass function, $M_\infty$, for $\Gamma=\frac43$, in Fig. \ref{15}.} }\label{ta3}}
\end{table}
\noindent
Also, the asymptotic value of the mass function is regarded for gravitationally decoupled hybrid stars with a fermionic core, in Tables \ref{ta4} -- \ref{ta6}, whose values are read off Fig. \ref{fig105}
\begin{table}[H]
\parbox{.45\linewidth}{
\centering
\medbreak
\begin{tabular}{||c||c|c|c||}
\hline\hline& \;$c_4=-0.5$\; & \;$c_4=0$ \;&\; $c_4 = 0.5$\; \\
 \hline\hline
\;$\upalpha = 0.1$\;&\;\;\;16.08 \;&\; 17.12\;&\; 18.01 \;\\ \hline
 \;$\upalpha = 0.5$\;&\; 13.10\;&\; 13.79\;&\; 14.23 \;\\ \hline
\hline
\end{tabular}
\caption{\footnotesize {Asymptotic value of the mass function, for $\Gamma=1$, in Fig. \ref{16}}.} \label{ta4}}
\hfill
\parbox{.45\linewidth}{
\centering
\medbreak
\begin{tabular}{||c||c|c|c||}
\hline\hline& \;$c_4=-0.5$\; & \;$c_4=0$ \;&\; $c_4 = 0.5$\; \\
 \hline\hline
 \;$\upalpha = 0.1$\;&\;\;\;17.77 \;&\; 18.57\;&\; 20.16 \;\\ \hline
 \;$\upalpha = 0.5$\;&\; 13.29\;&\; 14.02\;&\; 15.38 \;\\ \hline
\hline
\end{tabular}
\caption{\footnotesize {Asymptotic value of the mass function, for $\Gamma=2$, in Fig. \ref{17}.} }\label{ta5}}\vfill\medbreak
\parbox{.45\linewidth}{
\centering
\medbreak
\begin{tabular}{||c||c|c|c||}
\hline\hline& \;$c_4=-0.5$\; & \;$c_4=0$ \;&\; $c_4 = 0.5$\; \\
 \hline\hline
 \;$\upalpha = 0.1$\;&\;\;\;16.08 \;&\; 17.12\;&\; 18.01 \;\\ \hline
 \;$\upalpha = 0.5$\;&\; 13.10\;&\; 13.79\;&\; 14.23 \;\\ \hline
\hline
\end{tabular}
\caption{\footnotesize {Asymptotic value of the mass function, for $\Gamma=\frac43$, in Fig. \ref{18}.} }\label{ta6}}
\end{table}
\noindent
Figs. \ref{fig104} and \ref{fig105}, and their particular portrait displayed in Tables \ref{ta1} -- \ref{ta6}, show that the asymptotic value of the mass function for gravitationally decoupled hybrid stars with a fermionic core is one order of magnitude lower than their counterparts with a bosonic core.

When boson stars are investigated, among other species of stars, the effective radius $R_{99}$ of a self-gravitating compact distribution defines a region that encloses 99\% of the boson star total mass, namely, $M(R_{99})/M=0.99$. One can emulate this concept for determining the effective radius of gravitational decoupled hybrid stars, since analogously to boson stars, Figs. \ref{fig100} and \ref{fig102} showed that the scalar field profile decreases monotonically along the radial coordinate \cite{Roque:2021lvr}. Therefore one can similarly define the effective radius of gravitationally decoupled hybrid stars, also here denoted by $R_{99}$. It is worth emphasizing that the effective radius $R_{99}$ coincides to the standard classical radius, ${\rm R}$, at which $p({\rm R})=0$, to the case of hybrid stars with a fermionic core.
The dependence of $M_{99}=M(R_{99})$ to the effective radius of gravitational decoupled hybrid stars are numerically computed and displayed in Fig. \ref{fig106}, for $\upalpha = 0.1$, and in Fig. \ref{fig107}, for $\upalpha=0.5$. The values $\upsigma_0=0.06 M_{\scalebox{0.66}{$\textsc{p}$}}$ and $\upsigma_0=0.3 M_{\scalebox{0.66}{$\textsc{p}$}}$ of the scalar field central amplitude are adopted, and compared to the respective stellar configurations when $\upsigma=0$, for three values of the adiabatic index $\Upgamma$. 
 The results are shown in Figs.~\ref{fig106} and \ref{fig107}, respectively for $\upalpha = 0.1$ and $\upalpha=0.5$, for the generalized Hordenski theory, taking into account the Horndeski parameter $c_4=-0.5$ for comparison to the literature when $\upalpha=0$ \cite{Roque:2021lvr}. 
 Gravitational decoupled hybrid stellar configurations have effective radii and associated masses that increase up to an absolute supremum, where the effective radii and associated masses decrease.

Fig. \ref{fig106} shows the case $\upalpha=0.1$.
 The maximal mass portion $M_{99}$ as a function of the effective radius $R_{99}$ increases up to an absolute maximum, for $\upalpha=0.1$ and $\Uplambda= 1.5(M_{\scalebox{0.66}{$\textsc{p}$}}m^{2})^{1/3}$. The green dotted curve takes into account $\upsigma_0=0.3 M_{\scalebox{0.66}{$\textsc{p}$}}$ and the blue dot-dashed curve regards $\upsigma_0=0.06 M_{\scalebox{0.66}{$\textsc{p}$}}$, whereas $\upsigma_0=0$ regards the case for the fermionic [bosonic] core indicated by the black [light-gray] curve.
Table \ref{ta7} displays the maximal mass portion $M_{99}$ as a function of the effective radius $R_{99}$, for $\upalpha=0.1$, regarding gravitationally decoupled hybrid stars with a bosonic core, for different values of the involved parameters. The values can be read off Fig. \ref{fig106}.
\begin{table}[H]
\centering
--------------------- $(M_{99}^{\scalebox{0.66}{$\textsc{max}$}}, R_{99}^{\scalebox{0.66}{$\textsc{max}$}})$ ---------------------\medbreak
\begin{tabular}{||c||c|c|c||}
\hline\hline& \;$\upsigma_0=0$\; & \;$\upsigma_0=0.06M_{\scalebox{0.66}{$\textsc{p}$}}$ \;&\; $\upsigma_0=0.3 M_{\scalebox{0.66}{$\textsc{p}$}}$\; \\
 \hline\hline
\; $\Upgamma = 1$\;&\;\!\,\,\,(204.47, 38.21) \;&\; (198.05, 35.28)\;&\; (179.88, 36,69) \;\\ \hline
\; $\Upgamma = 2$\;&\; (238,03, 34.17)\;&\; (225.32, 32.81)\;&\; (203.75, 30.07) \;\\ \hline
\; $\Upgamma = \frac43$\;&\; (246.86, 32.79)\;&\;\;\;(232.90, 27.93) \;&\; (213.67, 28.27) \;\\ \hline
\hline\hline
\end{tabular}
\caption{\footnotesize {The maximal mass portion $M_{99}$ as a function of the effective radius $R_{99}$, for $\upalpha=0.1$}.} \label{ta7}
\end{table}
\noindent One can realize from Table \ref{ta7} that 
for each fixed value of the scalar field central density $\upsigma_0$, as the adiabatic index $\Upgamma$ increases, the greater the $M_{99}$ and the smaller the effective radius $R_{99}$ are. By fixing the adiabatic index $\Upgamma$, both the mass portion $M_{99}$ and the effective radius $R_{99}$ decrease as $\upsigma_0$ increases.

 A similar behavior holds for the case $\upalpha=0.5$, illustrated in Table \ref{ta8}, which displays the maximal mass portion $M_{99}$ as a function of the effective radius $R_{99}$ for $\upalpha=0.5$, regarding gravitationally decoupled hybrid stars with a bosonic core, for different values of the involved parameters. The values can be read off Fig. \ref{fig107}. Also for each fixed value of the scalar field central density $\upsigma_0$, the higher the adiabatic index $\Upgamma$, the greater the $M_{99}$ and the smaller the effective radius $R_{99}$ are. When the adiabatic index $\Upgamma$ is fixed, both the mass portion $M_{99}$ and the effective radius $R_{99}$ decrease as $\upsigma_0$ increases.

\begin{table}[H]
\centering
--------------------- $(M_{99}^{\scalebox{0.66}{$\textsc{max}$}}, R_{99}^{\scalebox{0.66}{$\textsc{max}$}})$ ---------------------\medbreak
\begin{tabular}{||c||c|c|c||}
\hline\hline& \;$\upsigma_0=0$\; & \;$\upsigma_0=0.06M_{\scalebox{0.66}{$\textsc{p}$}}$ \;&\; $\upsigma_0=0.3 M_{\scalebox{0.66}{$\textsc{p}$}}$\; \\
 \hline\hline
\; $\Upgamma = 1$\;&\;\;\;(330.09, 43.84) \;&\; (280.15, 34.60)\;&\; (220.18, 33,55) \;\\ \hline
\; $\Upgamma = 2$\;&\; (336.78, 35.72)\;&\; (301.34, 30.94)\;&\; (240.48, 31.48) \;\\ \hline
\; $\Upgamma = \frac43$\;&\; (340.01, 30.96)\;&\;\;\;(336.90, 27.62) \;&\; (310.56, 30.23) \;\\ \hline
\hline
\end{tabular}
\caption{\footnotesize {The mass portion $M_{99}$ as a function of the effective radius $R_{99}$, for $\upalpha=0.5$}.} \label{ta8}
\end{table}
\noindent
Increasing the value of $c_4$ makes the results to be qualitatively similar, with the mass portion $M_{99}$ also increasing, for each fixed the effective radius $R_{99}$. Besides, the rescaling \begin{equation}\label{conv3}
 M_{99} \mapsto \frac{5.31}{m} {M}_{99}\times 10^{-12}M_{\odot} ,\quad
 R_{99} \mapsto \frac{1.97}{m}{R}_{99}\times 10^{-10}\,\rm{km}
\end{equation}
in Ref. \cite{Roque:2021lvr} recovers the observable values of $M_{99}$ and $R_{99}$, in Figs. \ref{fig106} and \ref{fig107}, where the scalar field mass parameter mass 
parameter $m$ in Eq. (\ref{conv3}) must be input in electron-volt.

\begin{figure}[H]
 \centering
 \begin{subfigure}[b]{0.32\textwidth}
 \centering
 \!\!\!\!\!\!\!\!\!\!\!\!\!\!\!\includegraphics[width=1.45\textwidth]{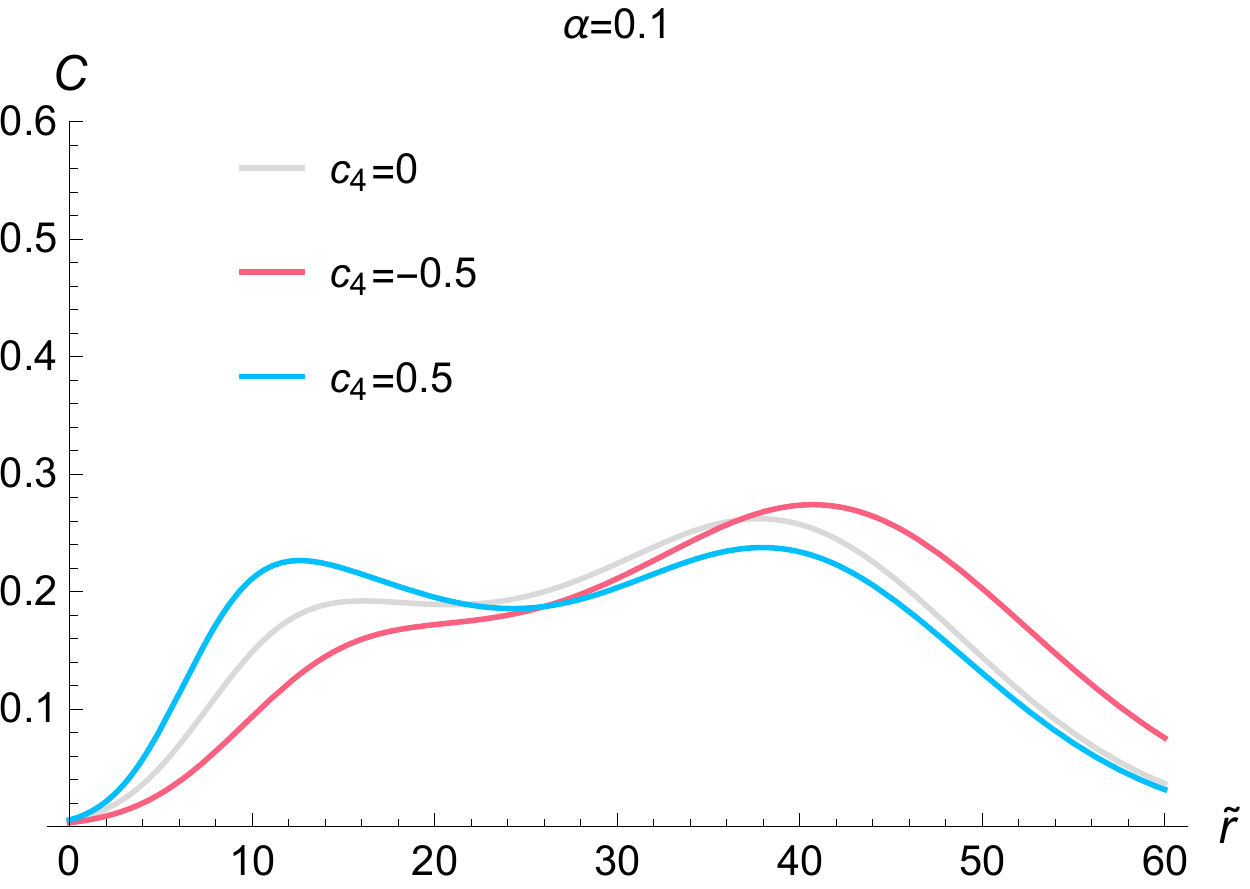}
 \caption{\footnotesize {$\Upgamma=1$.}}
 \label{28}
 \end{subfigure}\qquad\qquad\qquad
 \begin{subfigure}[b]{0.32\textwidth}
 \centering
 \includegraphics[width=1.5\textwidth]{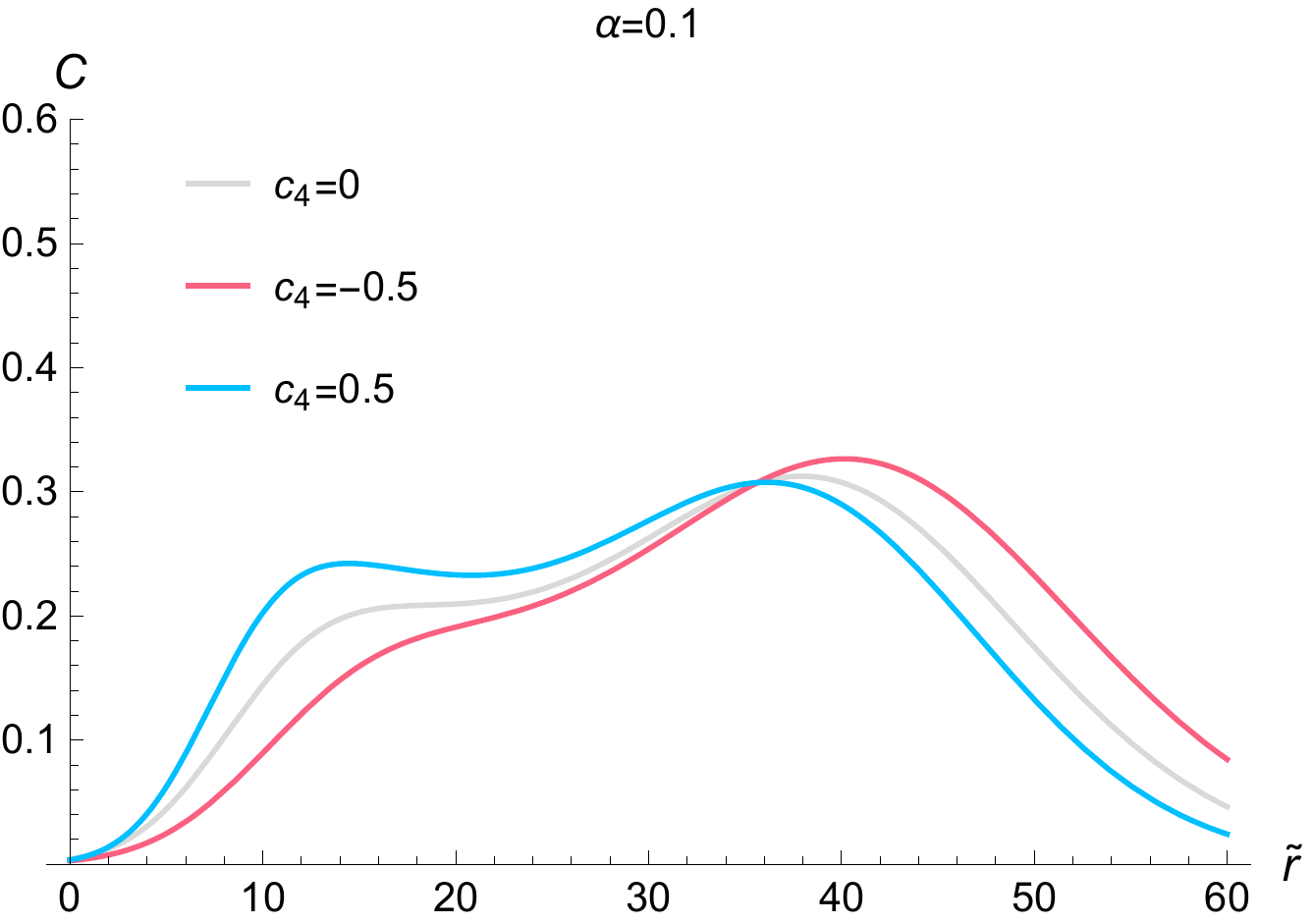}
 \caption{\footnotesize {$\Upgamma=2$.}}
 \label{29}
 \end{subfigure}\newline
 \hfill
 \begin{subfigure}[b]{0.325\textwidth}
 \centering
 \includegraphics[width=1.5\textwidth]{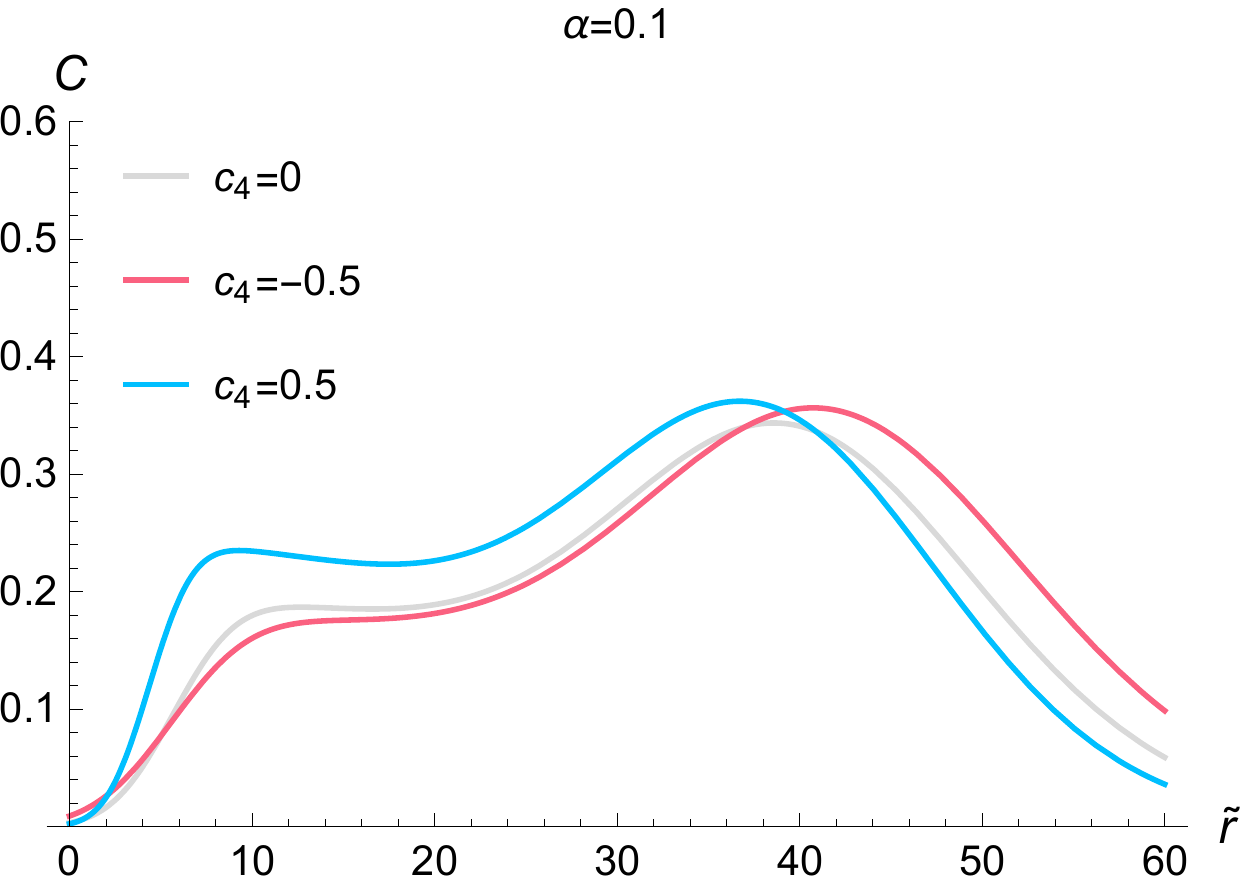}
 \caption{\footnotesize {$\Upgamma=\frac43$.}}
 \label{30}
 \end{subfigure}
	\caption{\footnotesize {Compactness of gravitational decoupled hybrid stars as a function of the radial coordinate, for $\upalpha=0.1$ and three values of $c_4$, and $\upsigma_0=0.26M_{\scalebox{0.66}{$\textsc{p}$}}$. The cyan [red; light-gray] curve represents the case $c_4=0.5$ [$c_4=-0.5; c_4=0$], for $\Uplambda= (M_{\scalebox{0.66}{$\textsc{p}$}}m^{2})^{1/3}$, $K=10^2\, m^{-2}M_{\scalebox{0.66}{$\textsc{p}$}}^{-2}$, $\upepsilon_0=10^{-2}\, (M_{{\scalebox{0.66}{$\textsc{p}$}}}m)^{2}$, and $p_0=3\times 10^{-3}\, (M_{{\scalebox{0.66}{$\textsc{p}$}}}m)^{2}$. Fig. \ref{28} shows the case $\Upgamma=1$ and Fig. \ref{29} illustrates the case where $\Upgamma=2$, whereas Fig. \ref{30} shows the case $\Upgamma=\frac43$.}}\label{fig108}
\end{figure}

\begin{figure}[H]
 \centering
 \begin{subfigure}[b]{0.32\textwidth}
 \centering
 \!\!\!\!\!\!\!\!\!\!\!\!\!\!\!\includegraphics[width=1.45\textwidth]{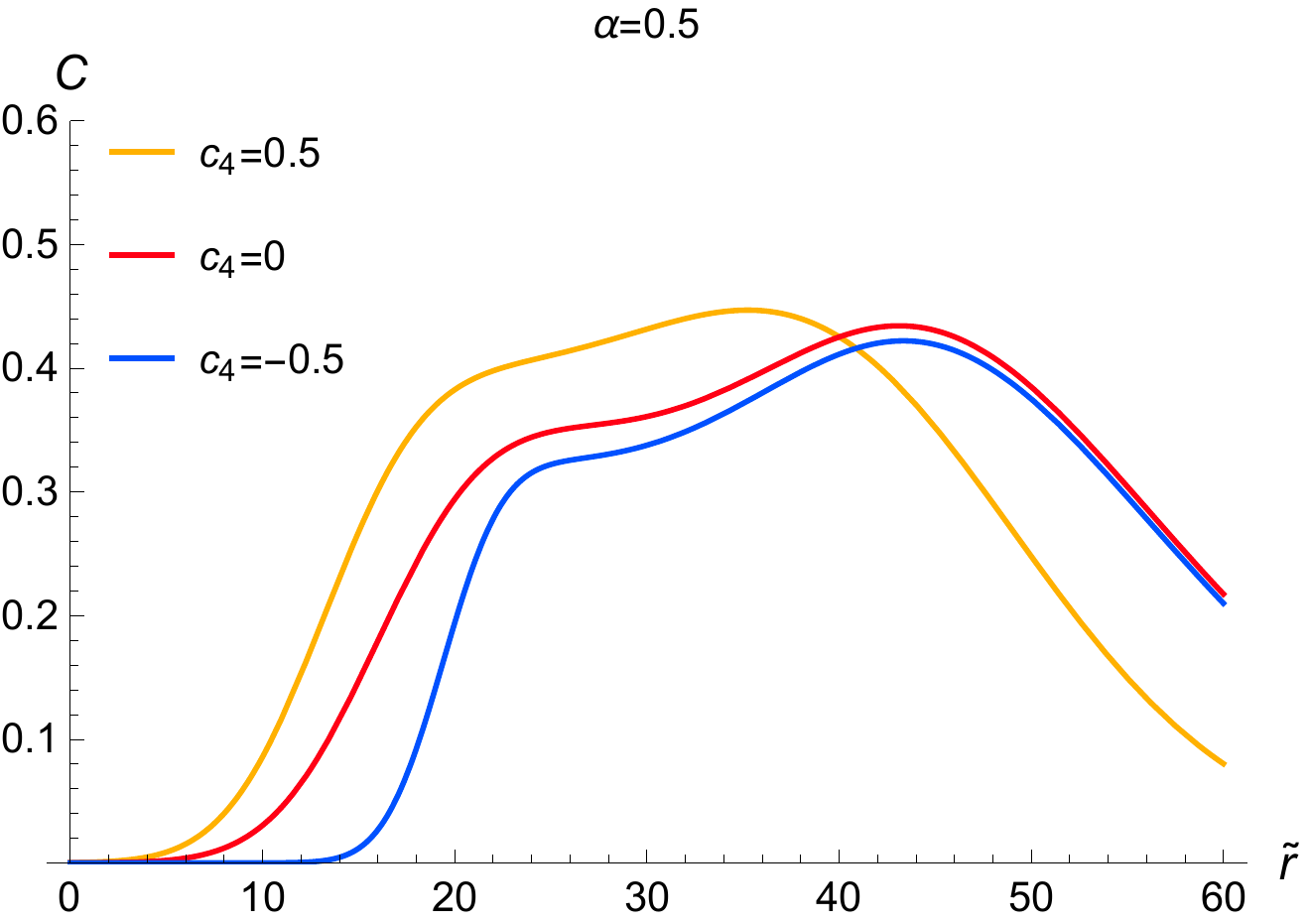}
 \caption{\footnotesize {$\Upgamma=1$.}}
 \label{31}
 \end{subfigure}\qquad\qquad\qquad
 \begin{subfigure}[b]{0.32\textwidth}
 \centering
 \includegraphics[width=1.5\textwidth]{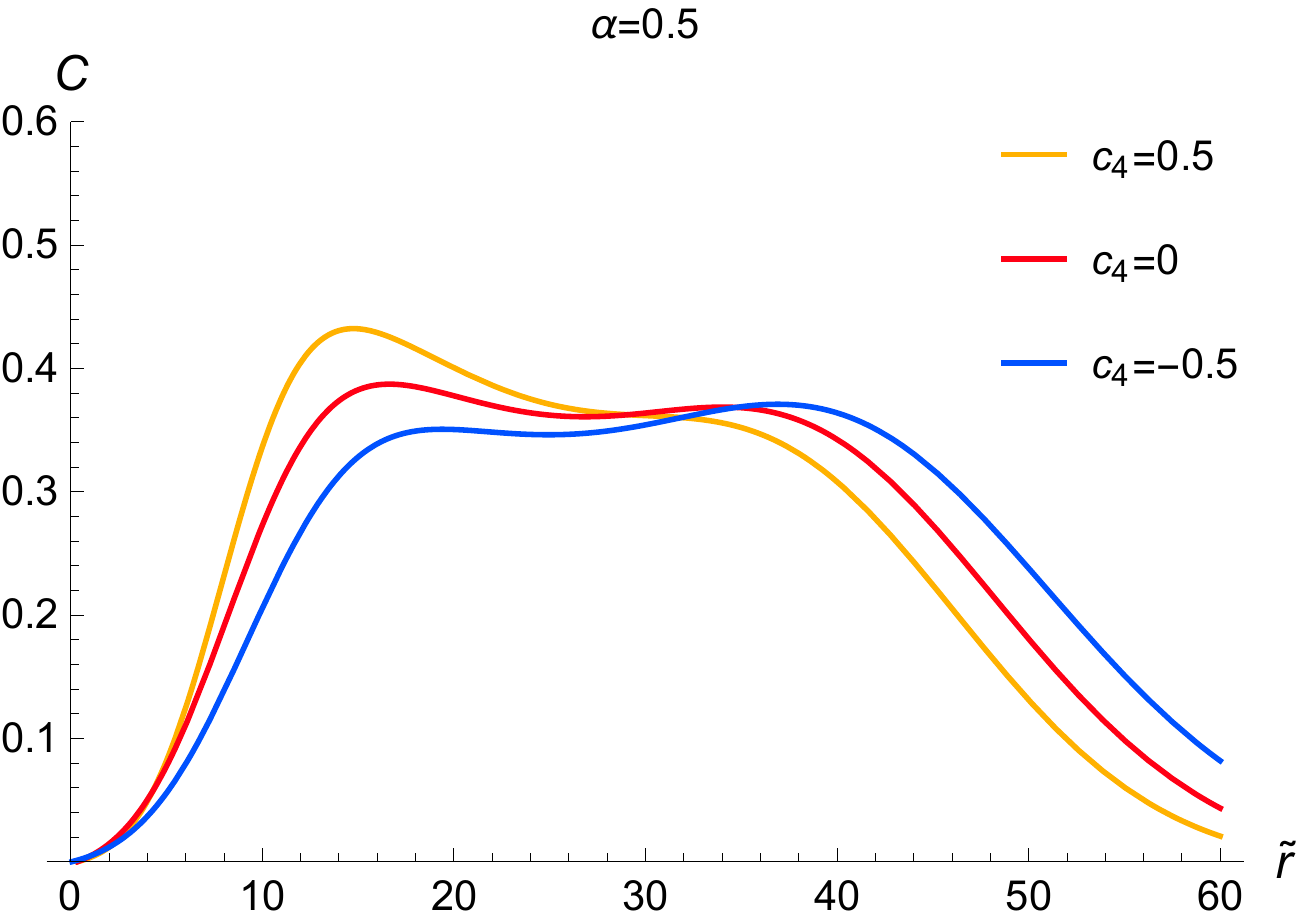}
 \caption{\footnotesize {$\Upgamma=2$.}}
 \label{32}
 \end{subfigure}\newline
 \hfill
 \begin{subfigure}[b]{0.325\textwidth}
 \centering
 \includegraphics[width=1.5\textwidth]{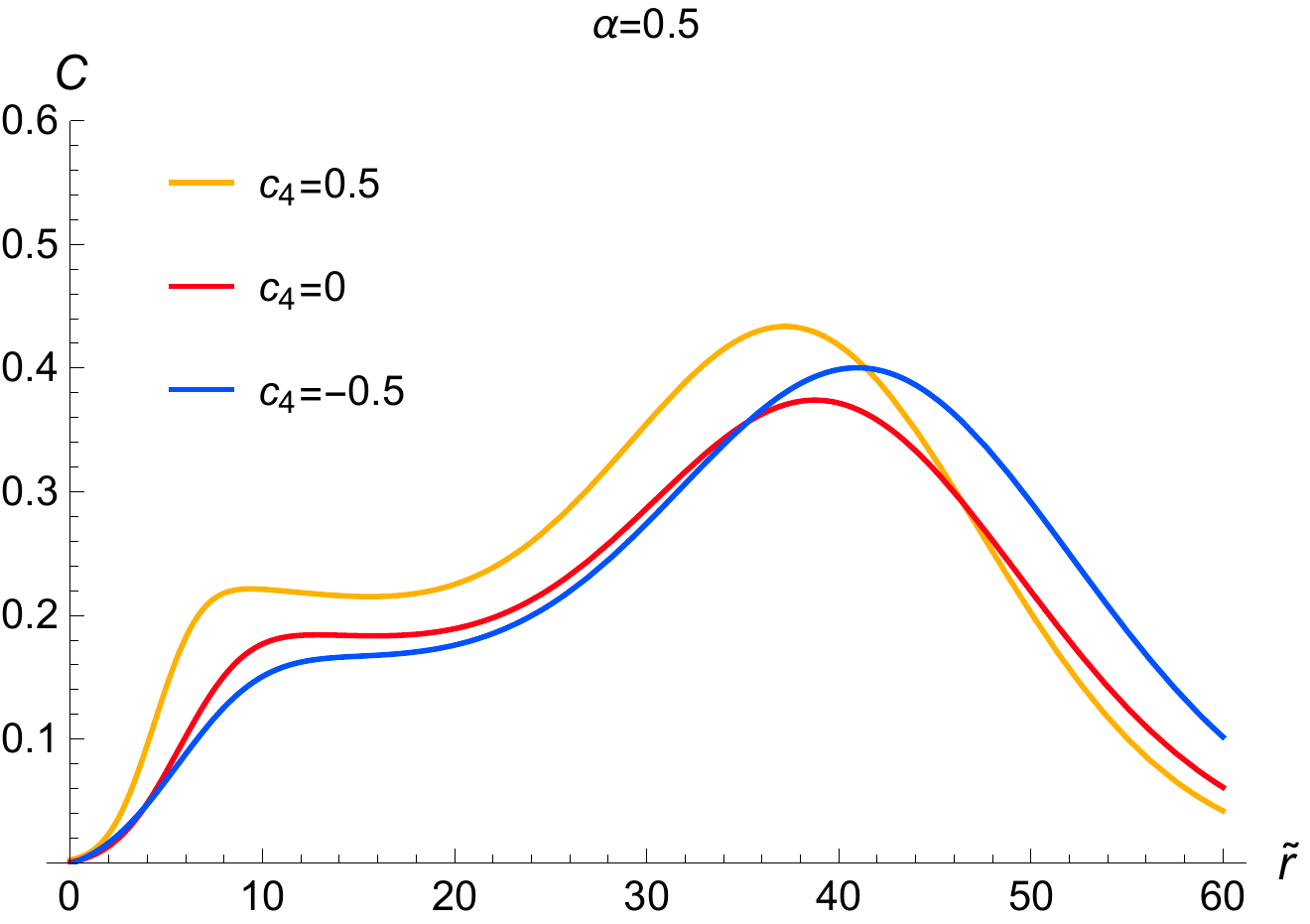}
 \caption{\footnotesize {The ultrarelativistic degenerate Fermi gas case, $\Upgamma=\frac43$.}}
 \label{33}
 \end{subfigure}
	\caption{\footnotesize {Compactness of gravitational decoupled hybrid stars as a function of the radial coordinate, for $\upalpha=0.5$ and three values of $c_4$, and $\upsigma_0=0.26M_{\scalebox{0.66}{$\textsc{p}$}}$. The orange [red; blue] curve represents the case $c_4=0.5$ [$c_4=-0.5; c_4=0$], for $\Uplambda= (M_{\scalebox{0.66}{$\textsc{p}$}}m^{2})^{1/3}$, $K=10^2\, m^{-2}M_{\scalebox{0.66}{$\textsc{p}$}}^{-2}$, and $\upepsilon_0=10^{-2}\, (M_{{\scalebox{0.66}{$\textsc{p}$}}}m)^{2}$. Fig. \ref{31} shows the case $\Upgamma=1$ and Fig. \ref{32} illustrates the case where $\Upgamma=2$, whereas Fig. \ref{33} shows the case $\Upgamma=\frac43$.}}\label{fig109}
\end{figure}

Now the compactness of gravitational decoupled hybrid stars 
can be addressed. The compactness $C$ of hybrid stars can be defined as the ratio between the mass portion $M_{99}$ and its effective radius, as $C = \frac{M_{99}}{R_{99}}$, emulating the gravitational
potential at the surface of the star\footnote{Remembering that relaxing the use of natural units, the compactness is defined as $C = \frac{GM_{99}}{c^2R_{99}}$}. Figs. \ref{fig108} and \ref{fig109} show the compactness of gravitational decoupled hybrid stellar configurations as a function of the radial coordinate, respectively for $\upalpha=0.1$ and $\upalpha=0.5$, for different values of $c_4$ and the scalar field central density. 
For both the cases $\upalpha=0.1$ and $\upalpha=0.5$, Figs. \ref{fig108} and \ref{fig109} respectively show that for each fixed $c_4$, the higher the value of the adiabatic index $\Upgamma$, the higher the absolute maximum of the gravitational decoupled hybrid star compactness is. Also, the higher the value of $c_4$, the sharper the compactness increases along the radial coordinate, whereas the faster it decreases after the maxima, along the radial coordinate, irrespectively of the value of the adiabatic index.  

The differences in the compactness profiles for gravitational decoupled hybrid stars yield alterations in their several properties, with special importance to the gravitational radiation emitted by the particular case of gravitational decoupled neutron stars. The gravitational decoupled hybrid stars compactness is in general thrice larger than the neutron stars compactness and can be around four times higher than conventional boson stars.
Gravitational decoupled hybrid stars have been here described as solutions to Einstein's effective field equations, with scalar matter confined to the star effective radius and the fermionic fluid described by a perfect fluid. However, bosonic matter described by a scalar field is here implemented by a polytropic fluid, which \emph{a priori} might circumvent the Buchdahl bound. Even though gravitational decoupled hybrid stars evade Buchdahl's theorem, the compactness of gravitational decoupled hybrid stars in Figs. \ref{fig108} and \ref{fig109} have values that are lower than the Buchdahl bound ${M<{\frac {4Rc^{2}}{9G}}}$ or, equivalently in natural units, $C<\frac49$.

\section{Conclusions and perspectives}\label{cppp}
Gravitational decoupled hybrid stars were scrutinized and described by anisotropic polytropic stellar configurations, that are self-gravitating bound regular stellar structures constituted by scalar bosons and fermionic matter. Gravitational decoupled hybrid stars were studied in a low-energy effective theory of the Gleyzes--Langlois--Piazza--Vernizzi theory, generalizing Horndeski scalar-tensor gravity with infrared modifications of the gravitational sector, coupled to a scalar field that comprises bosonic matter. 
As solutions of the resulting equations of motion, the three most important cases, involving realistic choices of the adiabatic index in the polytropic Lane--Emden equation of state, were analyzed. Each one of these cases encloses several gravitational decoupled astrophysical compact distributions that include: a) isothermal self-gravitating spheres of gas that encompasses collisionless systems of compact stellar configurations in globular clusters; b) gravitational decoupled neutron stars, and c) gravitational decoupled white dwarfs and ultrarelativistic degenerate Fermi gases. Each case was further considered in appropriate limits, also encompassing gravitational decoupled boson stars and solutions of the Einstein--Klein--Gordon system 
using the gravitational decoupling method. Several properties of the scalar field that generates hybrid stars were addressed and discussed. Also, the way the decoupling parameter affects the fermionic pressure field and the asymptotic values of the Misner--Sharp--Hernandez mass function was scrutinized and comprehensively presented in Sec. \ref{220}. The decoupling parameter was shown to make a remarkable difference in the high compactness regime of gravitational decoupled hybrid stars. It provides the possibility of more compact and more massive self-gravitating compact hybrid stars when compared to the general-relativistic case, for 
the same polytropic indexes, central pressure, and density.
The effective radius of gravitational decoupled compact hybrid stars was and its relationship to hybrid star masses was discussed, for two subcases involving a bosonic and a fermionic dominant core. 
Although some results regarding hybrid stars are qualitatively analogous to the ones obtained heretofore, the important results obtained here in the context of the gravitational decoupling devise new possibilities. The synergy among the adiabatic index, the Horndeski parameter, and the decoupling parameter has produced compelling physical results, weaving new features of hybrid stars. It also yields more realistic predictions for the emission of gravitational wave radiation, as the asymptotic value of the mass function increases at a higher rate than 
their general-relativistic counterparts. 

Figs. \ref{fig108} and \ref{fig109} represent the numerical computations of the compactness of gravitational decoupled hybrid stellar configurations, respectively with a bosonic and a fermionic dominant core. The gravitational decoupled setup emulates its general-relativistic limit of the generalized Horndeski scalar-tensor theory \cite{Roque:2021lvr}. The compactness of gravitational decoupled hybrid stars, along the radial coordinate, presents absolute minima and maxima, for each value of the adiabatic index and each value of both the Horndeski parameter and the decoupling parameter, in the range here studied. This feature differs from neutron and boson stars in general relativity and can be probed by astrophysical observations and detection of gravitational waves from realistic mergers of gravitational decoupled hybrid stars, as well as the subcase of gravitational decoupled neutron stars. 
 Gravitational decoupled hybrid stars were shown to present a variation in their compactness, that is compatible with the emission of gravitational radiation with more energy than
the corresponding black hole system. When regarding a coalescent binary constituted by gravitational decoupled neutron stars or gravitational decoupled hybrid stars, they can radiate more energy than black
hole mergers \cite{Hanna:2016uhs}, since the compactness reaches higher values when compared to Horndeski hybrid stars without gravitational decoupling. 
Figs. \ref{fig108} and \ref{fig109} respectively showed that for each fixed $c_4$, the higher the value of the adiabatic index $\Upgamma$, the higher the absolute maximum of the gravitational decoupled hybrid star compactness that is. Also, the higher the value of $c_4$, the sharper the compactness increases, whereas the faster it decreases after the maxima, along the radial coordinate, irrespectively of the value of the adiabatic index.

Gravitational decoupled hybrid stars were successfully studied in a gravitational equilibrium state, wherein the scalar field interacts with the fermionic field only by the gravitational force. One can further consider a self-interacting potential in Eq. (\ref{lagra}), and initial numerical results corroborate to a significant increment of the mass of gravitational decoupled hybrid stars, that can even surpass the order
of magnitude of general-relativistic neutron stars masses, for appropriate ranges of the decoupling and the Horndeski parameter. These results have been known for boson stars \cite{Burikham:2016cwz} and, although still an incipient numerical indication, one may continue to study extensions of the results presented heretofore. The setup considered here focused on gravitational decoupled compact hybrid stellar distributions generated by scalar fields, constituting boson stars that can interact with surrounding fermionic matter. An alternative procedure for the investigation of hybrid stars consists of considering the formation of fermionic stars by a dynamical mechanism, which is then encompassed by a bosonic cloud. This case was considered in Ref. \cite{DiGiovanni:2020frc}, discussing a cloud accretion of a self-interacting massive scalar field, and can be also studied in the framework of gravitational decoupling. Finally, recent results establish a bound on compactness due to quantum effects \cite{Casadio:2021cbv} and the results for the compactness of gravitational decoupled hybrid stellar configurations can be also studied in the context of quantum corrections.

\subsection*{Acknowledgements}
RdR thanks grants No. 2017/18897-8 and No. 2021/01089-1, São Paulo Research Foundation (FAPESP); and grants No. 303390/2019-0, No. 406134/2018-9, and No. 402535/2021-9,  National Council for Scientific and Technological Development -- CNPq, for partial financial support.
\bibliography{bib_DSS}

\end{document}